%% file: softF.tex
\documentclass[a4paper,11pt]{article}
\pdfoutput=1
\usepackage{jheppub}
\usepackage{amsmath}
\usepackage{amssymb}
\usepackage{color}
\usepackage{graphicx}
\usepackage{hyperref}
\usepackage{xspace,slashed}
\usepackage[utf8]{inputenc}
\usepackage{subcaption}
\usepackage{multirow}
\usepackage{algorithm}
\usepackage{algorithmic}
\usepackage{stackengine}
\usepackage{enumitem}
\usepackage{pdflscape}
\allowdisplaybreaks

\newcommand{\ep}{\epsilon}

\usepackage{soul}

\usepackage{booktabs}

\makeatletter
\g@addto@macro\bfseries{\boldmath}
\makeatother

\newcommand{\be}{\begin{equation}}
\newcommand{\ee}{\end{equation}}

\newcommand{\per}{\,}
\newcommand{\oS}{{\overline {\cal S}}}
\newcommand{\pp}[2]{p_{#1}\hspace{-0.ex} p_{#2}}
\newcommand{\pq}[2]{p_{#1}\hspace{-0.ex} q_{#2}}
\newcommand{\qq}[2]{q_{#1}\hspace{-0.ex} q_{#2}}
\newcommand{\dt}{\bigg)} 
\newcommand{\sq}{\bigg[} 
\newcommand{\dq}{\bigg]} 
\newcommand{\sg}{\bigg\{} 
\newcommand{\dg}{\bigg\}} 
\newcommand{\pst}{\Big(} 
\newcommand{\pdt}{\Big)} 
\newcommand{\psq}{\Big[} 
\newcommand{\pdq}{\Big]} 
\newcommand{\psg}{\Big\{} 
\newcommand{\pdg}{\Big\}} 

\usepackage{xcolor}
\definecolor{light-gray}{gray}{0.8}

\definecolor{azure}{rgb}{0.0, 0.5, 1.0}

\definecolor{semiblue}{rgb}{0.3,0.3,0.8}
\newcommand{\logbook}[2]{}

\newcommand{\TTPaff}{Institute for Theoretical Particle Physics,
  KIT, 76128 Karlsruhe, Germany}
\newcommand{\TUaff}{Physik Department, Technische Universität M{\"u}nchen, James-Franck-Stra{\ss}e 1, 85748 Garching, Germany}

\preprint{
  \begin{flushright}
    TTP-21-053\\  P3H-21-097\\ TUM-HEP-1376/21
  \end{flushright}
}

\title{
  On phase-space integrals with Heaviside functions
}
\author[a]{Daniel Baranowski,}
\author[a,b]{Maximilian Delto,}
\author[a]{Kirill Melnikov,}
\author[a]{Chen-Yu Wang}
\affiliation[a]{\TTPaff}
\affiliation[b]{\TUaff}

\abstract{We  discuss peculiarities that arise in the computation of   real-emission contributions
  to observables that contain Heaviside functions. A prominent example of such a case is 
   the zero-jettiness soft function in SCET, whose calculation at 
   next-to-next-to-next-to-leading order in perturbative QCD is an interesting problem.
   Since the  zero-jettiness soft function distinguishes between emissions into different hemispheres, 
   its definition  involves $\theta$-functions of light-cone components of emitted soft partons. This 
   prevents   a direct use of multi-loop methods,  based on  reverse unitarity, for computing
      the zero-jettiness soft function in high orders of perturbation theory.  We propose a  way to bypass this problem 
 and illustrate its effectiveness   by computing  various non-trivial  contrbutions to
 the zero-jettiness soft function
  at NNLO and N3LO in perturbative QCD. 
 }

 \input{misc/makros.tex}

\begin{document}

\maketitle 
\input{sections/01_intro.tex}
\input{sections/02_definitions_new.tex}
\input{sections/03_ibp.tex}
\input{sections/04_N2LO.tex}
\input{sections/05_N3LO_1.tex}
\input{sections/05_N3LO_2_a.tex}
\input{sections/05_N3LO_3_b.tex}
\input{sections/05_N3LO_4_c.tex}
\input{sections/05_N3LO_5_d.tex}

\input{sections/06_conclusion.tex}

\appendix
\input{sections/07_appendix.tex}

\bibliographystyle{JHEP}
\bibliography{misc/bib}


\end{document}

%% file: misc/makros.tex
\newcommand{\VEC}[1]{\boldsymbol{#1}}
\newcommand{\SUM}[2]{\sum\limits_{#1}^{#2}}

\newcommand{\INT}[2]{\int_{#1}^{#2}}
\newcommand{\FEYNINT}[1]{\INT{0}{1}\DIFFL #1}
\newcommand{\DIFFL}{\mathop{}\!\mathrm{d}}

\newcommand{\DIFFSPnopi}[1]{\mathop{}\!\mathrm{d}\Omega^{(#1)}}
\newcommand{\DIFFSPinopi}[2]{\mathop{}\!\mathrm{d}\Omega_{#2}^{(#1)}}

\newcommand{\DD}[1]{\mathop{}\!\delta\!\left(#1\right)}

\newcommand{\HT}[1]{\mathop{}\!\theta\!\left(#1\right)}

\newcommand{\GF}[1]{\mathop{}\!\Gamma\!\left(#1\right)}
\newcommand{\GFP}[2]{\mathop{}\!\Gamma^#1\!\left(#2\right)}
\newcommand{\GENHYPGF}[5]{\mathop{}\!_{#1}F_{#2}\left[\{#3\},\{#4\};#5\right]}
\newcommand{\HYPGF}[4]{\GENHYPGF{2}{1}{#1,#2}{#3}{#4}}


%% file: sections/01_intro.tex
\section{Introduction}
\label{sec:intro}

Precision studies of hadron collisions moved into the focus of the particle physics community after no clear evidence
for physics beyond the Standard Model has been  found during the first two LHC runs.
A pre-requisite for such studies
is a solid theoretical framework that allows one to describe hadron
collisions using quark and gluon degrees
of freedom and,  in this way,  connect experimental data to the
Standard Model Lagrangian without the  need for additional modeling.  The current theoretical
framework is based on the concept of collinear factorization~\cite{Collins:1989gx}  that, for processes with large
momentum transfer, relates hadronic  cross sections to convolutions of partonic  cross sections,
computable  in perturbation theory, with universal non-perturbative parton distribution functions.
Further refinements and practical advancements  of such  a  framework are  currently among  the central   topics in theoretical collider physics.

Improvements in  an existent framework may be provided by the  soft-collinear effective
theory (SCET)~\cite{Bauer:2000ew,Bauer:2000yr,Bauer:2001ct,Bauer:2001yt,Bauer:2002nz} that seeks to
establish a general pattern of factorization in collider processes  including both perturbative
and non-perturbative physics.  This effective theory
defines objects  that are sensitive to particular momenta ``modes'' such
as e.g. soft, collinear etc.  These objects can be calculated independently of
each other and then combined   to provide predictions
for physical quantities such as cross sections and kinematic distributions.
Key to this effort  are factorization theorems that both define these objects precisely and  also
provide information on how physical predictions should be assembled once these objects have
been computed. 

There are four  types of objects  that appear in SCET; they  are known as hard, beam, jet and soft functions. 
Theoretical predictions for these functions
are of interest since they can be used to re-sum logarithmically-enhanced terms that
appear in perturbative expansion in QCD. In addition, they  can also be used
to set up a slicing  method for deriving  fixed-order predictions for fully-differential calculations  in QCD~\cite{Catani:2007vq,Bonciani:2015sha,Grazzini:2017mhc,Catani:2019iny,Catani:2019hip,Kallweit:2020gcp,Catani:2020kkl,Catani:2021cbl,Boughezal:2015dva,Boughezal:2015aha,Gaunt:2015pea,Boughezal:2016wmq}. 

The slicing method requires an observable  which can be used to separate  the real-emission phase spaces into ``singular'' and
``regular'' parts.  Although, in principle, any observable can be used to do that, if an
observable is chosen in a way that does not violate factorization into collinear and soft modes,
a cross section, differential with respect to  such an observable, should satisfy a particular factorization theorem.
Such a factorization theorem  would then contain soft and beam functions whose computation
enables both, the construction of subtraction terms for perturbative computations and the re-summation of large logarithms
that arise in theoretical predictions once the slicing variable  becomes small. 

In this paper we will deal with the so called zero-jettiness variable
\cite{Stewart:2009yx,Stewart:2010tn} that can be used as a slicing
variable for processes where a color-less final state is produced in hadron collisions. 
Calculations of the corresponding soft and beam functions
have been performed during the past decade.
The zero-jettiness soft function has been computed through NNLO QCD in
Refs.~\cite{Monni:2011gb,Kelley:2011ng} (see also
 Ref.~\cite{Hornig:2011iu,Baranowski:2020xlp}). The  zero-jettiness beam function has been calculated
 through NNLO QCD in Refs.~\cite{Gaunt:2014xga,Gaunt:2014cfa,Boughezal:2017tdd}.
 Studies of zero-jettiness   beam functions at N3LO QCD  were initiated in Refs.~\cite{Baranowski:2020xlp,Melnikov:2018jxb,Melnikov:2019pdm,Behring:2019quf} and were recently completed in Refs.~\cite{Ebert:2020lxs}.\footnote{Jettiness soft functions for more complicated final states were studied in Refs.~\cite{Boughezal:2015eha,Li:2016tvb,Alioli:2021ggd,Campbell:2017hsw}.
 Beam functions  for $q_T$ factorization were calculated in~Refs.~\cite{Li:2016ctv,Angeles-Martinez:2018mqh,Luo:2019szz}.
 Automated approaches to computation of soft, beam and jet functions are discussed   in Refs.~\cite{Bell:2018oqa,Basdew-Sharma:2020wva}.}
 
In this paper, we  focus on certain technical aspects that arise in  the computation
of the zero-jettiness soft function in higher orders of perturbative QCD. 
 Due to algebraic 
complexity of  computations in high orders of perturbation theory,  specialized 
tools and methods are usually employed.
Chief among them is the integration-by-parts (IBP) method for loop integrals introduced  in Ref.~\cite{Chetyrkin:1981qh} and adapted to
real-emission integrals in Ref.~\cite{Anastasiou:2002yz}. However, application of 
the IBP method  to the computation of the zero-jettiness soft function is not straightforward, as the
zero-jettiness observable contains Heaviside functions that depend on the light-cone components of four-momenta
of the emitted partons.\footnote{We note that Heaviside functions can also appear in the construction
  of NNLO subtraction terms that need to be integrated over unresolved phase spaces of final state particles,
  see e.g. Refs.~\cite{Caola:2018pxp,Delto:2019asp,Bizon:2020tzr}.}
  This fact, as well
as the need to compute large number of complicated integrals,
makes the calculation of N3LO soft function non-trivial.
Our goal in this paper is to discuss possible ways to overcome these technical difficulties paving the
way  for  the computation of the triple-real  and real-virtual  contributions to the N3LO soft function.
As a proof of concept,  we  compute a number of  non-trivial contributions to the N3LO 
soft function, that describe emission of three gluons into the same hemisphere. 

The remainder of this paper is organized as follows. In Section~\ref{sec:defs} we introduce the zero-jettiness soft function.
In Section~\ref{sec:mod_ibp} we explain  how integration by parts  can be used to simplify computation of 
integrals with Heaviside functions. In Section~\ref{sec:n2lo}, we apply this method 
to compute the
soft function at next-to-next-to-leading order in QCD. We then discuss
various aspects of the N3LO calculation in Section~\ref{sec:n3lo}.
We conclude in Section~\ref{sec:conclusion}.  Some useful formulas are collected in  Appendices \ref{app:master_integrals} and \ref{app:deq}.

%% file: sections/02_definitions_new.tex
\section{Heaviside functions in the zero-jettiness soft function}
\label{sec:defs}

In this section we will  briefly discuss the zero-jettiness soft function to explain how 
Heaviside functions appear in phase-space integrals. Consider a process where two partons
with (normalized) momenta $n$ and $\bar n$ collide and produce a color-neutral final state together with
$m$ 
QCD partons with momenta $k_{1,2,.,m}$.    The jettiness variable ${\cal T}$ 
is  defined  as follows
\be
   {\cal T} = \sum \limits_{j=1}^{m} {\rm min}_{q \in \{ n,\bar n \}} \left [ \frac{2 q k_j}{ n \bar n }  \right ].
   \label{eq2.2}
\ee
To compute the minima in Eq.~(\ref{eq2.2}), we 
use the Sudakov parameterization  of the parton's  momenta and write 
\be
k_i= \frac{\alpha_i}{2} n + \frac{\beta_i}{2} \bar n + k_{\perp,i}, \;\;\; i=1,\dots,m,
\label{eq2.3}
\ee
where $k_{\perp,i} n = k_{\perp,i} \bar n = 0$, $n^2 = {\bar n}^2 = 0$ and $n \bar n = 2$.  
It follows that 
\be
{\cal T} = \sum \limits_{i=1}^{m} {\rm min} \left \{ \alpha_i,\beta_i \right \}.
\ee
To enable the
choice between $\alpha_i$ and $\beta_i$ in the above equation,  we can  partition the phase space by writing 
\be
1 = \theta(\alpha_i - \beta_i) + \theta(\beta_i - \alpha_i),
\label{eq2.5}
\ee
for each of the $m$ partons.\footnote{In what follows, we will focus on gluon emissions.}
It is clear  that  the first term will contribute $\beta_i$  and the second one $\alpha_i$ to
the jettiness variable ${\cal T}$.

Inserting the partition Eq.~(\ref{eq2.5}) into integrals over momenta of final-state gluons  we find
phase-space integrals with Heaviside functions. In general, for $m$-emitted gluons 
 we can obtain $2^m$ different terms however, using 
the symmetry between  $m$ gluons and the fact that the result is invariant
under the simultaneous replacement of all $\alpha_i$'s with $\beta_i$'s and vice versa, the number of
different terms is dramatically reduced.  For example, for   $m \le 3$ which covers NLO, NNLO and N3LO
cases,  only two independent contributions to the soft-function need  to be considered.
They  are 1)
all $m$ gluons are emitted into the same hemisphere and  2)
all but one gluons are emitted into the
same hemisphere.

These different cases  can be described as integrals over the following phase spaces
\begin{align}
\label{eqn:ps_withf_general}
\mathrm{d} \Phi^{h_1,\dots,h_m}_{f_1,\dots,f_m} = (N_\ep)^{-m} \left( \prod \limits_{j=1}^{m} [{\rm d} k_j] f_j(\Delta_{j,h_j}) \right) \;  \delta (1 - \sum_{j=1}^{m} \kappa_{j,h_j})\,,
\end{align}
where
\be
N_\ep = \frac{(4\pi)^{-\ep}}{16 \pi^2 \Gamma(1-\ep)},
\ee
and 
\begin{align}
\Delta_{j,h_j} ={} &
\begin{cases} 
\alpha_j-\beta_j \,, & h_j = n \,, \\
\beta_j-\alpha_j \,, & h_j = \bar{n} \,,
\end{cases} 
\;\;\;\;\;\; {\rm and}\;\;\;\;\;\; \kappa_{j,h_j} ={} 
\begin{cases} 
\beta_j \,, & h_j = n \,, \\
\alpha_j \,, & h_j = \bar{n} \,.
\end{cases}
\end{align}
Thanks to the definition of the jettiness variable, we associate 
 functions $f_j$  in Eq.~(\ref{eqn:ps_withf_general}) with Heaviside functions but, as we will see later,
we will also need to consider cases where one or several of these functions $f$ are $\delta$-functions. 

To explicitly see how these phase spaces are used, we consider an  example of the real-emission contribution
to the soft function. Then, for NNLO and N3LO computations we require the following integrals
\begin{align}
\begin{split} 
  & S_{nn} =
    \int \mathrm{d} \Phi^{nn}_{\theta\theta} \; {\rm Eik}_{2g}(\{k_i\},n,\bar n), ~~~ S_{n\bar n} =
     \int  \mathrm{d} \Phi^{n\bar{n}}_{\theta\theta}  \; {\rm Eik}_{2g}(\{k_i\},n,\bar n),
\end{split} 
\label{eqn:n2lo_hemi_def}
\end{align}
and
\begin{align}
\begin{split} 
  S_{nnn} =
  \int \mathrm{d} \Phi^{nnn}_{\theta\theta\theta}   \; {\rm Eik}_{3g}(\{k_i\},n,\bar n), ~~~
  S_{nn\bar n} = \int \mathrm{d} \Phi^{nn\bar{n}}_{\theta\theta\theta}  \; {\rm Eik}_{3g}(\{k_i\},n,\bar n),
\end{split} 
\label{eqn:n3lo_hemi_def}
\end{align}
where ${\rm Eik}_{3g}(\{k_i\}$ and ${\rm Eik}_{2g}(\{k_i\}$ are properly rescaled three-gluon and
two-gluon  eikonal functions. We will specify these functions later. For now, suffice  it to say
that they depend on the scalar products of gluon four-momenta $k_i k_j$ and on scalar products of gluon four-momenta
with external vectors  $n$ and $\bar n$. 

A standard way to simplify computation of complicated phase-space integrals is
to use reverse unitarity~\cite{Anastasiou:2002yz}
to map such integrals onto cut loop integrals for which IBP identities can be derived in a straightforward manner.
This is achieved by using the formula
\be
\delta(P(\vec x)) \to \frac{i}{2\pi} \left[ \frac{1}{P(\vec x)+i0} - \frac{1}{P(\vec x)-i0}  \right ],
\label{eq2.9new}
\ee
where $P(\vec x)$ is a polynomial in variables $\vec x$ that can be e.g   certain components of gluons' momenta.
Once all $\delta$-function constraints in phase-space integrals are removed using Eq.~(\ref{eq2.9new}) and,
provided, that there are no other non-polynomial constraints in the integrands, one can make use of the
powerful  integration-by-parts technology~\cite{Chetyrkin:1981qh} to reduce computation of a large number
of phase-space integrals to a few master integrals. 

Unfortunately, if integrands contain  Heaviside functions, this approach fails since the last condition
mentioned in the previous paragraph is not  fulfilled.  A possible solution to this problem was pointed out 
by one of us in  Ref.~\cite{Baranowski:2020xlp}, where it was suggested to rewrite all $\theta$-functions
that appear in relevant integrals as
follows\footnote{See also Section 4.2.2 in Ref.~\cite{Angeles-Martinez:2018mqh}.}
\begin{align}
\label{eqn:theta_as_delta}
\theta (b_i-a_i) = \int_{0}^{1} \mathrm{d} z_i \, \delta(z_i \, b_i -a_i ) \, b_i \,,  ~~~ a_i,b_i >0 \,.
\end{align}
While this representation yields an integrand whose dependence on  auxiliary variables $z_i$ 
can be computed using reverse unitarity, it also
introduces one additional parametric integral per $\theta$-function, which can become quite cumbersome.
For this reason, in this paper  we would like to investigate how to derive and use
IBP relations for phase-space integrals with $\theta$-functions \textit{directly}, i.e.~without the need to introduce additional variables.

%% file: sections/03_ibp.tex
\section{Applying  integration-by-parts technology to   integrals with $\theta$-functions}
\label{sec:mod_ibp}
  
The goal of reverse unitarity~\cite{Anastasiou:2002yz} is to turn phase-space integrals into loop integrals.
We explained the main idea of the method in the previous section; we will now make this discussion more
specific considering the zero-jettiness  soft function. 

To remove all $\delta$-function constraints from the integration measure, 
we  start with a phase-space element of a gluon $i$  with momentum $k^\mu_i = (E_i,\vec k_i)$ and write it as
  \be
     [{\rm d} k_i] = \frac{ {\rm d}^{d-1} \vec k_i}{(2 \pi)^{d-1} 2 E_{i}} =
     \frac{ {\rm d}^d k_i}{(2\pi)^d} \; 2\pi \; \delta_+(k_i^2).
  \ee
We then re-write the $\delta$-function as in Eq.~(\ref{eq2.9new})
  \be
   \delta(k_i^2) =  \frac{i}{2\pi} \left ( \frac{1}{k_i^2 +  i0}  - \frac{1}{k_i^2 -  i0} \right )
   =  \frac{1}{[k_i^2]_c}.
  \ee
  In addition, to deal with soft functions  shown in   Eqs.~(\ref{eqn:n2lo_hemi_def},\ref{eqn:n3lo_hemi_def}) we
  need to  re-write all $\delta$-functions that define the re-scaled jettiness as  cut propagators.
  For example, in case of the $nn \bar n$  kinematic configuration, we write 
  \be
  \delta(1 - k_{12} n - k_3 \bar n  ) = \frac{1}{[1-k_{12}n-k_3 \bar n]_c}.
   \label{eq:map}
  \ee
  where $k_{12} = k_1 + k_2$.

  Hence, if we ignore $\theta$-functions in the integrands of 
  $S_{nn},S_{n \bar n}, S_{nnn}, S_{nn\bar n}$,
  we  immediately recognize that we need to compute a collection of
  ``loop'' integrals with conventional and
  unconventional cut ``propagators''. To do that, 
  we can apply integration-by-parts identities~\cite{Chetyrkin:1981qh}
  to reduce the number of independent
  integrals that need to be calculated.
  Furthermore,  there are powerful public
  programs such as \texttt{Fire}~\cite{Smirnov:2008iw,Smirnov:2019qkx}, \texttt{Kira}~\cite{Maierhoefer:2017hyi,Klappert:2020nbg}, \texttt{LiteRed}~\cite{Lee:2012cn,Lee:2013mka}, and \texttt{Reduze}~\cite{Studerus:2009ye,vonManteuffel:2012np} that can perform  reductions to master integrals  in a highly  automated and efficient fashion. 

  However, if relevant integrands   contain  a collection
  of $\theta$-functions, as  is indeed the case for the soft function, this procedure can not be applied.
  An obvious problem is
  that $\theta$-functions cannot be turned into ``propagators'' since  the mapping similar to the one shown 
  in Eq.~(\ref{eq:map}) does  not exist.   However, we  would like to understand what happens if we ignore
  this problem   and attempt to derive 
  IBP identities
  for integrands with  $\theta$-functions. 

  To study this question, we consider the following integral
  \be
I[\theta(f),g] = \int {\rm d}^d k  \; \theta(f(k)) \;  g(k), 
  \ee
  where $f(k)$ is a polynomial in   momentum
  $k$
  and $g(k)$ is a function that allows a standard derivation
  of integration-by-parts identities.\footnote{In case of the soft functions, $f(k)$  reads $f(k) = \pm(k \bar{n}- k n) $.} To derive integration-by-parts identities for the integral $I[\theta(f),g]$, we write the standard equation 
  \be
  0 = \int {\rm d}^d k  \frac{\partial }{\partial k^\mu } \left [ v^\mu \theta(f(k)) \; g(k) \right  ],
  \label{eq3.5}
\ee
that is
valid for dimensionally-regularized integrals.  Vector $v^\mu$ in Eq.~(\ref{eq3.5}) is an arbitrary 
vector that
we do not need to specify further. 
Calculating the derivative, we obtain
  \be
  \frac{\partial }{\partial k^\mu } \big \{ v^\mu \HT{f(k)}  g(k) \big \}
   = \HT{f(k)}  \frac{\partial }{\partial k^\mu } \big \{ v^\mu g(k) \big \} + \big \{ g(k) \DD{f(k)} v^\mu \big \} \frac{\partial f(k) }{ \partial k_\mu }.
  \ee
  It follows that 
  \be
  0 = I[\theta(f),  \partial_\mu (v^\mu g) ] + I[\delta(f), g v^\mu (\partial_\mu f)  ].
  \label{eq3.7}
  \ee
  The first term on the right-hand side
  belongs to the same class of integrals as the original one $I[\theta(f), g]$ because 
  it involves the same $\theta$-function; we call this term the \textit{homogeneous} part of the IBP relation.
  Since, by  assumption, integrals of $g(k)$ can be studied using standard integration-by-parts technology,
  it follows that the homogeneous terms produce
  a closed set of linear equations when studied on their  own.
  
  The second term in Eq.~(\ref{eq3.7}) involves $\delta(f)$; we call this term the \textit{inhomogeneous} part of the IBP
  relation. Since we can use the generalized unitarity trick
  to write $\delta(f) \to 1/f$ and since $f$ is a polynomial in $k$,  $I[\delta(f), g v^\mu \partial_\mu f  ]$ defines
  a class of integrals that can be studied on their own {\it independent} of  integrals with  $\theta$-functions.
  In fact, obtaining integration-by-parts identities for this class of integrals can be done with    standard methods.
  The only subtlety that we have to deal with
  when working with inhomogeneous terms is that the function $g(k)/f(k)$ may contain linearly-dependent ``propagators'' that
  will have to be re-mapped onto properly-defined integral families.  Although this, by itself, is not a crucial issue, it does
  not allow   us to derive  IBP relations for integrals with arbitrary powers of propagators and forces us to produce
  IBP relations for each of the seed integrals individually.

  We thus conclude that it is possible to establish useful integration-by-parts identities for integrals with multiple $\theta$-functions by iteratively using Eq.~(\ref{eq3.7}). It follows from that
  equation that the derivative of an integrand produces inhomogeneous terms,
  where a  $\theta$-function is replaced by a $\delta$-function.
    
  Therefore, by using Eq.~(\ref{eq3.7}) repeatedly,
  we obtain  a hierarchical sequence of IBP relations containing integrals with a decreasing number of $\theta$-functions
  and an increasing number of $\delta$-functions. The  IBP relations
   can be used to express all relevant  integrals through a set of master integrals.
  When choosing  master integrals, we try to select those that contain 
  fewer $\theta$-functions since they are easier to compute. 
  We will illustrate the construction of IBP relations and their usage 
  in
  the next section where we will calculate  the real-emission contribution to the
  zero-jettiness soft function at NNLO.

%% file: sections/04_N2LO.tex
\section{IBP identities and the  NNLO QCD contribution  to the zero-jettiness soft function}
\label{sec:n2lo}

In this Section, we show how to use reverse unitarity and
modified IBP relations to compute  the maximally non-abelian contribution to zero-jettiness
soft function at NNLO.  We define it as  
\begin{align}
\label{eqn:S2g_mnab_def}
S^{2g}_{\rm NA} = \frac{1}{\tau^{1+4\ep}} \left[ \int \mathrm{d}\Phi^{nn}_{\theta\theta} \omega^{(2)}_{n\bar{n}}(k_1,k_2) + \int \mathrm{d}\Phi^{n\bar{n}}_{\theta\theta} \omega^{(2)}_{n\bar{n}}(k_1,k_2) \right] \,,
\end{align}
where~\cite{Catani:1999ss}
\begin{align}
\omega^{(2)}_{n\bar{n}}(k_1,k_2) = S_{n\bar{n}}(k_1,k_2) +  S_{\bar{n}n}(k_1,k_2) - S_{nn}(k_1,k_2) - S_{\bar{n}\bar{n}}(k_1,k_2) \,,
\end{align}
with
\begin{align}
\begin{split}
    \label{eqn:ds_gg_nab}
    & \mathcal{S}_{p_i p_j}(k_1,k_2)  =   \frac{(1-\ep)}{(k_1\cdot k_2)^2} \frac{\left[(p_i\cdot k_1)(p_j\cdot k_2) + i\leftrightarrow j \right]}{(p_i\cdot k_{12})(p_j\cdot k_{12})} \\
    & - \frac{(p_i\cdot p_j)^2}{2(p_i\cdot k_1)(p_j\cdot k_2)(p_i\cdot k_2)(p_j\cdot k_1)} \bigg[ 2 - \frac{\left[(p_i\cdot k_1)(p_j\cdot k_2) + i\leftrightarrow j \right]}{(p_i\cdot k_{12})(p_j\cdot k_{12})} \bigg] \\
    & + \frac{(p_i\cdot p_j)}{2(k_1\cdot k_2) } \bigg[ \frac{2}{(p_i\cdot k_1)(p_j\cdot k_2)} + \frac{2}{(p_j\cdot k_1)(p_i\cdot k_2)} - \frac1{(p_i\cdot k_{12})(p_j\cdot k_{12})}  \\
    & ~~ \times \left( 4 + \frac{\left[(p_i\cdot k_1)(p_j\cdot k_2) + i\leftrightarrow j \right]^2}{(p_i\cdot k_1)(p_j\cdot k_2)(p_i\cdot k_2)(p_j\cdot k_1)} \right) \bigg] \,.
    \end{split}
\end{align}

In the next section we  explicitly construct a  few examples of IBP
equations  with $\theta$-functions that are relevant for this case.
We   discuss the computation of the two terms
in Eq.~(\ref{eqn:S2g_mnab_def}) after that. 

\subsection{An example of an  IBP relation}

In this section, we  explain how to employ modified integration-by-parts identities
discussed in Section~\ref{sec:mod_ibp}. As the first step, we map all integrals that appear in 
Eq.~(\ref{eqn:S2g_mnab_def}) onto integral families.  These families 
are defined by sets  of linearly-independent
propagators\footnote{When referring to  propagators, we imply both 
   cut and ordinary ones. In  case of the real-emission contribution to NNLO soft function, we
  have three cut propagators and four ordinary ones.}  and \textit{additionally} contain
 two $\theta$-functions from phase-space measures in Eq.~(\ref{eqn:S2g_mnab_def}).
 Cut propagators are constructed from $\delta$-functions  $\delta(k_i^2)$
 that enforce the on-shell conditions for  the emitted gluons,
and also from  jettiness-dependent $\delta$-functions that appear in the corresponding  phase spaces
\begin{align}
\mathrm{d}\Phi^{nn}_{\theta\theta} &  \sim \delta(1-\beta_1-\beta_2  ) = \delta(1-k_1 n - k_2 n ) \,, \\
\mathrm{d}\Phi^{n\bar{n}}_{\theta\theta}  & \sim \delta(1-\beta_1-\alpha_2 ) = \delta(1-k_1 n - k_2 \bar{n} ).
\end{align}

After partial fractioning, we find that we need  several independent integral families in this case. 
For example, one integral family  that is required  to describe the $nn$-configuration reads 
\begin{align}
\mathcal{T}^{\text{ex}}_{a_1\dots a_7} =  \int  \frac{{\rm d}^d k_1 {\rm d}^d k_2 \, \theta(k_1\bar{n} - k_1 n)\theta(k_2\bar{n} - k_2 n)}{\left[(k_1^2)^{a_1}(k_2^2)^{a_2}(1-k_{12}n )^{a_3} \right]_c (k_1 k_2)^{a_4} \,  (k_2 n )^{a_5} \, (k_1 \bar{n})^{a_6} \, (k_{12}\bar{n})^{a_7}} \,,
\label{eqn:n2lo_topo_ex}
\end{align}
where the subscript $c$ denotes cut propagators.

In order to construct  an explicit example of the modified IBP relations discussed in Section~\ref{sec:mod_ibp},  we consider the integral
\begin{align}
\mathcal{I}_\text{ex} =  \int  \frac{\mathrm{d}\Phi^{nn}_{\theta\theta}}{ (k_2 n ) \, (k_1 \bar{n}) \, (k_{12}\bar{n})} = \mathcal{T}^{\text{ex}}_{1,1,1,0,1,1,1} \,.
\end{align}

Starting with  this ``seed integral''
and following the discussion around  Eq.~(\ref{eq3.5}), we derive eight different equations
by computing derivatives  w.r.t.~$k_1$ and $k_2$ and by using  vectors $v \in \{k_1,k_2,n,\bar{n}\}$.
For example, differentiating  w.r.t.~$k_1$ and choosing $v=k_1$, we find
\begin{align}
\begin{split}
0 ={} & \int {\rm d}^d k_1 {\rm d}^d k_2 \,  \frac{\partial}{\partial k_1^\mu}   \frac{k_1^\mu \theta(k_1\bar{n} - k_1 n)\theta(k_2\bar{n} - k_2 n)}{(k_1^2)(k_2^2)(1-k_{12} n) \, (k_2 n ) \, (k_1 \bar{n}) \, (k_{12}\bar{n})} \\
= {} & (d-4) \mathcal{T}^{\text{ex}}_{1,1,1,0,1,1,1} - \mathcal{T}^{\text{ex}}_{1,1,1,0,1,0,2} -   \mathcal{T}^{\text{ex}}_{1,1,2,0,0,1,1} + \mathcal{T}^{\text{ex}}_{1,1,2,0,1,1,1} \\
& + \int {\rm d}^d k_1 {\rm d}^d k_2 \,    \frac{k_1^\mu (\bar{n}-n)_\mu \,
  \delta(k_1\bar{n} - k_1 n)\theta(k_2\bar{n} - k_2 n)}{(k_1^2)(k_2^2)(1-k_{12}n) (k_2 n ) \,  (k_1 \bar{n}) \, (k_{12}\bar{n})} \,.
\end{split}
\label{eqn:ibp_ex_dk1_vk1}
\end{align}
In writing this equation, we have used the fact that  homogeneous terms, which arise when the derivative
does not act on the $\theta$-function,  belong to the same family of integrals as the  seed integral.

As we already mentioned,  \textit{inhomogeneous} terms  arise when   derivatives
act   on the $\theta$-functions. In our example the last term in 
Eq.~(\ref{eqn:ibp_ex_dk1_vk1}) is inhomogeneous.
This term vanishes because  $k_1^\mu (\bar{n}-n)_\mu \, \delta(k_1\bar{n} - k_1 n)=0$.
This  is a general feature; indeed, we find that integration-by-parts identities that
involve a differential operator $k_i^\mu\partial/\partial k_{i,\mu}$
do  not produce inhomogeneous terms. To prove that assertion, consider 
\begin{align}
  \frac{\partial}{\partial k_i^\mu} \left[  k_i^\mu \theta(k_i^\mu q_\mu) g(k) \right]
  =  \theta(k_i^\mu q_\mu) \frac{\partial}{\partial k_i^\mu} \left[  k_i^\mu  g(k) \right]
  + \left[ g(k) \right] k_i^\mu q_\mu \DD{k_i^\mu q_\mu} \,.
\label{eqn:partder_dki_vki}
\end{align}
The second term on the right hand side
of Eq.~(\ref{eqn:partder_dki_vki}) vanishes which implies that differential
operators of the form $k_i^\mu\partial/\partial k_i^\mu$ do not produce inhomogeneous contributions.

As a second example, we consider the derivative w.r.t.~$k_1$ and use $v=n$. We find
\begin{align}
\begin{split}
0 ={} & \int {\rm d}^d k_1 {\rm d}^d k_2 \,  n^\mu  \frac{\partial}{\partial k_1^\mu}   \frac{ \theta(k_1\bar{n} - k_1 n)\theta(k_2\bar{n} - k_2 n)}{(k_1^2)(k_2^2)(1-k_{12} n)(k_1 \bar{n}) \, (k_2 n ) \, (k_{12}\bar{n})} \\
= {} & 2 \mathcal{T}^{\text{ex}}_{1,1,1,0,1,1,2} - 2 \mathcal{T}^{\text{ex}}_{1,1,1,0,1,2,1} + 2   \mathcal{T}^{\text{ex}}_{2,1,1,0,0,1,1} -2 \mathcal{T}^{\text{ex}}_{2,1,1,0,1,1,1} \\
& + (n \bar{n}) \int {\rm d}^d k_1 {\rm d}^d k_2 \,
\frac{\delta(k_1\bar{n} - k_1 n)\theta(k_2\bar{n} - k_2 n)}{(k_1^2)(k_2^2)(1-k_{12}n)(k_1 \bar{n}) \, (k_2 n ) \, (k_{12}\bar{n})} \,. 
\end{split}
\label{eqn:n2lo_tt_ex_2}
\end{align}
Similar to the previous case,  we have expressed homogeneous terms through integrals of the family
$\mathcal{T}^{\text{ex}}$. However, in this second example, the inhomogeneous term does not vanish and requires
further treatment in the IBP setup.

The reason for that is the fact that the propagators in that  term are linearly dependent due to the new
$\delta$-function in the integrand 
$\delta(k_1\bar{n} - k_1 n)$ that becomes a  rational function if reverse unitarity is used. 
To see this explicitly, we make the replacement $\delta(k_1\bar{n} - k_1 n) \rightarrow 1/[k_1 \bar n - k_1 n]_c $,  multiply
the result with the partial fractioning identity
\begin{align}
1 = \frac{(k_1\bar{n})-\left[k_1\bar{n}-k_1n\right]_c}{(k_1n)} \times \left\{ \left[ 1-k_{12} n \right]_c + (k_1n) + (k_2n) \right\}\,,
\end{align}
and obtain
\begin{align}
\begin{split}
 & \int   \frac{ {\rm d}^d k_1 {\rm d}^d k_2 \, \theta(k_2\bar{n} - k_2 n)}{\left[(k_1^2)(k_2^2)(1-k_{12} n)(k_1\bar{n} - k_1 n)\right]_c(k_1 \bar{n}) \, (k_2 n ) \, (k_{12}\bar{n})} \\
 & = \int  \frac{{\rm d}^d k_1 {\rm d}^d k_2 \, \theta(k_2\bar{n} - k_2 n)}{\left[(k_1^2)(k_2^2)(1-k_{12} n)(k_1\bar{n} - k_1 n)\right]_c(k_{12}\bar{n})} \left[ \frac1{(k_1 n) \, } +  \frac1{(k_2 n)} \right] \, .
\end{split}
\end{align}
We note that terms  that  do not contain the complete set of cut propagators were set to zero in
the above equation.  Furthermore, we note that it is the  partial fractioning step that prevents us from
writing  IBP relations for arbitrary powers of propagators, as it is usually done when traditional
IBP relations are derived for integrals \textit{without} $\theta$-functions. Because of this, we need to generate equations by selecting
seed integrals, deriving  IBP relations for them and explicitly mapping  all inhomogeneous terms to  simpler
integral families. We further elaborate on this point in the next section. 

In summary, we found that \textit{modified} IBP relations in the considered case
relate integrals with two  $\theta$-functions
to integrals  with one $\theta$-function.
Written as phase-space integrals, the modified IBP relations read 
\begin{align}
0 ={} & \SUM{i}{} a_i^{\text{hom}}(\ep) \int  \mathrm{d}\Phi^{h_1h_2}_{\theta\theta} g_i(n,\bar{n},k_1,k_2) + \SUM{i}{} a_i^{\text{inhom}}(\ep) \int  \mathrm{d}\Phi^{h_1h_2}_{\delta\theta} g_i(n,\bar{n},k_1,k_2) \,, \\
0 ={} & \SUM{i}{} b_i^{\text{hom}}(\ep) \int  \mathrm{d}\Phi^{h_1h_2}_{\theta\theta} g_i(n,\bar{n},k_1,k_2) + \SUM{i}{} b_i^{\text{inhom}}(\ep) \int  \mathrm{d}\Phi^{h_1h_2}_{\theta\delta} g_i(n,\bar{n},k_1,k_2) \,,
\end{align}
where the two equations arise  from derivatives w.r.t.~$k_1$ and $k_2$, respectively.

We can   apply the same logic to  integrals with
a  $\theta$-function  and a $\delta$-function. Inhomogeneous terms in this case will 
contain two $\delta$-functions and \textit{no} $\theta$-function. IBP relations for such integrals
can be derived using  conventional methods since $\delta$-function constraints can be immediately
mapped onto  rational functions of parton momenta using reverse unitarity.

\subsection{Reduction to master integrals }

In the previous section, we presented an explicit example of a modified IBP relation that we derived 
starting with a  seed integral with  two $\theta$-functions. We note that modified IBP relations naturally
form a hierarchical structure since  the smaller the number of $\theta$-functions that
a particular  integral contains, the easier it is to compute it.
Hence, in the course of the  reduction, we try to express integrals with larger number of $\theta$-functions
through integrals with some of the $\theta$-functions  replaced by the $\delta$-functions. 
Integrals that belong to the same ``hierarchical level'' are organized into distinct  integral families.

In practice, we implement derivation of   modified IBP relations in 
\texttt{Mathematica} and solve them using the ``user-defined system'' functionality of  \texttt{Kira}. This requires us to define 
integral families with different numbers  of $\theta$-
and $\delta$-functions and to derive relations among  integrals that belong to
different families. We note that   \emph{for conventional integrals} all these steps
are done automatically by publicly-available  reduction programs,  but we have to take care of them ourselves 
in the present  case. 

As we already mentioned, it is straightforward to derive homogeneous relations for integrals in each family
for  arbitrary powers of propagators. 
For example, for the integral family defined  in Eq.~(\ref{eqn:n2lo_topo_ex}),
the derivative with respect to $k_2$, contracted with $v=\bar{n}$ yields\footnote{We neglect inhomogeneous terms, represented by the ellipses.}
\begin{align}
\begin{split}
0 ={} &\int {\rm d}^d k_1 {\rm d}^d k_2 \, \bar{n}^\mu \frac{\partial}{\partial k_2^\mu}  \frac{ \theta(k_1\bar{n} - k_1 n)\theta(k_2\bar{n} - k_2 n)}{\left[\dots \right]_c (k_1 k_2)^{a_4} \, (k_2 n )^{a_5} \,(k_1 \bar{n})^{a_6}  \, (k_{12}\bar{n})^{a_7}} \\
& = \left[ 2 a_2 \hat{2}^+ ( \hat{6}^- - \hat{7}^-) + 2 a_3 \hat{3}^+ - a_4 \hat{4}^+ \hat{6}^- - 2 a_5 \hat{5}^+  \right] \mathcal{T}^{\text{ex}}_{a_1,a_2,a_3,a_4,a_5,a_6,a_7} + \dots \,.
\end{split}
\end{align}
In writing this equation, we have defined operators $\hat{i}^{+(-)}$ that raise (lower) the index $a_i$ of the
integral $\mathcal{T}^{\text{ex}}_{a_1,a_2,a_3,a_4,a_5,a_6,a_7}$ by one.

Inhomogeneous terms, such as the one that appeared  in Eq.~(\ref{eqn:n2lo_tt_ex_2}) and discussed after that equation, 
are generated on an integral-by-integral  basis; they require  partial fraction decomposition and
``on-the-fly'' matching to topologies with fewer $\theta$-functions. Although this step is straightforward, 
it needs to be implemented in a separate \texttt{Mathematica} code whose output is then  fed  to \texttt{Kira}.

Using this setup, we express the maximally non-abelian contribution to the zero-jettiness soft function defined in
Eq.~(\ref{eqn:S2g_mnab_def}) in terms of eleven master integrals
\begin{align}
S^{2g}_{\rm NA} ={} & \tau^{-1-4\ep} \Bigg \{ \bigg[   \frac{\left(192 \ep^5+48 \ep^4-736
    \ep^3+1336 \ep^2-376 \ep+33\right) }{3 \ep^3 (2 \ep-3) (2 \ep-1)} \mathcal{I}^{nn}_1
  \nonumber \\
  &  -\frac{8 (4 \ep-1) \left(12 \ep^4-25 \ep^2+41 \ep-3\right)}{3 \ep^2 (2 \ep-3) (2 \ep-1)} \mathcal{I}^{nn}_2 +\frac{3}{\ep}\mathcal{I}^{nn}_3+\frac{2 }{\ep} \mathcal{I}^{nn}_4 \bigg]
\nonumber \\
& + \bigg[ \frac{128 \ep^7+864 \ep^6-848 \ep^5-1680 \ep^4+152 \ep^3+770 \ep^2-163 \ep+3}{\ep^3 (\ep+1) (2 \ep-1) (2 \ep+1) (2 \ep+3)} \mathcal{I}^{n\bar{n}}_1
  \label{eqn:N2LO_reduced}  \\
   &  -\frac{8 \left(64 \ep^7+120 \ep^6-164 \ep^5-246 \ep^4+69 \ep^3+126 \ep^2-46 \ep+3\right)}{\ep^2 (\ep+1) (2 \ep-1) (2 \ep+1) (2 \ep+3)} \mathcal{I}^{n\bar{n}}_2
  \nonumber \\
  & + \frac{\left(16 \ep^5+16 \ep^3+36 \ep^2+11 \ep-9\right)}{\ep (\ep+1) (2 \ep+1) (2 \ep+3)} \mathcal{I}^{n\bar{n}}_3 + \frac{2}{\ep} \mathcal{I}^{n\bar{n}}_4
  \nonumber \\
& -\frac{8 (4 \ep-1) \left(2 \ep^3+3 \ep^2+3 \ep-3\right)}{\ep (2 \ep+1) (2 \ep+3)}  \mathcal{I}^{n\bar{n}}_5 + 2 \,  \mathcal{I}^{n\bar{n}}_6 + 4  \mathcal{I}^{n\bar{n}}_7 \bigg] \Bigg\} \,.
\nonumber
\end{align}

The master integrals that appear in Eq.~(\ref{eqn:N2LO_reduced}) read 
\begin{equation}
\begin{alignedat}{4}
\mathcal{I}^{nn}_1 ={} & \int \mathrm{d}\Phi^{nn}_{\delta\delta} \,,
&~~ \mathcal{I}^{nn}_2 ={} &  \int \frac{\mathrm{d}\Phi^{nn}_{\delta\theta}}{(k_{12}\bar{n})}\,, \\
\mathcal{I}^{nn}_3 ={} & \int \frac{\mathrm{d}\Phi^{nn}_{\delta\theta}}{(k_1k_2)(k_2\bar{n})}\,,
&~~ \mathcal{I}^{nn}_4 ={} &  \int \frac{\mathrm{d}\Phi^{nn}_{\delta\theta}}{(k_1k_2)(k_2n)(k_{12}\bar{n})}\,, \\
\end{alignedat}
\label{eqn:mi_2g_nn}
\end{equation}
and
\begin{equation}
\begin{alignedat}{4}
\mathcal{I}^{n\bar{n}}_1 ={} & \int  \mathrm{d}\Phi^{n\bar{n}}_{\delta\delta}  \,,
&\mathcal{I}^{n\bar{n}}_2 ={} &  \int \frac{\mathrm{d}\Phi^{n\bar{n}}_{\delta\theta}}{(k_{12}n)}  \,, \\
\mathcal{I}^{n\bar{n}}_3 ={} & \int \frac{\mathrm{d}\Phi^{n\bar{n}}_{\delta\theta}}{(k_1k_2)(k_2n)}\,, 
&\mathcal{I}^{n\bar{n}}_4 ={} &\int \frac{\mathrm{d}\Phi^{n\bar{n}}_{\delta\theta}}{(k_1k_2)(k_2\bar{n})(k_{12}n)} \,, \\
\mathcal{I}^{n\bar{n}}_5 ={} & \int \frac{\mathrm{d}\Phi^{n\bar{n}}_{\theta\theta}}{(k_{12}n)(k_{12}\bar{n})} \,,  
&\mathcal{I}^{n\bar{n}}_6 ={} & \int \frac{\mathrm{d}\Phi^{n\bar{n}}_{\theta\theta}}{(k_1k_2)(k_1\bar{n})(k_2n)} \,, \\
\mathcal{I}^{n\bar{n}}_7 ={} & \int \frac{\mathrm{d}\Phi^{n\bar{n}}_{\theta\theta}}{(k_1k_2)(k_2n)(k_{12}\bar{n})} \,.
\end{alignedat}
\label{eqn:mi_2g_nnb}
\end{equation}

It is interesting that the above set  of master integrals is actually redundant since for $i = 1,2,3,4,$ 
$\mathcal{I}^{n\bar{n}}_i = \mathcal{I}^{n n}_i$.
This happens because in certain cases integrals 
depend on  $\alpha_i$ and $\beta_i$ (or on $n$ and $\bar n$) in a symmetric way.  As an example, consider $\mathcal{I}^{n\bar{n}}_2$.
We find 
\begin{align}
\begin{split}
 \mathcal{I}^{n\bar{n}}_2 ={} & N_\ep^{-2} \int \frac{[{\rm d} k_1] [{\rm d} k_2] \delta(1-k_1n-k_2\bar{n}) \delta(k_1\bar{n}-k_1n) \theta(k_2n-k_2\bar{n}) }{(k_{12}n)} \\
\stackrel{n\leftrightarrow\bar{n}}{=} {} &  N_\ep^{-2} \int \frac{[{\rm d} k_1] [{\rm d} k_2] \delta(1-k_1\bar{n}-k_2n) \delta(k_1n-k_1\bar{n}) \theta(k_2\bar{n}-k_2n) }{(k_{12}\bar{n})} \\
\stackrel{\delta(k_1n-k_1\bar{n})}{=} {} & N_\ep^{-2} \int \frac{[{\rm d} k_1] [{\rm d} k_2] \delta(1-k_1n-k_2n) \delta(k_1\bar{n}-k_1n) \theta(k_2\bar{n}-k_2n) }{(k_{12}\bar{n})} =  \mathcal{I}^{nn}_2\, .
\end{split}
\end{align}

Another interesting feature of the above set of master integrals is that
only three master integrals $I_{5\dots7}$ contain
two $\theta$-functions;  all these  integrals correspond
to the $n \bar n$ configuration. 
For all other master integrals, either one or both  $\theta$-functions are replaced by $\delta$-functions;
these  integrals are simpler to compute than the original ones.

To understand why there are no $nn$ master integrals with two $\theta$-functions,  we note that
homogeneous parts of IBP relations are unaffected  by  $\theta$-functions; hence, by solving homogeneous
parts of the IBP relations we should find master integrals that would be present if all $\theta$-functions
are removed from the integrand. It is then easy to see that, in the case of $nn$ integrals, removal of $\theta$-functions
from the integrand leads to scaleless integrals since the jettiness constraint only depends on $\beta_{1,2}$ in this case.

\subsection{Computation of master integrals}
\label{sec:n2lo_mi}
Having discussed the reduction to master integrals, in this Section we explain how to compute the master integrals that appear
in Eqs.~(\ref{eqn:mi_2g_nn},\ref{eqn:mi_2g_nnb}).
Since the NNLO soft function is required for the computation of N3LO soft function,
we will compute  $S^{2g}_{NA}$ through weight six. 

We begin by calculating  the four integrals that are needed to describe  the $nn$ configuration.
To this end,  we combine  the phase-space
parameterization in Eq.~(\ref{eqn:ps_withf_general}) and 
the Sudakov parametrization of the  phase-space element of gluon $i$ 
\begin{align}
\begin{split} 
&  [{\rm d} k_i] = \frac{{\Omega}^{(d-2)}
}{4 (2 \pi)^{d-1} } \;  {\rm d} \alpha_i \; {\rm d} \beta_i \; (\alpha_i \beta_i )^{-\ep} \;
\frac{ {\rm d} {\Omega}_{i,\perp}^{(d-2)}}{{\Omega}^{(d-2)}} \,, ~~~ \alpha_i\,,\beta_i \in [0,\infty) \,,
\end{split} 
\label{eq2.8a}
\end{align}
to write 
\begin{align}
\begin{split}
\mathcal{I}^{nn}_1 ={} & \int \DIFFL\Phi^{nn}_{\delta\delta}
= \int \prod_{i=1}^{2} \DIFFL \alpha_i \DIFFL \beta_i (\alpha_i \beta_i)^{-\ep} \DD{1-\beta_{12}} \DD{\alpha_1-\beta_1}\DD{\alpha_2-\beta_2} \\
={} &   \INT{0}{1} \DIFFL \beta_1 \DIFFL \beta_2 ( \beta_1\beta_2)^{-2\ep} \DD{1-\beta_{12}}
=    \frac{\GFP{2}{1-2\ep}}{\GF{2-4\ep}} \,,
\end{split}
\end{align}
where we have used $\beta_{12} = \beta_1 + \beta_2$. 

The computation of $\mathcal{I}^{nn}_2 $ proceeds in a similar way. We find 
\begin{align}
\begin{split}
\mathcal{I}^{nn}_2 ={} & \int \frac{\mathrm{d}\Phi^{nn}_{\delta\theta}}{(k_{12}\bar{n})}
= \INT{0}{\infty} \DIFFL \alpha_2 \DIFFL \beta_1 \DIFFL \beta_2 \,  \frac{(\alpha_2 \beta_1^2 \beta_2)^{-\ep} \DD{1-\beta_{12}} \HT{\alpha_2-\beta_2} }{\beta_1+\alpha_2} \\
\stackrel{\alpha_2\to\beta_2/\xi_2}{=} {} &   \FEYNINT{\xi_2} \FEYNINT{\beta_1}  \,  \frac{ \beta_1^{-2\ep} (1-\beta_1)^{1-2\ep} \xi_2^{\ep-1}  }{1-\beta_1(1-\xi_2)} \\
={} & \frac{\GF{1-2\ep}\GF{2-2\ep}}{\ep \GF{3-4\ep}} \GENHYPGF{3}{2}{1,1,1-2\ep}{3-4\ep,1+\ep}{1}.
\end{split}
\end{align}
The integral $\mathcal{I}^{nn}_3 $ involves the scalar product of the two gluon momenta. To facilitate its
computation,  
we write 
\begin{align}
\begin{split}
\mathcal{I}^{nn}_3 = {} & \int \frac{\mathrm{d}\Phi^{nn}_{\delta\theta}}{(k_1k_2)(k_2\bar{n})}  
\stackrel{\alpha_2\to\beta_2/\xi_2}{=} {}  \int \frac{
\DIFFSPinopi{d-2}{12} }{\left[ \Omega^{(d-2)} \right]^2 } \int   \frac{ \DIFFL \xi_2 \DIFFL \beta_1 \; \xi_2^\ep \; [\beta_1 (1-\beta_1)]^{-1-2\ep} }{
  \left[ \xi_2+1-2\sqrt{\xi_2} \cos \varphi_{12}\right]},
\end{split}
\label{eqn:I_2g_nn_3}
\end{align}
where $\DIFFSPinopi{d-2}{12} = \DIFFSPnopi{d-2}_1 \DIFFSPnopi{d-2}_2$ and 
$\varphi_{12}$ is the relative angle between transverse components of $k_1$ and $k_2$.
To integrate over this angle, we introduce a new variable $\eta$ defined as 
\be
\eta = \frac{1-\cos \varphi_{12}}{2} \,,
\ee
and   write 
\begin{align}
\DIFFSPinopi{d-2}{12} =   2 \DIFFSPnopi{d-2} \DIFFSPnopi{d-3} \DIFFL \eta \left[ 4\eta(1-\eta) \right]^{-1/2-\ep} \,.
\end{align}

We integrate over $\eta$ using the formula
\begin{align}
2 \frac{\Omega^{d-3}}{\Omega^{(d-2)}} \FEYNINT{\eta}  \frac{ \left[ 4\eta(1-\eta) \right]^{-1/2-\ep}}{\left[ \xi_2+1-2\sqrt{\xi_2} (1-2\eta)\right]} = \frac{ \HYPGF{1}{1/2-\ep}{1-2\ep}{\frac{-4\sqrt{\xi_2}}{(1-\sqrt{\xi_2})^2}}}{(1-\sqrt{\xi_2})^2} \,,
\end{align}
apply the following  hypergeometric identity\footnote{See, e.g.~Eq.~(15.3.16) in Ref.~\cite{Abramowitz:1974}.}
\begin{align}
\label{eqn:abramowitz15316}
\HYPGF{a}{b}{2b}{z} = \left( 1 - \frac{z}{2} \right)^{-a} \HYPGF{\frac{a}{2}}{\frac{a+1}{2}}{b+\frac{1}{2}}{\frac{z^2}{4(1-z/2)^{2}}}\,,
\end{align}
and  obtain 
\begin{align}
\begin{split}
\mathcal{I}^{nn}_3
= {} & - \frac{2\GFP{2}{1-2\ep}}{\ep(1+\ep)\GF{1-4\ep}} \GENHYPGF{3}{2}{1,1+\ep,1+\ep}{1-\ep,2+\ep}{1} \,.
\end{split}
\end{align}

Following similar steps, we derive 
\begin{align}
\begin{split}
\mathcal{I}^{nn}_4 = \int \frac{\mathrm{d}\Phi^{nn}_{\delta\theta}}{(k_1k_2)(k_2n)(k_{12}\bar{n})} 
={} &  -2   \frac{\GFP{2}{1-2\ep}}{\ep\GF{1-4\ep}} \FEYNINT{\xi_2} \; \xi_2^{-1-\ep} (1-\xi_2)^{-1-2\ep}  \mathcal{X}_4(\xi_2) \,,
\end{split}
\label{eqn:I_nn_4_xi2}
\end{align}
where the function $\mathcal{X}_4$ reads 
\begin{align}
\mathcal{X}_4(\xi_2) = \HYPGF{-1-4\ep}{-2\ep}{-4\ep}{1-\xi_2} \HYPGF{-\ep}{-2\ep}{1-\ep}{\xi_2} \,.
\end{align}
We note that the integral in Eq.~(\ref{eqn:I_nn_4_xi2}) diverges logarithmically at the integration
boundaries $\xi_2=0,1$ but the function $\mathcal{X}_4(\xi_2)$ is regular at these points 
\begin{align}
\mathcal{X}_4(0) = \frac{\GF{1-4\ep}\GF{1+2\ep}}{2\GF{1-2\ep}} \,, ~~~ \mathcal{X}_4(1) = \frac{\GF{1-\ep}\GF{1+2\ep}}{\GF{1+\ep}} \,.
\end{align}
Hence, we can compute the master integral $\mathcal{I}^{nn}_4$  by subtracting divergent  contributions at endpoints
and adding them back. Specifically, we write 
\begin{align}
\begin{split}
  \mathcal{I}^{nn}_4 ={} & -2  
  \frac{\GFP{2}{1-2\ep}}{\ep\GF{1-4\ep}} \bigg\{ \mathcal{X}_4(0)   \FEYNINT{\xi_2} \xi_2^{-1-\ep}  +  \mathcal{X}_4(1) \FEYNINT{\xi_2}
   (1-\xi_2)^{-1-2\ep}   \\
& ~~ + \FEYNINT{\xi_2} \xi_2^{-1-\ep} (1-\xi_2)^{-1-2\ep}  \mathcal{X}_4(\xi_2) - \xi_2^{-1-\ep}  \mathcal{X}_4(0)  - (1-\xi_2)^{-1-2\ep}  \mathcal{X}_4(1)  \bigg\} \,.
\end{split}
\label{eqn:I_nn_4_xi2_endptsubtr}
\end{align}
The first two integrals on the r.h.s~of Eq.~(\ref{eqn:I_nn_4_xi2_endptsubtr}) are trivial. The last integral is regular in
the integration domain  $\xi_2\in[0,1]$ and can be computed \textit{after} expansion in $\epsilon$.
We construct such an expansion using   the package
\texttt{HypExp}~\cite{Huber:2005yg} and use the program  \texttt{HyperInt}~\cite{Panzer:2014caa} to integrate the result over $\xi_2$. 
We arrive at
\begin{align}
\mathcal{I}^{nn}_4 = \frac{2}{\ep^2} + \frac{\pi^2}{3} - \frac{17\pi^4\ep^2}{90} + \ep^3\left[ -6 \pi^2 \zeta_3 -26 \zeta_5 \right] - \ep^4 \left[ \frac{193\pi^6}{810} + 64 \zeta_3^2 \right]   +\mathcal{O}(\ep^5) \,,
\end{align}
where we have discarded contributions of  weight seven and higher.

It remains to compute three additional integrals for the $n \bar n$ configuration. 
We begin with $\mathcal{I}^{n\bar{n}}_5$ and change variables  $\alpha_1 = \beta_1/\xi_1$, $\beta_2 =\alpha_2/\xi_2$. We find 
\begin{align}
\begin{split}
\mathcal{I}^{n\bar{n}}_5 ={} & \int \frac{\mathrm{d}\Phi^{n\bar{n}}_{\theta\theta}}{(k_{12}n)(k_{12}\bar{n})} = \frac{ 2  \GFP{2}{2-2\ep} }{\GF{4-4\ep}} \\
& ~~~  \times  \FEYNINT{\xi_1} \FEYNINT{\xi_2} \frac{(1-\xi_1)(\xi_1\xi_2)^{-1+\ep}}{1-\xi_1\xi_2} \HYPGF{1}{2-2\ep}{4-4\ep}{1-\xi_1} \\
={} & \frac{ 2  \GFP{2}{2-2\ep} }{\ep \GF{4-4\ep}}  \FEYNINT{\xi_1} (1-\xi_1) \xi_1^{-1+\ep} \mathcal{X}_5(\xi_1) \,,
\end{split}
\end{align}
where
\begin{align}
\begin{split}
\mathcal{X}_5(\xi_1) = \HYPGF{1}{\ep}{1+\ep}{\xi_1} \HYPGF{1}{2-2\ep}{4-4\ep}{1-\xi_1}\,.
\end{split}
\end{align}
We subtract the (only) logarithmic  singularity at $\xi_1 = 0$ and obtain  after the integration
\begin{align}
\begin{split}
\mathcal{I}^{n\bar{n}}_5 ={} &  \frac1{\ep^2} + \frac{2}{\ep} + \left[ 2+\frac{\pi ^2}{6} \right] + \ep \left[ 2 \zeta_3-8+\pi^2 \right] + \ep^2\left[ 16 \zeta_3-64+4 \pi ^2+\frac{\pi^4}{9} \right] \\
& + \ep^3 \left[ 64 \zeta_3-\frac{2 \pi ^2 \zeta_3}{3}+30 \zeta_5-256+\frac{32 \pi^2}{3}+\frac{26 \pi ^4}{45} \right] \\
& + \ep^4 \left[ 128 \zeta_3+4 \pi ^2 \zeta_3+8 \zeta_3^2+60 \zeta_5-512+\frac{92 \pi^4}{45}+\frac{44 \pi ^6}{945}  \right]+ \mathcal{O}\left(\ep^5\right) \,.
\end{split}
\end{align}

Following the same steps as described above,  we find   the result  for $\mathcal{I}^{n\bar{n}}_6$
\begin{align}
\begin{split}
\mathcal{I}^{n\bar{n}}_6
={} & -\frac{2\GFP{2}{1-2\ep}}{\ep(1+\ep)^2\GF{1-4\ep}}  \GENHYPGF{4}{3}{1,1+\ep,1+\ep,1+\ep}{1-\ep,2+\ep,2+\ep}{1}.
\end{split}
\end{align}

Finally,  for the integral $\mathcal{I}^{n\bar{n}}_7$ we obtain
\begin{align}
\begin{split}
\mathcal{I}^{n\bar{n}}_7 ={} & \int \frac{\mathrm{d}\Phi^{n\bar{n}}_{\theta\theta}}{(k_1k_2)(k_2n)(k_{12}\bar{n})} = - \frac{\GFP{2}{1-2\ep}}{\ep(1+\ep)\GF{1-4\ep}} \FEYNINT{\xi_1} \xi_1^\ep \\
& ~~~ \times \HYPGF{1}{-2\ep}{1-4\ep}{1-\xi_1} \, \GENHYPGF{3}{2}{1,1+\ep,1+\ep}{1-\ep,2+\ep}{\xi_1} \,.
\end{split}
\end{align}
The $\xi_1$-integral is finite; we expand it in powers of $\ep$ and integrate using \texttt{HyperInt}. The result reads
\begin{align}
\mathcal{I}^{n\bar{n}}_7 = {} & - \frac{\pi^2}{6\ep} + 2 \zeta_3 + \frac{\pi^4\ep}{12} + \ep^2 \left[ \frac{5\pi^2\zeta_3}{3} + 19 \zeta_5 \right] + \ep^3 \left[ \frac{937\pi^6}{3780} - 82 \zeta_3^2 \right] + \mathcal{O}(\ep^4)\,.
\end{align}
This concludes the computation of master integrals required for the  calculation
of the real-emission contribution to the
zero-jettiness soft function at NNLO. 

\subsection{Results for the real emission contribution at NNLO}

We use the master integrals computed in Section~\ref{sec:n2lo_mi}, insert them into Eq.~(\ref{eqn:N2LO_reduced}) and obtain
\begin{align}
\begin{split}
S^{2g}_{\rm NA} ={} & \tau^{-1-4\ep}  \Bigg\{ \frac{2}{\ep^3} + \frac{11}{3\ep^2} + \frac1{\ep} \left[ \frac{67}{9} - \pi^2 \right] + \left[\frac{404}{27} - \frac{11\pi^2}{9} - 18 \zeta_3  \right] \\
& + \ep \left[ \frac{2140}{81} + \frac{67\pi^2}{9} - \frac{59\pi^4}{90} -\frac{220\zeta_3}{3} \right] + \ep^2 \bigg[ \frac{12416}{243} - \frac{368\pi^2}{81} \\
& - \frac{143\pi^4}{45} + 268 \zeta_3 + 4\pi^2 \zeta_3 - 182 \zeta_5 \bigg] + \ep^3 \bigg[ \frac{67528}{729} + \frac{2416\pi^2}{81} \\
  & + \frac{469\pi^4}{45} -\frac{17\pi^6}{105} - \frac{7864\zeta_3}{27} + \frac{880\pi^2\zeta_3}{9}
  -122 \zeta_3^2 - \frac{6248 \zeta_5}{3}  \bigg]
 + {\cal O}(\ep^4)
\Bigg\} \,.
\end{split}
\label{eqn:n2lo_result}
\end{align}
The result agrees with the one derived earlier  in Ref.~\cite{Baranowski:2020xlp}.\footnote{We note that it also agrees with the results of Refs.~\cite{Monni:2011gb,Kelley:2011ng} which were computed through weight four. Partial results through weight six have been obtained in Ref.~\cite{Karo:ba}.}

To summarize, we have shown that  by constructing the  integration-by-parts identities for  phase-space integrals
with Heaviside functions  it is possible to express
the  real emission contribution to the zero-jettiness NNLO soft function through seven master integrals. The majority
of these integrals needs to be computed by integrating
over a \textit{simplified} phase space with all or some $\theta$-functions replaced by the $\delta$-functions.
This simplification is very striking in case of the $nn$ kinematic configuration where we find that
{\it no} master integrals with two $\theta$-functions need to be computed.
As we discussed earlier, this interesting feature can be readily understood 
if IBP technology
is applied to phase-space integrals with Heaviside functions.

%% file: sections/05_N3LO_1.tex
\section{Testing  the method with some N3LO contributions to the zero-jettiness soft function}
\label{sec:n3lo}

It appears from the discussion in the previous section that it is useful to construct
IBP relations for  integrals with
Heaviside functions.  However, given an enormous growth in computational complexity
with increase in the perturbative order, it is important to check this statement by considering
a more complex example.
Given our interest in the N3LO QCD contribution  to the zero-jettiness soft function,
it is natural to check if modified IBP relations  can be constructed and used to compute it. 

To this end, in this section 
we consider the maximally non-abelian part of the real-emission contribution
to the soft function with  all  gluons emitted
to the same hemisphere. We define it as follows 
\begin{align}
\label{eqn:3g_mnab_full}
S_{nnn}^{3g} = \tau^{-1-6\ep} 
\int \mathrm{d} \Phi^{nnn}_{\theta\theta\theta} \omega^{(3)}_{n\bar{n}}(k_1,k_2,k_3),
\end{align}
where 
the function $\omega^{(3)}_{n\bar{n}}(k_1,k_2,k_3)$ in Eq.~(\ref{eqn:3g_mnab_full}) reads~\cite{Catani:2019nqv}
\begin{align}
\label{eqn:def_w3_eikonal}
\omega^{(3)}_{n\bar{n}}(k_1,k_2,k_3) = \sum_{t \in \{ a,b,c,d \}}  \omega^{(3),t}_{n\bar{n}}(k_1,k_2,k_3) \,,
\end{align}
\begin{align}
\begin{split}
\omega^{(3),t}_{n\bar{n}} ={} &  \left[ \oS_{n\bar{n}}^{(t)}(k_1,k_2,k_3) + \oS_{\bar{n}n}^{(t)}(k_1,k_2,k_3) - \oS_{nn}^{(t)}(k_1,k_2,k_3) - \oS_{\bar{n}\bar{n}}^{(t)}(k_1,k_2,k_3)  \right] \\
& + \text{permutations}\{k_1,k_2,k_3\} \,.
\end{split}
\label{eqn:def_w3t_eikonal}
\end{align}
In Eq.~(\ref{eqn:def_w3t_eikonal}), ``$ \text{permutations}\{k_1,k_2,k_3\}$'' describes
all possible permutations  of the gluon momenta $k_i$. The four terms  $\oS_{ik}^{(t)}$, $t=a,b,c,d$
in Eq.~(\ref{eqn:def_w3t_eikonal})
are contributions to the soft eikonal function that are ordered according to the  structure of their collinear
singularities~\cite{Catani:2019nqv}.

A simple generalization of the discussion in  Section~\ref{sec:mod_ibp}
implies that, in order to set up IBP relations for integrals that appear in
$S_{nnn}^{3g}$, we require eight 
distinct classes of integrals. They correspond to integrations over the following phase spaces 
$I_{\theta\theta\theta} \sim \int \mathrm{d} \Phi^{nnn}_{\theta\theta\theta}$, $
I_{\theta\theta\delta} \sim \int \mathrm{d} \Phi^{nnn}_{\theta\theta\delta}$, $I_{\theta\delta\delta} \sim 
\int \mathrm{d} \Phi^{nnn}_{\theta\delta\delta}$, $I_{\delta\delta\delta} \sim \int \mathrm{d} \Phi^{nnn}_{\delta\delta\delta}$
and four  more cases where  the ordering of  $\theta$- and $\delta$-functions differs from the above examples. 

Similar to the two-gluon case,  these  classes of integrals possess hierarchical structure  that we  rely upon when 
solving    the integration-by-parts identities.
Indeed, a  closed set of IBP relations
  can be derived for  $I_{\delta \delta \delta}$ using reverse  unitarity.
  On the other hand,  IBP identities  for
  $I_{\theta \delta \delta}$ involve  $I_{\delta \delta \delta}$,  and IBP identities for $I_{\theta \theta \delta}$ involve  $I_{\theta \delta \delta}$
  and $I_{\delta \delta \delta}$. Finally, integration-by-parts identities for $I_{\theta \theta \theta}$ make use
  of all other integrals with $\theta$- and $\delta$-functions. We illustrate how this approach applies
  to various N3LO QCD contributions to the zero-jettiness soft function in the  next sections.

%% file: sections/05_N3LO_2_a.tex
\subsection{The $\omega_{n  \bar n}^{(3),a}$ contribution}
\label{sec:n3lo_Sa}

The simplest contribution to consider is an integral of  $\omega_{n  \bar n}^{(3),a}$.
According to Eq.~(\ref{eqn:def_w3t_eikonal}), this function is  
constructed from the function $\oS_{ik}^{(a)}$, which reads~\cite{Catani:2019nqv}\footnote{We note that we take the emitters $i,j$ to be massless.}
\begin{footnotesize}
\begin{align}
\begin{split}
&\oS_{ik}^{(a)}(q_1,q_2,q_3)=\frac{31}{144} \per \frac{(\pp{i}{k})^3}{\pq{i}{1} \per \pq{k}{1} \per \pq{i}{2} \per \pq{k}{2} \per \pq{i}{3} \per \pq{k}{3}} \\
& +\frac{(\pp{i}{k})^3}{32 \per \pq{k}{12} \per \pq{k}{3} \per \pq{i}{1} \per \pq{i}{3}} \per \left(\frac{1}{\pq{k}{1}}-\frac{1}{\pq{k}{2}}\right) \per \left(\frac{6}{\pq{i}{12}}+\frac{p_i(q_{3}-q_{1})}{\pq{i}{13} \per \pq{i}{2}}\right) \\
  & +\frac{(\pp{i}{k})^3}{288 \per \pq{k}{123} \per \pq{i}{123}} \per \left(\frac{1}{\pq{k}{1}}-\frac{1}{\pq{k}{2}}\right)
\times  \per \sg \frac{2}{\pq{i}{1}} \per \left( \frac{1}{\pq{i}{3}}-\frac{3}{\pq{i}{12}} \right) \per \left(\frac{1}{\pq{k}{3}}-\frac{3}{\pq{k}{12}}\right)+ \\
&\left(\frac{1}{\pq{i}{2}}-\frac{3}{\pq{i}{13}}\right) \per \left(\frac{1}{\pq{i}{1}}-\frac{1}{\pq{i}{3}}\right) \per \left(\frac{1}{\pq{k}{3}}-\frac{3}{\pq{k}{12}} \right)\dg, 
\end{split}
\end{align}
\end{footnotesize}%
where $p_i = n$, $p_k = \bar n$ and $q_i = k_i$, $i=1,2,3$ in our notations. 
As can be easily checked, $\oS_{ik}^{(a)}$ is not singular when any of the
two gluons become collinear to each  other.  This feature reduces the number of scalar products
that appear in the denominators of that  function, making  integration over 
three-gluon phase space simpler.

Applying integration-by-parts identities, we find that the integral of $\omega_{n  \bar n}^{(3),a}$
over the phase space ${\rm d} \Phi^{nnn}_{\theta \theta \theta}$ can be expressed through six master integrals.
The result of the reduction reads
\begin{align}
\begin{split}
& \int {\rm d} \Phi^{nnn}_{\theta \theta \theta} \, \omega^{(3),a}_{n\bar{n}}(k_1,k_2,k_3) \\
={} & \frac1{(1-4\ep)(1-5\ep)} \bigg \{ \left[ \frac{182}{5 \ep^5}-\frac{3392}{5 \ep^4}+\frac{23268}{5
   \ep^3}-\frac{69432}{5 \ep^2}+\frac{75728}{5 \ep} \right] I_1 \\
   &  + \left[ \frac{8}{5 \ep^3}-\frac{72}{5 \ep^2}+\frac{32}{\ep} \right] I_2 + \left[ -\frac{112}{5 \ep^4}+\frac{2016}{5 \ep^3}-\frac{13328}{5
   \ep^2}+\frac{38304}{5 \ep}-8064 \right] I_3 \\
   & + \left[ -\frac{4}{\ep^4}+\frac{356}{5 \ep^3}-\frac{2248}{5 \ep^2}+\frac{5968}{5
   \ep}-\frac{5664}{5} \right] I_4 + \left[ -\frac{8}{5 \ep^3}+\frac{104}{5 \ep^2}-\frac{448}{5 \ep}+128 \right] I_5 \\
   & + \left[ -\frac{36}{5 \ep^4}+\frac{648}{5 \ep^3}-\frac{4428}{5 \ep^2}+2592
   \ep+\frac{14328}{5 \ep}-\frac{22032}{5} \right] I_6  \bigg\}. 
\end{split}
\label{eq5.5}
\end{align}
Definitions of the  six master integrals are  given in Appendix~\ref{app:master_integrals}, Eq.~(\ref{eqn:mi_3g_nnn_a}). However, it is useful to
emphasize at this point
that  {\it none} of these integrals contains three $\theta$-functions.   We have explained why this happens 
when discussing the NNLO contribution to the soft function in Section~\ref{sec:n2lo}.

Calculation of master integrals that appear in Eq.~(\ref{eq5.5}) 
is rather straightforward. As an example, consider a phase-space integral
$I_1$  with three $\delta$-functions.  It reads
\be
I_1 ={}   \int \DIFFL \Phi^{nnn}_{\delta\delta\delta}\;.
\ee
Since the integrand does not depend on the relative  orientation of the three partons in the transverse plane,
we can integrate over ${\rm d} {\Omega}_{i,\perp}^{(d-2)}$ for $i=1,2,3$.   We then
remove all $\delta(\alpha_i-\beta_i)$-functions
by integrating over  $\alpha_i$, $i=1,2,3$,  and find 
\be
I_1 = \int \prod \limits_{i=1}^{3}  {\rm d} \beta_i \beta_i^{-2\ep} \;\; \delta(1-\beta_{123}) = 
\frac{\Gamma^3(1-2\ep) }{\Gamma(3-6\ep)}.
\label{eq4.4}
\ee

For a less trivial example, consider the integral $I_5$. It reads 
\be
I_5 = \int 
\frac{{\rm d} \Phi^{nnn}_{\delta \delta \theta} }{(k_{13} n) (k_{123} \bar n) }.
\ee
To compute this integral, we remove two $\delta$-functions by  integrating over $\alpha_{1,2}$ and then integrate over
$\beta_1$ to remove  $\delta(1-\beta_{123})$.  Then, writing $\beta_2 = (1-\beta_3) \xi$, we arrive at
\be
I_5 = \int {\rm d} \alpha_3 {\rm d} \beta_3 \alpha_3^{-\ep} \beta_3^{-\ep} \theta(\alpha_3 - \beta_3)
\frac{(1-\beta_3)^{1-4\ep} }{1-\beta_3 + \alpha_3}
\int \limits_{0}^{1} \frac{{\rm d} \xi \; \xi^{-2\ep} (1-\xi)^{-2\ep}}{1-(1-\beta_3) \xi}.
\ee
It is straightforward to express the integrals over $\xi$ and $\alpha_3$ through hypergeometric functions.
The result reads
\be
I_5 =  \frac{\Gamma^2(1-2\ep)}{\ep \Gamma(2-4\ep)}
\int \limits_{0}^{1} {\rm d} u  \; (1-u)^{1-2\ep} u^{1-4\ep}
\HYPGF{1}{1-2\ep}{2-4\ep}{u} \HYPGF{1}{1}{1+\ep}{u},
\label{eq4.8}
\ee
where we introduced a new variable $u=1-\beta_3$.
Although both hypergeometric functions in Eq.~(\ref{eq4.8}) are singular  at $u =1$, this singularity is made integrable  by an explicit factor
$(1-u)^{1-2\ep}$ in the integrand. Hence, we can directly
expand the integrand in Eq.~(\ref{eq4.8})  in powers of $\ep$ and integrate over $u$ using \texttt{HyperInt}~\cite{Panzer:2014caa}. We find
\be
\begin{split} 
I_5 & = \frac{1}{\ep} + 8 +\ep \left(2 \zeta_3-\frac{7 \pi ^2}{6}+46\right)
+\ep^2 \left(-26 \zeta_3+\frac{83 \pi ^4}{360}-\frac{23 \pi ^2}{3}+216\right)
\\
& +\ep^3 \left(-\frac{1}{6} 89 \pi ^2 \zeta_3-180 \zeta_3+\frac{585 \zeta_5}{2}+\frac{17 \pi ^4}{180}-\frac{86 \pi ^2}{3}+776\right)
\\
& +  \ep^4 \left(-271 \zeta_3^2-\frac{89 \pi ^2 \zeta_3}{3}-536 \zeta_3+729 \zeta_5+\frac{7739 \pi ^6}{7560}
+\frac{4 \pi ^2}{3}-\frac{149 \pi ^4}{90}+1024\right),
\end{split} 
\ee
where we retained all contributions through weight six.

The results for all other integrals
which  appear in Eq.~(\ref{eq5.5})  can be obtained
along similar lines; they are given  in an ancillary file. 
Using them 
in Eq.~(\ref{eq5.5}), we obtain  a remarkably simple result for the integral of  $\omega_{n  \bar n}^{(3),a}$
in the $nnn$ configuration 
\begin{align}
\begin{split}
\int {\rm d} \Phi^{nnn}_{\theta \theta \theta} \, \omega^{(3),a}_{n\bar{n}}(k_1,k_2,k_3) 
  & ={} \frac{8}{\ep^5} - \frac{20\pi^2}{\ep^3} - \frac{584\zeta_3}{\ep^2} - \frac{86\pi^4}{15\ep}
  +  \left( 1472 \pi^2 \zeta_3 - 12480 \zeta_5  \right )
\\
& 
+ \ep \left( 22288 \zeta_3^2 - \frac{3796\pi^6}{945} \right ),
\end{split}
\label{eqn:result_Sa}
\end{align}
showing the usefulness of applying  integration-by-parts technology to  phase-space integrals with Heaviside functions.

%% file: sections/05_N3LO_3_b.tex
\subsection{The $\omega_{n  \bar n}^{(3),b}$ contribution}
\label{sec:n3lo_Sb}

We consider the second contribution to the zero-jettiness soft function given by  an integral of the
function  $\omega_{n  \bar n}^{(3),b}$.
This function  is constructed from the function $\oS_{ik}^{(b)}$, which reads~\cite{Catani:2019nqv}
\begin{footnotesize}
\begin{align}
\begin{split}
&\oS_{ik}^{(b)}=\frac{1}{16 \per \qq{1}{2} \per \pq{i}{12}} \per \sg\frac{7 \per \pp{i}{k}}{\pq{i}{1} \per \pq{i}{3} \per \pq{k}{2} \per \pq{k}{3}} \per (-\pp{i}{k} \per \pq{i}{12}) \\
&+\frac{(\pp{i}{k})^2}{\pq{i}{3}} \per \left[\frac{12 \per p_i(q_{1}-q_{2})}{\pq{k}{12} \per \pq{i}{1} \per \pq{k}{3}}+\frac{1}{\pq{k}{13}} \per \left(\frac{1}{\pq{k}{3}}-\frac{1}{\pq{k}{1}}\right) \per \left(3+\frac{\pq{i}{1}}{\pq{i}{2}}-\frac{2 \per \pq{k}{1}}{\pq{k}{2}}\right)\right] \dg  \\
&+\frac{1}{48 \per \qq{1}{2} \per \pq{k}{123}} \per \sg \frac{3}{\pq{k}{1} \per \pq{i}{2}}\left(\frac{1}{\pq{k}{12}}-\frac{1}{\pq{k}{3}}\right) \per \left(\frac{\pp{i}{k}}{\pq{i}{3}}\right) \per (\pp{i}{k} \per \pq{k}{12})    \\
&+3 \per \left(\frac{1}{\pq{k}{3}}-\frac{1}{\pq{k}{12}}\right) \per (\pp{i}{k} \per \pq{k}{12}) \per \left[\frac{\pp{i}{k}}{\pq{i}{23} \per \pq{k}{1}} \per \left(\frac{1}{\pq{i}{2}}-\frac{1}{\pq{i}{3}}\right)\right]\dg  \\
&+\frac{\pp{i}{k}}{48 \per \qq{1}{2} \per \pq{k}{123} \per \pq{i}{123}} \per \left(\frac{1}{\pq{i}{12}}-\frac{1}{\pq{i}{3}}\right) \per \pst 2 \per \pp{i}{k} \per p_k(q_{2}-q_{1})\pdt \\
&\times  \per \left[\left(\frac{1}{\pq{k}{3}}-\frac{3}{\pq{k}{12}}\right) \per \left(\frac{1}{\pq{k}{1}}-\frac{1}{\pq{k}{2}}\right)+\left(\frac{1}{\pq{k}{2}}-\frac{3}{\pq{k}{13}}\right) \per \left(\frac{1}{\pq{k}{1}}-\frac{1}{\pq{k}{3}}\right)\right] \;,
\end{split}
\label{sikb}
\end{align}
\end{footnotesize}%
where $p_i = n$, $p_k = \bar n$ and $q_i = k_i$, $i=1,2,3$ in our notations. 
At variance with 
$\oS_{ik}^{(a)}$, the function $\oS_{ik}^{(b)}$  contains a scalar product of two gluon momenta $q_1 q_2$ which
causes a collinear singularity in the limit $\VEC{q}_1 \parallel \VEC{q}_2$ and makes it more difficult
to integrate the function $\omega_{n  \bar n}^{(3),b}$ compared to the discussion in the
previous section.   However, there is no reason to expect that things  may
work differently for this contribution.

We have, therefore, proceeded as before and  performed a reduction to master integrals.
Upon checking the results numerically,  we have found
that reduction to master integrals   produces wrong results because
\emph{not all integrals that appear in the integration-by-parts identities in the course of the reduction
  are regulated dimensionally}.

It is interesting  to point out that a)  integrals that
appear in the function $\omega^{(3),b}_{n\bar{n}}(k_1,k_2,k_3) $
are well-defined in dimensional regularization 
 and
b) all master integrals that appear in the expression for the amplitude obtained
{\it after} the reduction  do not exhibit
singularities that cannot be regulated dimensionally.  This means that the failure
of dimensional regularization is very well hidden in the internal dynamic of the reduction
process and, therefore, hard to detect. 

To remedy  this situation,  we introduced an analytic regulator in addition to the dimensional one.
The analytic regulator is introduced in such a way that the phase-space element is multiplied by 
a factor
\be
{\rm d} \Phi_{fff}^{nnn} \to {\rm d} \Phi_{fff}^{nnn} \; (k_1 n)^\nu (k_2 n)^\nu (k_3 n)^\nu.
\label{eq5.14aaa}
\ee
This modification changes the integration-by-parts equations making them $\nu$-dependent
and, therefore, significantly more complicated. However, the main steps that we have described earlier in connection
with deriving IBP equations and establishing a hierarchy of integrals is not affected by the analytic regulator.

Having performed the reduction for finite values of $\nu$ and $\ep$, we have found that  the 
integral of $\omega^{(3),b}_{n\bar{n}}(k_1,k_2,k_3)$ 
can be written  in a remarkably simple  form which, however, demonstrates very clearly  why the analytic
regulator is needed. We obtain
\be
\int {\rm d} \Phi^{nnn}_{\theta \theta \theta} \, (k_1 n)^\nu (k_2 n)^\nu (k_3 n)^\nu \omega^{(3),b}_{n\bar{n}}(k_1,k_2,k_3)  =
 \sum \limits_{i=1}^{23} C^{(b)}_i(\ep,\nu) J_i(\ep,\nu). 
\label{eq5.15}
 \ee
 As indicated in Eq.~(\ref{eq5.15}), both the reduction coefficients and the integrals $J_{1,..,23}$ are
 functions of $\nu$ but, studying the $\nu \to 0$ limit of Eq.~(\ref{eq5.15}), 
we find that for  $i=1,2,..,22$ the following holds
\be
\lim_{\nu \to 0} C_i(\ep,\nu) = C_i(\ep),\;\;\; \lim_{\nu \to 0} J_i(\ep,\nu) = I_i.
\ee
The integrals  $I_{1,..,6}$ have been already discussed in the previous section and $I_{7,..,22}$ are another sixteen
integrals that can be evaluated at $\nu = 0$.  These integrals are given in Eq.~(\ref{eqn:mi_3g_nnn_b}).

The
last integral, $J_{23}(\ep,\nu)$,  has a $1/\nu$ pole but $C_{23}(\ep,\nu)$ is proportional to $\nu$.
Therefore,  $C_{23}(\ep,\nu) J_{23}(\ep,\nu)$ produces a finite result in the $\nu \to 0$ limit that, however,
is  completely missed if the reduction at $\nu=0$  is performed. 

Following this discussion and
taking the $\nu \to 0$ limit where appropriate, we write the result of the reduction as follows 
\begin{align}
  & \int {\rm d} \Phi^{nnn}_{\theta \theta \theta} \, \omega^{(3),b}_{n\bar{n}}(k_1,k_2,k_3)  =
  \nonumber \\
& 
  \frac{\left(10249456 \ep^5-6479980 \ep^4+713856 \ep^3+268429 \ep^2
  -67966 \ep+4287\right) }{200 \ep^5 (2 \ep-1) (4 \ep-1) (5 \ep-1)} I_1
\nonumber \\
&     + \frac{7}{5 \ep^3} I_2 -\frac{38 \left(504 \ep^3-270 \ep^2+37 \ep-1\right) }{5 \ep^4 (2 \ep-1)} I_3
\nonumber \\
& -\frac{\left(4260192 \ep^5-3531008 \ep^4+674380 \ep^3+124140 \ep^2-49897 \ep
  +3923\right) }{50 \ep^4 \left(40 \ep^3-38 \ep^2+11 \ep-1\right)}I_4
\label{eq5.17}
\\
& +\frac{\left(704 \ep^2-280 \ep+26\right) }{5 \ep^3-10 \ep^4} I_5
+\frac{9 \left(9288 \ep^4-7308 \ep^3+858 \ep^2+347 \ep-55\right) }{5 \ep^4 (4 \ep-1)} I_6
-\frac{237 }{4 \ep^2}I_7 
\nonumber \\
& -\frac{6 (4 \ep+1) }{\ep^3} I_8 +\frac{10 }{3 \ep^2} I_{9} -\frac{10 }{\ep^2} I_{10}
  +\frac{6 }{\ep^2}I_{11}
  +\frac{6 }{5 \ep^2}I_{12}
  -\frac{6 }{5 \ep^2}I_{13}
  +\frac{18 }{\ep^2}I_{14}
  -\frac{9 }{5 \ep^2}I_{15} 
\nonumber \\
&
  +\frac{(14-84 \ep) }{5 \ep^2} I_{16}  +\frac{12 }{5 \ep} I_{17}  +\frac{93 }{5 \ep} I_{18}
-\frac{12 }{\ep}I_{19}-\frac{39 }{\ep} I_{20}
  +\frac{22 (4 \ep-1) }{\ep^2} I_{21}
\nonumber \\
&   -\frac{7 }{\ep} I_{22}
  -\frac{4}{3 \ep^2} \left[ \lim_{\nu \to 0} \nu J_{\nu}(\ep,\nu) \right] \,, \nonumber
\end{align}
where we renamed $J_{23}$ to $J_\nu$. 

As we already mentioned the  integrals $I_{1..22}$ that appear in the above expression
can be found in Appendix~\ref{app:master_integrals}, Eqs.~(\ref{eqn:mi_3g_nnn_a},\ref{eqn:mi_3g_nnn_b}).   The  integral that is singular in the $\nu \to 0$ limit is defined as follows 
\be
J_{\nu}(\ep,\nu)
= \int \frac{ \DIFFL     \Phi^{nnn}_{\theta\delta\theta}
(k_1 n)^{\nu} (k_2 n)^{\nu} (k_3 n)^{\nu}
}{(k_1k_3)(k_1 n)(k_{12}\bar{n})( k_3 \bar {n})}.
\label{eq5.18}
\ee
We will now explain how 
this  ``pathological'' integral can be computed.


To integrate over the relative azimuthal angle between $\vec{k}_{1,\perp}$ and $\vec{k}_{3,\perp}$, we use the following formula 
\be
\begin{split} 
 \int \frac{{\rm d} \Omega^{(d-2)}}{\Omega^{(d-2)}}  \frac{1}{k_i k_j} 
  = 2 \Bigg \{ &
 \frac{ \theta(\alpha_i \beta_j - \alpha_j \beta_i) }{\alpha_i \beta_j}
 \HYPGF{1}{1+\ep}{1-\ep}{\frac{\alpha_j \beta_i}{\alpha_i \beta_j}}
\\
+ &
\frac{ \theta(\alpha_j \beta_i - \alpha_i \beta_j) }{\alpha_j \beta_i}
\HYPGF{1}{1+\ep}{1-\ep}{\frac{\alpha_i \beta_j}{\alpha_j \beta_i}}
\Bigg \},
\end{split} 
\label{eq2.12}
\ee
where $i=1$ and $j=3$. To proceed further, we
change variables $\alpha_{1,3} \to \beta_{1,3}/\xi_{1,3}$ and
   obtain the following integral representation 
   \be
   \begin{split}
 J_{\nu}
 & =
 \; 2 \int \prod \limits_{i=1}^{3}
{\rm d} \beta_i \beta_i^{-2\ep+\nu} \delta(1-\beta_{123})
\frac{{\rm d} \xi_1 {\rm d} \xi_3 (\xi_1\xi_3)^{\ep-1}  }{  \beta_1 \beta_3 ( \beta_1  + \beta_2 \xi_1)}  \, \Bigg\{  \xi_3 \theta(\xi_1-\xi_3)   \\
 &  \times \HYPGF{1}{1+\ep}{1-\ep}{\frac{\xi_3 }{\xi_1}}  + \xi_1 \theta(\xi_3 - \xi_1) \HYPGF{1}{1+\ep}{1-\ep}{\frac{\xi_1 }{\xi_3}}
  \Bigg\} \,.
\end{split}
\ee
We make a further change of variables, $\xi_3 = r f$ and $\xi_1 = f$   in the first term in the square brackets
and $\xi_3 = f$ and $\xi_1 = r f$ in the second one. We find
\be
\begin{split} 
J_{\nu}
 & =
 \; 2 \int \prod \limits_{i=1}^{3}
    {\rm d} \beta_i \beta_i^{-2\ep+\nu} \delta(1-\beta_{123})
 \;    {\rm d} r \; {\rm d} f\; r^\ep \; f^{2\ep} \; \HYPGF{1}{1+\ep}{1-\ep}{r }
    \\
& \times   \frac{1}{\beta_1 \beta_3} \Bigg (   \frac{1 }{  ( \beta_1  + \beta_2 f )}  + \frac{1 }{  ( \beta_1  + \beta_2 fr  )} 
  \Bigg  ).
\end{split}
  \ee
  It is convenient to write $\beta_1 = x(1-y), \; \beta_2 = xy$ and $\beta_3 = 1-x$, $0 < x,y < 1$.  Upon doing that, 
we find   that integrations over $x$ and $y$ can be readily performed. We obtain
  \be
\label{eq5.36}
    J_{\nu}
   =
\; 2 \frac{\Gamma(-4\ep + 2\nu) \Gamma(-2\ep + \nu)}{\Gamma(-6\ep+3 \nu)}
\frac{\Gamma(1-2\ep + \nu) \Gamma(-2\ep+\nu)}{\Gamma(1-4\ep+2\nu)} \; \tilde J_{\nu},
\ee
where 
\be
\begin{split}
& \tilde J_{\nu}  = 
\int {\rm d} f \; {\rm d} r \; f^{2\ep} r^{\ep} 
\Bigg  [
   \HYPGF{1}{1-2\ep+\nu}{1-4\ep+2\nu}{1-f}
\\
   & +
   \HYPGF{1}{1-2\ep+\nu}{1-4\ep+2\nu}{1-r f}
   \Bigg  ] \HYPGF{1}{1+\ep}{1-\ep}{r}.
\end{split}
\label{eq5.23}
  \ee
  The two hypergeometric functions in the square brackets
  in the last equation can be conveniently re-written  using the following identity
  \be
  \begin{split} 
& \HYPGF{1}{1-2\ep+\nu}{1-4\ep+2\nu}{z} = \\
& \;\;\;\;\;\;\;\;\;\;\;\;\; (1-z)^{-1-2\ep+\nu}  \HYPGF{-4\ep+2\nu}{-2\ep+\nu}{1-4\ep+2\nu}{z}.
  \label{eq5.24}
\end{split}
  \ee
    Inserting Eq.~(\ref{eq5.24}) into  Eq.~(\ref{eq5.23}), we obtain
  \be
\begin{split}
  \tilde J_{\nu} & = 
\int \limits_{0}^{1} {\rm d} f \; {\rm d} r \; f^{\nu-1} r^{\ep}  
\Bigg\{
  \HYPGF{-4\ep+2\nu}{-2\ep+\nu}{1-4\ep+2\nu}{1-f}
 \\
   & +
  r^{-2\ep+\nu -1} \HYPGF{-4\ep+2\nu}{-2\ep+\nu}{1-4\ep+2\nu}{1-rf}
   \Bigg\} \\
   & ~~~~~ \times \HYPGF{1}{1+\ep}{1-\ep}{r}.
\end{split}
\label{eq5.25}
\ee
From the above equation we readily  see how the  $1/\nu$-singularity  appears;   it is
generated by the singularity at $f=0$ in Eq.~(\ref{eq5.25}) which is not regulated dimensionally. 

Although we can compute  $\tilde J_{\nu}$ by  expanding it in Laurent  series in  $\ep$ and $\nu$, 
we only require the $1/\nu$ singularity of this integral. 
We therefore compute  the residue at $f=0$  and find 
\begin{align}
  &  {\tilde J}_{\nu} |_{\nu \to 0} = \frac{\HYPGF{-4\ep}{-2\ep}{1-4\ep}{1}}{\nu} 
   \int \limits_{0}^{1}  {\rm d} r \;  r^{\ep} \HYPGF{1}{1+\ep}{1-\ep}{r}
\Bigg  [
  1 +   r^{-2\ep -1}    \Bigg  ] 
\nonumber \\
& = \frac{\HYPGF{-4\ep}{-2\ep}{1-4\ep}{1}}{\nu}  \Bigg (
\frac{\Gamma(1+\ep) }{\Gamma(2+\ep)} \GENHYPGF{3}{2}{1,1+\ep,1+\ep}{1-\ep,2+\ep}{1}
\label{eq5.39} \\
& +
\frac{\Gamma(-\ep)}{\Gamma(1-\ep)} \GENHYPGF{3}{2}{1,1+\ep,-\ep}{1-\ep,1-\ep}{1}
\Bigg  ). 
\nonumber 
\end{align}
We now use  Eqs.~(\ref{eq5.36},\ref{eq5.39}), expand the result in $\ep$ and obtain
\be
\begin{split} 
\lim_{\nu \to 0} (\nu J_{\nu}) = -\frac{3}{2 \ep^3}& +\frac{\pi ^2}{\ep}+75 \zeta_3+\frac{77 \pi ^4 \ep}{20}
+\ep^2 \left(2025 \zeta_5-50 \pi ^2 \zeta_3\right) 
\\
& 
+\ep^3 \left(\frac{1787 \pi ^6}{210}-1875 \zeta_3^2\right) + \mathcal{O}\left(\ep^4\right).
\end{split} 
\ee

Calculation of the  master integrals $I_{1,2,...,22}$  required for computing the 
integral of the function  $\omega^{(3),b}_{n\bar{n}}$ proceeds in full analogy to   the above case  and for this
reason we do not discuss it here.  Finally, if we
use the reduction to master integrals Eq.~(\ref{eq5.17}) 
and explicit results for the master integrals that can be found in an ancillary file, 
 we obtain the
   following result for the integral of the function  $\omega^{(3),b}_{n\bar{n}}$ in the $nnn$ configuration
\begin{align}
\begin{split}
& \int {\rm d} \Phi^{nnn}_{\theta \theta \theta} \, \omega^{(3),b}_{n\bar{n}}(k_1,k_2,k_3) 
  = \frac{8}{\ep^5} + \frac{32}{\ep^4} + \frac1{\ep^3} \left( 64 - \frac{41\pi^2}{3} \right)
\\
&   + \frac1{\ep^2} \left(128 - 64 \pi^2 -774\zeta_3 \right) 
+ \frac1{\ep} \left( 256 - 128 \pi^2 - \frac{581\pi^4}{10} - 2144 \zeta_3 \right)
\\
& + \bigg( 512 - 256 \pi^2 - \frac{1688\pi^4}{15} - 4288 \zeta_3 
+ \frac{1306\pi^2 \zeta_3}{3}  - 28770\zeta_5 \bigg)
\\
& + \ep \bigg( 1024 -512 \pi^2 - \frac{3376 \pi^4}{15} - \frac{616\pi^6}{5}  - 8576 \zeta_3 + 2304 \pi^2 \zeta_3
\\
& + 19480 \zeta_3^2 - 68736 \zeta_5 \bigg ).
\end{split}
\label{eqn:result_Sb}
\end{align}
Again, we see that, apart from an unexpected (and interesting) complication  related to the need to introduce
an analytic regulator, it is  beneficial to make use of the IBP identities to compute
non-trivial contributions to the zero-jettiness soft function
at N3LO.

%% file: sections/05_N3LO_4_c.tex
\subsection{The $\omega_{n  \bar{n}}^{(3),c}$ contribution}
\label{sec:n3lo_Sc}

The third contribution to the zero-jettiness soft function is associated with the
integral of the  function $\omega_{n  \bar n}^{(3),c}$.
Similar to the other two contributions,
we construct $\omega_{n  \bar n}^{(3),c}$ from the function $\oS_{ik}^{(c)}$, which reads~\cite{Catani:2019nqv}
\begin{footnotesize}
\begin{align}
\begin{split}
&\oS_{ik}^{(c)}=\frac{1}{8 \per (\qq{1}{2})^2 \per \pq{i}{12} \per \pq{k}{3}} \per \sg \pst (4-d) \per \pq{i}{1}+d \per \pq{i}{2}\pdt \per \sq\frac{\pp{i}{k} \per \pq{k}{1}}{2 \per \pq{k}{123} \per \pq{i}{123}} \per \left(\frac{\pq{k}{3}}{\pq{k}{12}}-1\right) \per \left(\frac{\pq{i}{12}}{\pq{i}{3}}-1\right)\\
&+\frac{\pp{i}{k}}{\pq{i}{3}} \per \left(\frac{\pq{i}{1}}{\pq{i}{12}}-\frac{3}{2} \per \frac{\pq{k}{1}}{\pq{k}{12}}\right)\dq\\
&+\frac{\pp{i}{k}}{\pq{i}{123}} \per \left(\frac{1}{\pq{i}{3}}-\frac{1}{\pq{i}{12}}\right) \per \pst (4-d) \per (\pq{i}{1})^2+d \per \pq{i}{1} \per \pq{i}{2}\pdt\dg\\
&+\frac{1}{32 \per \qq{1}{2} \per \qq{1}{3} \per \pq{i}{12}} \per \sg\frac{\pp{i}{k}}{\pq{k}{2}} \per \sq\frac{4 \per p_k(q_{2}-q_{1})}{\pq{k}{3}}+\frac{2 \per \pq{i}{12}}{\pq{i}{3}}+2 \per \frac{\pq{i}{2} \per \pq{i}{3}+\pq{i}{1} \per \pq{i}{123}}{\pq{i}{13} \per \pq{i}{3}}\\
&+\frac{1}{\pq{k}{13} \per \pq{i}{3}} \per \pst\pq{i}{1} \per p_k (5\per q_1 -8\per q_2 +2\per q_3)-3 \per \pq{i}{2} \per \pq{k}{3}+4 \per \pp{i}{k} \per \qq{2}{3}\pdt\dq\\
&+\frac{2}{\pq{i}{123} \per \pq{k}{3}} \left(\frac{1}{\pq{i}{2}}-\frac{1}{\pq{i}{13}}\right) \pst \pp{i}{k} \per \pq{i}{12} \per \pq{i}{13} \pdt\\
&+\frac{2}{\pq{k}{123} \per \pq{k}{3}} \per \left(\frac{1}{\pq{k}{13}}-\frac{1}{\pq{k}{2}}\right) \psq 2 \per \pp{i}{k} \per \pq{k}{13} \per p_k(q_{2}-q_{1})\pdq+\frac{1}{\pq{k}{123} \per \pq{i}{123}} \per \left(1-\frac{\pq{i}{12}}{\pq{i}{3}}\right) \per \\
& \times \left(\frac{1}{\pq{k}{13}}-\frac{1}{\pq{k}{2}}\right) \per \psq 4 \per (\pp{i}{k})^2 \per \qq{2}{3}+\pp{i}{k} \pst\pq{i}{1} \per p_k (5\per q_1-8\per q_2+2\per q_3)-3 \per \pq{i}{2} \per \pq{k}{3}\pdt \pdq\dg,
\end{split}
\label{sikbarc}
\end{align}
\end{footnotesize}%
where $p_i = n$, $p_k = \bar n$ and $q_i = k_i$, $i=1,2,3$ in our notations. 
At variance with  the functions $\oS_{ik}^{(a,b)}$, the function $\oS_{ik}^{(c)}$ contains propagators $1/(q_1 q_3)$ and
$1/(q_2 q_3)$ and, therefore, exhibits a  more complicated singularity  structure.  However, 
no additional issues 
arise when modified IBP relations are constructed and used to reduce required integrals,  so that 
working with the analytic regulator introduced in the previous section, see Eq.~(\ref{eq5.14aaa}),
suffices. 

Performing the reduction to master integrals and taking the $\nu \to 0$ limit where appropriate, we obtain 
\begin{align}
& \int {\rm d} \Phi^{nnn}_{\theta \theta \theta} \, \omega^{(3),c}_{n\bar{n}}(k_1,k_2,k_3)  = \notag \\
& \bigg( -\frac{5152}{675 \ep^5}+\frac{60883}{1350\ep^4}+\frac{2218663}{5400\ep^3}-\frac{33423797}{10800\ep^2}-\frac{49850253233\ep}{466560}+\frac{44313583}{12960  \ep}+\frac{40023347}{15552} \bigg) I_1 \notag \\
& +\bigg(-\frac{262}{135 \ep^3}+\frac{8}{135 \ep^2}-\frac{1280 \ep}{9}-\frac{16}{9 \ep}+\frac{160}{9}\bigg)  I_2 +\bigg(\frac{2242}{135 \ep^4} -\frac{23954}{135 \ep^3}+\frac{74967872 \ep^2}{1215} \notag \\
& +\frac{14596}{45  \ep^2}+\frac{2845312 \ep}{405}+\frac{69832}{45  \ep}-\frac{733168}{135}\bigg) I_3 +\bigg(-\frac{47683}{675 \ep^4}+\frac{27427}{25 \ep^3} -\frac{7899529}{1350  \ep^2} \notag \\
& +\frac{11094485353  \ep^2}{12960}-\frac{119650151 \ep}{2160}+\frac{7137263}{900  \ep}+\frac{5243129}{360}\bigg) I_4  +\bigg(\frac{334}{135 \ep^3}-\frac{600703 \ep^2}{81} \notag \\
& -\frac{794}{45 \ep^2}+\frac{38854  \ep}{27}+\frac{5036}{135 \ep}-\frac{2024}{9}\bigg)   I_5+\bigg( -\frac{1202}{15 \ep^4}+\frac{4022}{3\ep^3}+\frac{6077660177 \ep^2}{6480}-\frac{121631}{15 \ep^2} \notag \\
& -\frac{76894303  \ep}{1080}+\frac{170291}{10 \ep}+\frac{534641}{180} \bigg)  I_6 \notag +\bigg(-3211 \ep^4+1409 \ep^3-151 \ep^2-\frac{9}{2 \ep^2}+119 \ep \notag \\
& +\frac{3}{2 \ep}+44\bigg) I_7+\bigg(\frac{16640 \ep^4}{3}-\frac{3968  \ep^3}{3}+\frac{1088  \ep^2}{3}-\frac{14}{3 \ep^2}-\frac{224  \ep}{3}-\frac{8}{3 \ep}+\frac{80}{3}\bigg)   I_9 +\frac{65 }{9 \ep^2} I_{10} \notag \\
& +\bigg(-\frac{430971 \ep^4}{32}+\frac{238589 \ep^3}{48}-\frac{21817 \ep^2}{24}-\frac{47}{9 \ep^2}+\frac{1527 \ep}{4}+\frac{337}{9 \ep}+\frac{47}{6}\bigg) I_{11} +\bigg(\frac{1613075 \ep^4}{96} \notag \\
& -\frac{118055 \ep^3}{48}+\frac{8195\ep^2}{24}-\frac{13}{15 \ep^2}-\frac{515 \ep}{12}+\frac{25}{6}\bigg) \bigg[ I_{12} - I_{13} \bigg] +\bigg(-\frac{127360 \ep^4}{3}+\frac{18560 \ep^3}{3} \notag \\
& -\frac{2560  \ep^2}{3}  -\frac{578}{45 \ep^2}+\frac{320 \ep}{3}+\frac{16}{45 \ep}-\frac{32}{3}\bigg) I_{14}+\bigg(-\frac{377080 \ep^4}{9}+\frac{54680\ep^3}{9}-\frac{7480 \ep^2}{9} \notag \\
& +\frac{19}{15 \ep^2}+\frac{920 \ep}{9}+\frac{8}{45 \ep}-\frac{88}{9}\bigg) I_{15} +\bigg(-\frac{110384 \ep^3}{3}+\frac{21808 \ep^2}{3}-\frac{92}{15 \ep^2}-\frac{4496 \ep}{3} \notag \\
& +\frac{212}{15 \ep}+\frac{832}{3}\bigg) I_{16}  +\bigg(-\frac{1075877 \ep^5}{16}+\frac{114497 \ep^4}{8}-\frac{10397 \ep^3}{4}+\frac{1217 \ep^2}{2}-17 \ep \notag \\
& +\frac{68}{5 \ep}+14\bigg) I_{17}+\bigg(\frac{1111653 \ep^5}{16}-\frac{127713  \ep^4}{8}+\frac{9933 \ep^3}{4}-\frac{1593 \ep^2}{2}-87 \ep+\frac{12}{5\ep}-36\bigg) I_{18} \label{eq5.30} \\
& +\bigg(\frac{2546205 \ep^5}{16}-\frac{275865 \ep^4}{8}+\frac{22005 \ep^3}{4}-\frac{3465 \ep^2}{2}-15 \ep+\frac{18}{\ep}-150\bigg) I_{19} \notag \\
& +\bigg(-172 \ep^3+44 \ep^2+32 \ep+\frac{2}{\ep}-14\bigg) I_{20}  +\bigg(64 \ep^3+32 \ep^2-\frac{4}{3 \ep^2}+16  \ep+\frac{4}{3 \ep}+8\bigg) I_{21}  \notag \\
&+\bigg(1968 \ep^3-320 \ep^2+\frac{172 \ep}{3}+\frac{40}{9 \ep}-\frac{68}{9}\bigg) I_{22} +\bigg(-312 \ep^3+\frac{151}{60 \ep^3}-\frac{4532  \ep^2}{3}+\frac{73}{6 \ep^2} \notag \\
& -270 \ep-\frac{69}{10 \ep}-\frac{353}{3}\bigg)  I_{23} +\bigg(3840 \ep^4-672 \ep^3+96 \ep^2-24 \ep-\frac{2}{\ep}\bigg) I_{24} +\bigg(69719 \ep^4 \notag \\
&-9413 \ep^3+2419  \ep^2+\frac{1}{3 \ep^2}-107 \ep-\frac{125}{18 \ep}+\frac{920}{9}\bigg) I_{25} +\bigg(\frac{83680 \ep^4}{3}-\frac{13280 \ep^3}{3}+\frac{2080 \ep^2}{3} \notag \\
& +\frac{4}{45 \ep^2}-\frac{320 \ep}{3}-\frac{104}{45 \ep}+16\bigg) I_{26}  +\bigg(-96 \ep^4-48 \ep^3-24 \ep^2-12 \ep-\frac{3}{\ep}-6\bigg)    I_{27} \notag \\
& +\bigg(\frac{4480 \ep^5}{3}+\frac{2240 \ep^4}{3}+\frac{1120 \ep^3}{3}+\frac{560 \ep^2}{3}+\frac{4}{3 \ep^2}+\frac{280 \ep}{3}+\frac{91}{6 \ep}+\frac{140}{3}\bigg) I_{28} \notag \\
& +\bigg(117120 \ep^5-19200 \ep^4+3360  \ep^3-480 \ep^2-\frac{2}{\ep^2}+120  \ep+\frac{2}{3 \ep}\bigg) I_{29} +\bigg(34560 \ep^5 \notag \\
& -5760 \ep^4+960 \ep^3-160 \ep^2+\frac{80 \ep}{3}+\frac{20}{9 \ep}-\frac{40}{9}\bigg) I_{30} +\bigg(-96576 \ep^5+15216 \ep^4 \notag \\
& -2976 \ep^3+276 \ep^2+\frac{2}{3 \ep^2}-156 \ep-\frac{9}{\ep}-29\bigg) I_{31} +\bigg(-\frac{40832 \ep^5}{3}+\frac{7232  \ep^4}{3}-\frac{992 \ep^3}{3} \notag \\
& +\frac{272 \ep^2}{3}+\frac{7}{18 \ep^2}+\frac{8 \ep}{3}+\frac{29}{18 \ep}+\frac{76}{9}\bigg) I_{32} +\bigg(-5728 \ep^4+1048 \ep^3-128 \ep^2-\frac{37}{18 \ep^2} \notag \\
& +\frac{134 \ep}{3}-\frac{9}{2 \ep}+\frac{38}{9}\bigg) I_{33} +\bigg(\frac{93152 \ep^5}{3}-\frac{15632 \ep^4}{3}+\frac{2552 \ep^3}{3}-\frac{452 \ep^2}{3}-\frac{7}{18 \ep^2}+\frac{62 \ep}{3} -\frac{1}{6 \ep} \notag  \\
& -\frac{17}{3}\bigg)  I_{34} +\bigg(34695 \ep^5-5805 \ep^4+975 \ep^3-165  \ep^2+\frac{85 \ep}{3}+\frac{5}{3  \ep}-5\bigg) \bigg[ I_{35} - I_{36} + I_{37} \bigg] \notag \\
& +\frac{4 }{3 \ep} \bigg[ I_{38} - I_{39} \bigg]   + \bigg(-\frac{16640 \ep^4}{3}+\frac{3968 \ep^3}{3}-\frac{1088 \ep^2}{3}+\frac{8}{3 \ep^2} +\frac{224 \ep}{3}+\frac{8}{3 \ep} \notag \\
& -\frac{80}{3}\bigg) \left[ \lim_{\nu \to 0} \nu J_\nu \right] + \mathcal{O}(\ep^2)  \,.   \notag
\end{align}

All the master integrals $I_{1,..,39}$ that appear in Eq.~(\ref{eq5.30}) are  computed at $\nu = 0$; their definition can be
found in Appendix~\ref{app:master_integrals}, Eqs.~(\ref{eqn:mi_3g_nnn_a},\ref{eqn:mi_3g_nnn_b},\ref{eqn:mi_3g_nnn_c}).  The only $\nu$-dependent integral $J_\nu$ present in Eq.~(\ref{eq5.30}) is the same
integral which we have seen
(and computed)  in the previous section. 

In principle, computation of the $I$-integrals that appear in the calculation of the case ``c''
does not differ from what we already
discussed for cases ``a'' and ``b''.  However,   integrals with two separate scalar products of
parton's four-momenta are significantly more complicated than what we have discussed  so far and,  to illustrate this
point,  we will discuss how one  of them can be computed. 

We consider integral  $I_{30}$ defined as 
   \be
   I_{30} = \int  \frac{\DIFFL     \Phi^{nnn}_{\theta\delta\theta }}{(k_1k_2)(k_1 k_3 ) (k_{12} n) (k_{13} \bar{n})}.   \ee
   We can use Eq.~(\ref{eq2.12}) to perform integration over azimuthal angles of $k_2$ and $k_3$.
   We note, however, that
   the momentum $k_2$ is special because the $\delta$-function in the definition of $I_{30}$ implies
   that $\alpha_2 = \beta_2$. This condition removes one of the two hypergeometric
   functions in Eq.~(\ref{eq2.12}).
   Integrating over azimuthal angles and changing variables $\alpha_{1,3} = \beta_{1,3}/\xi_{1,3}$, we arrive at
   the following representation for $I_{30}$
   \be
\begin{split} 
   & I_{30} = 4
   \int \prod \limits_{i=1}^{3} {\rm d} \beta_i \; \beta_i^{-2\ep-1} \; \delta(1-\beta_{123}) \;
  \frac{{\rm d} \xi_1 {\rm d} \xi_3 \;  \xi_1^{\ep} \xi_3^{\ep -1} \; \beta_3 
   }{(\beta_1 + \beta_2) (\beta_1 \xi_3 + \beta_3 \xi_1 )}
   \\
   & \times  \HYPGF{1}{1+\ep}{1-\ep}{ \xi_1}
   \Bigg  [ \xi_1 \theta(\xi_3 - \xi_1)
     \HYPGF{1}{1+\ep}{1-\ep}{\frac{\xi_1 }{\xi_3}}
  \\
   & + \xi_3 \theta(\xi_1 - \xi_3) \HYPGF{1}{1+\ep}{1-\ep}{\frac{\xi_3}{\xi_1}}
  \Bigg  ].
   \end{split}
     \ee
     As the next step, we remove $\delta(1-\beta_{123})$ by integrating over $\beta_2$ and change variables
     $(\beta_1, \beta_3) \to (x,y)$  according to the following formula     $\beta_1 = x(1-y), \beta_3 = xy$. 
     Integration over $x$ leads to a hypergeometric function. We obtain
     \begin{align}
     I_{30} & = 
     \frac{4 \Gamma(-4\ep)\Gamma(-2\ep)}{\Gamma(-6\ep) }
     \int {\rm d} y \; {\rm d} \xi_1 {\rm d} \xi_3
     \frac{y^{-2\ep} (1-y)^{-2\ep-1} \xi_1^{\ep} \xi_3^{\ep}  }{(1-y) \xi_3 + y \xi_1 }
   \nonumber   \\
   & \times \HYPGF{1}{-4\ep}{-6\ep}{y} \; \HYPGF{1}{1+\ep}{1-\ep}{\xi_1}
   \label{eq4.29}
     \\
     &  \times \Bigg  [ \frac{\xi_1}{\xi_3} \theta(\xi_3 - \xi_1)
       \HYPGF{1}{1+\ep}{1-\ep}{\frac{\xi_1 }{\xi_3}}
       + \theta(\xi_1- \xi_3)
      \HYPGF{1}{1+\ep}{1-\ep}{\frac{\xi_3}{\xi_1}}
  \Bigg  ]. \nonumber 
     \end{align}

     The integral naturally splits into two parts. To proceed,  we change variables
     $\xi_1 = r \xi,\; \xi_3 = \xi$,  and  $\xi_3 = r \xi, \;\; \xi_1 = \xi$, in the first and
     the second integral in Eq.~(\ref{eq4.29}), respectively.
     We also rewrite the hypergeometric functions that appear in square brackets in the above equation 
     using the transformation
     \be
     \HYPGF{a}{b}{c}{z} =(1-z)^{c-a-b}\HYPGF{c-a}{c-b}{c}{z}.
\ee
     We then write 
 \be 
     I_{30}  =      \frac{4 \Gamma(-4\ep)\Gamma(-2\ep)}{\Gamma(-6\ep) } \left (
  {\cal I}_{30}^{(a)} + {\cal I}_{30}^{(b)}  
     \right ),
     \ee
     where
\begin{align} 
        {\cal I}_{30}^{(a)}
        & = \int \DIFFL y \DIFFL \xi \DIFFL r \;
        \frac{y^{-2\ep} (1-y)^{-4\ep-2} \; \xi^{2\ep} \; r^{1+\ep} \left[(1-r)(1-r \xi)\right]^{-1-2\ep} }{1 - y(1-r)} \nonumber \\
&  ~~~~~~~~~~~\times  \HYPGF{-1-6\ep}{-2\ep}{-6\ep}{y} \; \HYPGF{-\ep}{-2\ep}{1-\ep}{r \xi} \label{eq4.31} \\
& ~~~~~~~~~~~ \times  \HYPGF{-\ep}{-2\ep}{1-\ep}{r}.
  \nonumber 
\end{align}
        and
             \begin{align}
        {\cal I}_{30}^{(b)}
        & = \int {\rm d} y {\rm d} \xi  {\rm d} r \;
        \frac{y^{-2\ep} (1-y)^{-4\ep-2} \; \xi^{2\ep} \; r^{\ep}(1-r)^{-2\ep-1} \; }{r + y(1-r)}  \nonumber        \\
& ~~~~~~~~~~~  \times  \HYPGF{-1-6\ep}{-2\ep}{-6\ep}{y} \; \HYPGF{1}{1+\ep}{1-\ep}{\xi} \\
& ~~~~~~~~~~~  \times \;\HYPGF{-\ep}{-2\ep}{1-\ep}{r}. \nonumber
\end{align}

             The difficulty with computing these integrals is a power-like singularity at $y=1$. To extract and
             isolate it,
        we transform the $y$-dependent hypergeometric function in the following way
        \be
\begin{split} 
 &  \HYPGF{-1-6\ep}{-2\ep}{-6\ep}{y}  = \frac{\Gamma(-6\ep) \Gamma(1+2\ep)}{\Gamma(-4\ep)} y^{1+6\ep}
  \\
  & \;\;\;\;\;\;\;
  +(1-y)^{1+2\ep} \frac{\Gamma(-1-2\ep) \Gamma(-6\ep)}{\Gamma(-1-6\ep) \Gamma(-2\ep)} \HYPGF{1}{-4\ep}{2+2\ep}{1-y}.
\end{split}
\label{eq2.22}
        \ee
        This representation is helpful because the $y$-dependence of the first term in Eq.~(\ref{eq2.22})
        is simple so that integration 
        over $y$ can be immediately performed,  and the second term in   Eq.~(\ref{eq2.22})
        leads to integrals with only a logarithmic singularity at $y=1$.

        To proceed further, we consider integral ${\cal I}_{30}^{(a)}$ and write it as 
        \be
{\cal I}_{30}^{(a)} = {\cal I}_{30}^{(a,1)} + {\cal I}_{30}^{(a,2)},
        \ee
        where the two terms correspond to the two terms on the right hand side of Eq.~(\ref{eq2.22}). To compute
        ${\cal I}_{30}^{(a,1)}$ we integrate over $y$ and obtain 
        \be
 \int \DIFFL y \frac{y^{1+4\ep} (1-y)^{-4\ep-2} }{1-y(1-r)} = \Gamma(2+4\ep) \Gamma(-1-4\ep) \; r^{-2-4\ep}.
        \ee
We then use this result in the expression for ${\cal I}_{30}^{(a,1)}$ and write 
\be
\begin{split} 
           {\cal I}_{30}^{(a,1)} & = -\frac{\Gamma(2+4\ep)  \Gamma(-6\ep) \Gamma(1+2\ep)}{
             (1+4\ep)}
           \int {\rm d} \xi \; {\rm d} r \; \xi^{2\ep} r^{-1-3 \ep} (1-r)^{-2\ep-1} (1-\xi r)^{-2\ep-1}
           \\
           & \;\;\;\;\;\;\;\;\;\; \times \HYPGF{-\ep}{-2\ep}{1-\ep}{r \xi} \; \HYPGF{-\ep}{-2\ep}{1-\ep}{r}.
           \end{split}
\ee
This integral has a singularity at $r=0$ and another (overlapping) singularity at $r=1,\xi=1$.
The two singularities  can be separated by multiplying the integrand with  $1 = (1-r) +r$.
The first term in the sum  removes the $r=1$ singularity.
Since the $r=0$ singularity does not overlap with any other singularity, it can be easily extracted.
On the contrary, the $r=1$ singularity overlaps with $\xi=1$ singularity and for this reason a slightly
more sophisticated treatment is needed.  To this end, we subtract the product of hypergeometric functions at
$r=1$ and add it back.  When the difference 
\be
\begin{split} 
& \HYPGF{-\ep}{-2\ep}{1-\ep}{r \xi} \; \HYPGF{-\ep}{-2\ep}{1-\ep}{r}
\\
& -  \HYPGF{-\ep}{-2\ep}{1-\ep}{\xi} \; \HYPGF{-\ep}{-2\ep}{1-\ep}{1}, 
\end{split}
\ee
is used in the integrand, it becomes non-singular at $r =1$,  so that we can expand it in $\ep$ and integrate. On the other
hand, hypergeometric functions in the subtraction term do not depend on $r$ so that integration over $r$ becomes straightforward.
The resulting one-dimensional integration over $\xi$ contains
a logarithmic singularity at $\xi =1$ that can be easily isolated and extracted.
Putting everything together, we find
\be
\begin{split}
{\cal I}_{30}^{(a,1)} & =     -\frac{1}{18 \ep^2}+\frac{1}{48 \ep^3} + \frac{1}{\ep}\left ( \frac{5}{18}+\frac{\pi ^2}{6} \right ) 
+\frac{43 \zeta_3}{12}+\frac{5 \pi ^2}{108}-\frac{25}{18}
\\
&
+\ep \left(\frac{61 \zeta_3}{6}+\frac{437 \pi ^4}{540}
+\frac{125}{18}
-\frac{13 \pi ^2}{108}   \right)
+\ep^2 \left(\frac{89 \pi ^2 \zeta_3}{9}-\frac{85 \zeta_3}{2}
+\frac{845 \zeta_5}{4}+\frac{209 \pi ^4}{162}
\right.
\\
& \left.
+\frac{17 \pi ^2}{108}-\frac{625}{18}\right)
+\ep^3 \left(-32 \zeta_3^2+\frac{304 \pi ^2 \zeta_3}{9}+\frac{1075 \zeta_3}{6}
+\frac{3059 \zeta_5}{6}+\frac{57637 \pi ^6}{17010}
\right.
\\
& \left.
+\frac{107 \pi ^2}{108}+\frac{3125}{18}
  -\frac{2194 \pi ^4}{405}\right).
  \end{split} 
\ee

The computation of ${\cal I}_{30}^{(a,2)}$ proceeds in the following way. First, we observe that
in this case   there are
two singularities, $y=1$ and
$r=1$. We note that the latter  overlaps with the $\xi=1$ singularity.
To disentangle overlapping singularities, we replace  
\be
\begin{split}
& \frac{r^{1+\ep}}{1-y(1-r)} \HYPGF{-\ep}{-2\ep}{1-\ep}{\xi r} \HYPGF{-\ep}{-2\ep}{1-\ep}{r}  \to 
\\
 & \HYPGF{-\ep}{-2\ep}{1-\ep}{\xi} \HYPGF{-\ep}{-2\ep}{1-\ep}{1}, 
\end{split}
\label{eq5.44}
\ee
in Eq.~(\ref{eq4.31}) and add the difference of the two terms back. In the difference, the $r=1$ singularity
is regulated,  so that
one only needs to extract a logarithmic $y=1$ singularity.   To compute the contribution of the subtraction term (the r.h.s. of Eq.~(\ref{eq5.44})),
we can integrate over $y$ and  over $r$ to obtain yet another hypergeometric function of $\xi$. The resulting integral over $\xi$ has a logarithmic 
singularity at $\xi=1$ which can be easily extracted.  We obtain
\be
\begin{split} 
{\cal I}_{30}^{(a,2)} & =
-\frac{1}{16 \ep^3} -\frac{1}{4 \ep^2} +\frac{1}{\ep} \left ( 1-\frac{\pi ^2}{4} \right )
-\frac{15 \zeta_3}{4}-\frac{2 \pi ^2}{3}-4 +\ep \left(-23 \zeta_3+\frac{8 \pi ^2}{3}
\right.
\\
& \left.
-\frac{23 \pi ^4}{36}+16\right)
+\ep^2 \left(\frac{23 \pi ^2 \zeta_3}{6}+104 \zeta_3-\frac{595 \zeta_5}{4}
-\frac{32 \pi ^2}{3}
-\frac{47 \pi ^4}{30}-64\right)
\\
& +\ep^3 \left(249 \zeta_3^2+\frac{2 \pi ^2 \zeta_3}{3}-476 \zeta_3-577 \zeta_5+\frac{67 \pi ^4}{10}+\frac{128 \pi ^2}{3}-\frac{4756 \pi ^6}{2835}+256\right).
\end{split}   
\ee

Computation of the integral ${\cal I}_{30}^{(b)}$  proceeds in a similar way except that the integration over $\xi$ can be performed right away.
We find
\be
\begin{split} 
        {\cal I}_{30}^{(b)}
        & =   \frac{1}{1+2\ep} \GENHYPGF{3}{2}{1,1+\ep,1+2\ep}{1-\ep,2+2\ep}{1}
        \\
& \times         \int {\rm d} y \;  {\rm d} r \;
        \frac{y^{-2\ep} (1-y)^{-4\ep-2}  \; r^{\ep}(1-r)^{-2\ep-1} \; }{r + y(1-r)}
        \\
&        \times  \HYPGF{-1-6\ep}{-2\ep}{-6\ep}{y} \; 
        \;\HYPGF{-\ep}{-2\ep}{1-\ep}{r}.
\end{split} 
        \ee
        We then rewrite $\HYPGF{-1-6\ep}{-2\ep}{-6\ep}{y} $ using 
Eq.~(\ref{eq2.22})
        and integrate the two terms that appear in that
        equation separately. This integration is relatively straightforward since integration
        of the first term in  Eq.~(\ref{eq2.22})
        leads to yet another $_3F_2$-function and integration of the second term does not require resolution of any overlapping singularities.

        Putting everything together, we arrive at the
        following result for the integral $I_{30}$
        \be
\begin{split} 
        I_{30} & =  
        \frac{3}{8 \ep^4}
        +\frac{13}{6 \ep^3}
        +\frac{1}{\ep^2} \left ( -\frac{109}{12}-\frac{\pi ^2}{12} \right )
          +\frac{1}{\ep} \left ( -31 \zeta_3+\frac{5 \pi ^2}{9}+\frac{461}{12} \right ) -76 \zeta_3
\\
   &  -\frac{1969}{12} -\frac{23 \pi ^2}{18}-\frac{1211 \pi ^4}{360}
+\ep \left(-\frac{311}{6}  \pi ^2 \zeta_3+292 \zeta_3-\frac{3111 \zeta_5}{2}+\frac{7 \pi ^2}{18}+\frac{8501}{12}
\right.
\\
&
\left.
-\frac{355 \pi ^4}{36}  \right)
+ \ep^2 \left(18 \zeta_3^2-\frac{727 \pi^2 \zeta_3}{3}-1108 \zeta_3-4035 \zeta_5+\frac{1147 \pi ^4}{30}
\right.
\\
&
\left.
+\frac{397 \pi ^2}{18}
        -\frac{37129}{12}-\frac{10729 \pi ^6}{810}\right).
\end{split} 
  \ee

  Finally, using  reduction to master integrals and explicit expressions for
  integrals given in the ancillary file, we arrive at the following result for the
  integral of $\omega_{n  \bar n}^{(3),c}$ over the $nnn$ phase space
\begin{align}
\begin{split}
& \int {\rm d} \Phi^{nnn}_{\theta \theta \theta} \, \omega^{(3),c}_{n\bar{n}}(k_1,k_2,k_3) \\
={} & -\frac{4}{\ep^5}  + \frac{70}{3\ep^4} + \frac1{\ep^3} \bigg( \frac{920}{9} + \frac{19\pi^2}{3}  \bigg) + \frac1{\ep^2} \bigg( \frac{8527}{27} + \frac{122\pi^2}{3} + 162\zeta_3 \bigg) \\
& + \frac1{\ep} \bigg(\frac{67193}{81} + \frac{1280\pi^2}{9} + \frac{197 \pi^4}{90} + \frac{2732 \zeta_3}{3} \bigg) + \bigg( \frac{558745}{243} + \frac{10990\pi^2}{27} + \frac{439\pi^4}{9} \\
& + \frac{32032 \zeta_3}{9} - \frac{1604\pi^2 \zeta_3}{3} + 4204\zeta_5 \bigg) + \ep \bigg( \frac{4074557}{729} + \frac{89138\pi^2}{81} + \frac{28024 \pi^4}{135} \\
& - \frac{23029\pi^6}{2835} + \frac{288992\zeta_3}{27} - \frac{9224\pi^2\zeta_3}{3} - 7604\zeta_3^2 + 50296 \zeta_5 \bigg).
\end{split}
\label{eqn:result_Sc}
\end{align}

%% file: sections/05_N3LO_5_d.tex
\subsection{Differential equations and  $\omega_{n  \bar n}^{(3),d}$ contribution}
\label{sec:n3lo_Sd}

Application of integration-by-parts identities to phase-space integrals with
Heaviside functions opens up an opportunity to compute complicated
integrals using differential equations. We will illustrate this by considering the 
 contribution of the function $\omega_{n  \bar n}^{(3),d}$,  constructed using the function  $\oS_{ik}^{(d)}$ defined
in Ref.~\cite{Catani:2019nqv}, to the zero-jettiness soft function. The function  $\oS_{ik}^{(d)}$ reads
\begin{footnotesize}
\begin{align}
&\oS_{ik}^{(d)}=\frac{-(\pp{i}{k})^2}{4 \per q_{123}^2 \per \pq{k}{123} \per \pq{k}{1} \per \pq{i}{2} \per \pq{i}{3}}+\frac{1}{4 \per q_{123}^2 \per \pq{k}{123}} \per \sg\frac{1}{\qq{1}{2}} \per \sq \frac{1}{\pq{k}{1} \per \pq{i}{2}} \per \bigg( \pp{i}{k} \per (\qq{2}{3}-\qq{1}{3}) \per \pst\frac{\pp{i}{k}}{\pq{i}{3}}\pdt\nonumber\\
&+2 \per \frac{\pp{i}{k}}{\pq{i}{3}} \per (\pq{i}{12} \per \pq{k}{123}-\pq{i}{3} \per \pq{k}{2}-\pq{i}{2} \per \pq{k}{3})\dt\nonumber\\
&+\frac{1}{2 \per \pq{i}{13} \per \pq{k}{2}} \psg\pst\frac{1}{\pq{i}{1}}-\frac{1}{\pq{i}{3}}\pdt\nonumber\\
&\times \psq \pp{i}{k} \pst \pq{k}{1} \per p_i(4\per q_{3}-3\per q_{12})-3 \per \pq{k}{2} \per \pq{i}{12}+\pq{k}{3} \per p_i(q_{1}-3 q_{2})+2 \per \pp{i}{k} \per (\qq{2}{3}-\qq{1}{3})\pdt\pdq\nonumber\\
&+ 4 \per \pp{i}{k} \per p_k(q_{3}-q_{1})\pdg\nonumber\\
&+\frac{1}{\pq{i}{12} \per \pq{k}{3}} \per \psg\pst \frac{1}{\pq{i}{1}}-\frac{1}{\pq{i}{2}}\pdt \per \pst -2 \per \pp{i}{k} \per \pq{i}{1} \per \pq{k}{123}\pdt\nonumber\\
&+\pp{i}{k} \per p_k(q_{1}-3\per q_{2}-q_{3})\pdg +\frac{1}{\pq{k}{13} \per \pq{i}{2}} \psg \pp{i}{k} \per \pq{k}{13}\nonumber\\
&+\pst\frac{1}{\pq{k}{3}}-\frac{1}{\pq{k}{1}}\pdt \per \psq \frac{\pp{i}{k}}{2} \per \pq{k}{123} \per \pq{k}{12}\pdq\pdg +\frac{1}{\pq{k}{12} \per \pq{i}{3}} \psg\pp{i}{k} \per p_k(3 \per q_{1}+q_{2}-3 \per q_{3})\nonumber\\
&+\pst\frac{1}{\pq{k}{2}}-\frac{1}{\pq{k}{1}}\pdt \per \psq 2 \per \pp{i}{k} \per \pq{k}{1} \per \pq{k}{13}\pdq \pdg \nonumber \\
&+\frac{2}{\pq{i}{123}} \psg\pst\frac{\pq{k}{13}}{\pq{k}{2}}-1\pdt \pst\frac{\pp{i}{k}}{2}\pdt + \pst\frac{1}{\pq{k}{12}}-\frac{1}{\pq{k}{3}}\pdt \pst-\pp{i}{k} \per p_k (q_{2}+2 \per q_{3})\pdt\nonumber\\
&+\frac{1}{3} \per \pst\frac{1}{\pq{k}{3}}-\frac{3}{\pq{k}{12}}\pdt \pst\frac{1}{\pq{k}{1}}-\frac{1}{\pq{k}{2}}\pdt \pst 2 \per \pp{i}{k} \per \pq{k}{2} \per p_k(q_{12}-q_{3})\pdt\nonumber\\
&+\frac{1}{12} \per \pst\frac{1}{\pq{k}{2}}-\frac{3}{\pq{k}{13}}\pdt \pst\frac{1}{\pq{k}{1}}-\frac{1}{\pq{k}{3}}\pdt \psq 4 \per \pp{i}{k} \per p_k(q_{1}-q_{2}) \per p_k(q_{3}-q_{12})\pdq\pdg\dq\nonumber\\
& +\frac{1}{(\qq{1}{2})^2} \sq \frac{2}{\pq{i}{123}} \per \left(\frac{1}{\pq{k}{3}}-\frac{1}{\pq{k}{12}}\right) \label{eqn:skid} \\
&\times  \per \psq\pp{i}{k} \per \qq{1}{3} \pst(d-4) \per \pq{k}{1}-d \per \pq{k}{2}\pdt +2 \per (d-2) \per \pq{i}{2} \per (\pq{k}{1})^2+\pq{k}{1} \per \pq{k}{2} \per \pst (4-d) \per \pq{i}{1}+\frac{d}{2} \per \pq{i}{3}\pdt\pdq\nonumber\\
&+\frac{1}{\pq{i}{12} \per \pq{k}{3}} \per \psg (d-2) \per p_i(q_{2}-q_{1}) \per \pq{k}{1} \per \pq{k}{13}\pdg\nonumber\\
&+\frac{1}{\pq{k}{12} \per \pq{i}{3}} \psg 2 \per \pp{i}{k} \per \qq{1}{3} \per \pst (4-d) \per \pq{k}{1}+d \per \pq{k}{2}\pdt +(d-2) \per \pq{i}{1} \per p_k(q_{1}-q_{2}) \per p_k(q_{13}-3 \per q_{2})\pdg\dq\nonumber\\
&+\frac{1}{2 \per \qq{1}{2} \per \qq{1}{3}} \sq \frac{1}{\pq{i}{12} \per \pq{k}{3}} \psg 4 \per \pp{i}{k} \per \qq{2}{3} \per p_k(2 \per q_{1}+q_{3})\nonumber\\
&+(\pq{k}{1})^2 \per \pst(7-2 \per d) \per \pq{i}{1}+(2 \per d+1) \per \pq{i}{2}-4 \per \pq{i}{3}\pdt + 2 \per \pq{k}{1} \per \pq{k}{3} \per \pst(5-d) \per \pq{i}{1}+(d-5) \per \pq{i}{2}-2 \per \pq{i}{3}\pdt\nonumber\\
&+\pq{k}{1} \per \pq{k}{2} \per \pst(2 \per d-3) \per \pq{i}{1}+(9-2 \per d) \per \pq{i}{2}-2 \per \pq{i}{3}\pdt+3 \per (\pq{k}{3})^2 \per p_i(q_{1}-q_{2})\nonumber\\
&+\pq{k}{2} \per \pq{k}{3} \per p_i(9 \per q_{1}-3 \per q_{2}+2 \per q_{3})+ 2 \per(\pq{k}{2})^2 \per p_i(q_{3}-q_{1})\pdg\nonumber\\
&+\frac{1}{\pq{k}{12} \per \pq{i}{3}} \per \psg 2 \per \pp{i}{k} \per \qq{2}{3} \per p_k(q_{12}+2 \per q_{3})+(\pq{k}{1})^2 \pst (2 \per d-7) \per \pq{i}{1}+2 \per \pq{i}{2}+\pq{i}{3}\pdt\nonumber\\
&+\pq{k}{1} \per \pq{k}{3} \pst (2 \per d-7) \per \pq{i}{1}+4 \per \pq{i}{2}-3 \per \pq{i}{3}\pdt +2 \per \pq{k}{1} \per \pq{k}{2} \per (\pq{i}{2}-2 \per d \per \pq{i}{1})+2 \per (\pq{k}{3})^2 \per p_i(2 \per q_{1}-q_{2})\nonumber\\
&+\pq{k}{2} \per \pq{k}{3} \pst 2 \per \pq{i}{2}+3 \per \pq{i}{3}-(2 \per d+1) \per \pq{i}{1} \pdt +(\pq{k}{2})^2 \pst (2 \per d-9) \per \pq{i}{1}-\pq{i}{3}\pdt\pdg\nonumber\\
&+\frac{4}{\pq{i}{123}} \left(\frac{1}{\pq{k}{12}}-\frac{1}{\pq{k}{3}}\right) \psg \qq{2}{3} \psq \frac{\pp{i}{k}}{2} \per p_k ( 5\per q_{1}-3 \per q_{2}+4 \per q_{3} ) \pdq +(\pq{k}{2})^2 \per \pq{i}{3}\nonumber\\
&+(\pq{k}{1})^2 \pst \pq{i}{1}+(3-d) \per \pq{i}{2}\pdt +(d-2) \per \pq{k}{1} \per \pq{k}{2} \per \pq{i}{2}+\pq{k}{1} \per \pq{k}{3} \per p_i (q_{1}-3 \per q_{2})+\pq{k}{2} \per \pq{k}{3} \per p_i(3 \per q_{1}-q_{2})\pdg\dq\dg\nonumber\\
&+\frac{1}{2 \per (q_{123}^2)^2 \per \pq{i}{123} \per \pq{k}{123}} \sg (3 \per d-10) \per \pp{i}{k} +\frac{2 \per \pp{i}{k} \per \qq{1}{3}}{(\qq{1}{2})^2} \pst (d-4) \per \qq{1}{3}-d \per \qq{2}{3}\pdt\nonumber\\
&+\frac{1}{\qq{1}{2}} \per \left[\pq{k}{1} \pst (8-3 \per d) \per \pq{i}{1}+(16-7 \per d) \per \pq{i}{2} \pdt -\frac{d}{2} \per \pq{k}{3} \per \pq{i}{3}-2 \per \pp{i}{k} \per (2 \per \qq{1}{3}+3 \per \qq{2}{3})\right]\nonumber\\
&+\frac{\qq{2}{3}}{\qq{1}{2} \per \qq{1}{3}} \sq\pp{i}{k} \per \qq{2}{3}+4 \per (d-4) \per \pq{k}{1} \per \pq{i}{1}-16 \per \pq{k}{1} \per \pq{i}{2}+4 \per (2-d) \per \pq{k}{2} \per \pq{i}{2}\dq\dg\;, \nonumber
\end{align}
\end{footnotesize}%
where $p_i = n$, $p_k = \bar n$ and $q_i = k_i$, $i=1,2,3$ in our notations. 
A distinct feature of the function $\oS_{ik}^{(d)}$ 
is that it  contains a  propagator $1/k_{123}^2$ which  depends on the relative
orientation of all three gluons in the plane transverse to the collision axis.
This feature makes it very difficult to integrate  the function $\omega_{n  \bar n}^{(3),d}$
analytically over the phase space with $\theta$-functions.  On the other hand, the  ability to
write down the IBP equations and perform reduction to master
integrals should allow us to use differential equations to compute even the most complicated
master integrals. This is what  we would like to discuss in this section. 

We begin by   expressing  the integral of
$\omega_{n  \bar n}^{(3),d}$ through master integrals. These master integrals
can be found in Appendix~\ref{app:master_integrals};  they can be split into two groups,
depending on whether   they contain the $1/k_{123}^2$ propagator or not.
The master integrals from the first group are similar to what we have discussed earlier; they can be calculated 
in a similar manner and the  results are provided  in the  ancillary file.  Master
integrals that belong to the second group involve a  propagator  $1/k_{123}^2$. These are the most complicated
integrals and we decided to compute them numerically.

To this end, we replace  a propagator $1/k_{123}^2$ with $1/( k_{123}^2 + m^2)$ in all master
integrals that contribute to the integral of $\omega_{n  \bar n}^{(3),d}$ and derive differential equations
for these integrals  w.r.t. the mass parameter $m^2$. We emphasize that the very possibility to use   differential
equations for  phase-space integrals with Heaviside functions depends on our ability to set  up 
integration-by-parts  identities and reduce the derivatives of master integrals back to basis  integrals.

It is quite clear that   additional mass parameter  makes 
integration-by-parts identities more complex and requires us to introduce more 
integrals to close them. However,  once the differential equations
are derived, it is in principle straightforward to compute the required integrals by solving them
numerically. We do this  by using the method described in Ref.~\cite{Liu:2017jxz}.
In what follows, we first discuss calculation of boundary conditions and then explain how to solve 
differential equations  by considering a (relatively) simple example.

\subsubsection{Calculation of the boundary conditions}

We have seen  in the previous sections that, although IBP reductions have to be performed for a non-vanishing
value of the analytic regulator, the limit $\nu \to 0$ can be taken in an absolute  majority of integrals after the
reduction is performed.  The same applies to integrals that contain  propagators $1/k_{123}^2$ that we modify
by introducing the mass parameter $m$.   We will therefore discuss calculation of boundary conditions
for such integrals,  setting the analytic regulator to zero. 

The complexity of the boundary conditions computation strongly depends on the type of constraints that
a particular integral is subject to.
As we  explain below, the more $\delta$-functions a particular integral has, the easier
it is to compute the boundary conditions.   To understand this, consider 
integrals with  three $\delta$-functions 
\be
I_{\delta \delta \delta }(m) = \int  \frac{\DIFFL     \Phi^{nnn}_{\delta \delta \delta }}{(k_{123}^2 + m^2)^i \dots}.
\label{eqbc.1}
\ee
In Eq.~(\ref{eqbc.1}) the  ellipses stand for mass-parameter-independent scalar products and $i$ is an integer number.
In   all  integrals that refer to 
the $nnn$ configuration,  all $\beta$-variables are restricted because
$\beta_{i} > 0$, $i=1,2,3$ and $\beta_1+\beta_2+\beta_3 = 1$. Moreover, for
integrals with three $\delta$-functions, 
all $\alpha$-variables are equal to $\beta$-variables and $k_{i,\perp}^2 = \alpha_i \beta_i = \beta_i^2$, $i=1,2,3$. 
Hence,  integration in Eq.~(\ref{eqbc.1}) is performed over a \emph{finite} region
of the three-particle phase space,  so that the approximate form of the integral in  the $m^2 \to \infty$ limit
is simply obtained by expanding the propagator
$1/(k_{123}^2 + m^2)$ in $k_{123}^2/m^2$.   It follows that all $I_{\delta \delta \delta}$ integrals
have a particularly simple asymptotic mass dependence
\be
\lim_{m \to \infty} I_{\delta \delta \delta}(m) \sim m^{-2i}.
\ee
It is obvious that a Taylor expansion of $I_{\delta \delta \delta}$-integrals in $k_{123}^2/m^2$ produces integrals where a ``massive''
propagator is absent. As the result, once the expansion is performed, we can use integration by parts for $k_{123}^2$-independent 
integrals to express any expanded $I_{\delta \delta \delta}$ integral
through master  integrals  computed in the previous sections.

Consider now an integral that contains two
$\delta$-functions  and one $\theta$-function. We choose four--momenta $k_{1,2,3}$
in such a way that the $\theta$-function depends on $k_1$ (more precisely on
$\alpha_{1}$ and $\beta_1$) and write 
\be
I_{\theta \delta \delta }(m) = \int
\frac{\DIFFL     \Phi^{nnn}_{\theta \delta \delta }}{
(k_{123}^2 + m^2)^i \dots},
\label{eqbc.2}
\ee
At variance with  $I_{\delta \delta \delta}$ integrals,   the  integration over $\alpha_1$ in $I_{\theta \delta \delta}$ integrals
is \emph{not restricted} from above. This implies that, in the limit $m \to \infty$, there are 
contributions from the region $\alpha_1 \sim m$ as well as $\alpha_1 \sim 1$.  Since the phase-space element scales as
$\alpha_1^{-\ep}$ and since in the limit $\alpha_1 \to \infty$, 
$k_{123}^2 + m^2 \sim \alpha_1 (\beta_2+\beta_3)  \dots + m^2$, we find that all
integrals $I_{\theta \delta \delta }$ have  the following
asymptotic dependence
on the mass parameter $m$ in the $m \to \infty$ limit 
\be
\lim_{m \to \infty} I_{\theta \delta \delta }(m)  \sim m^{-2i_1 -2\ep} A_2 + m^{-2i} A_1.
\label{eqbc.3}
\ee
In this formula $i_1,i$ are two integers that are particular to
the  integral under consideration, and $A_{1,2}$ are independent of the mass. 
To compute $A_1$ we need to Taylor-expand the integrand in $k_{123}^2/m^2$ and then use integration by parts
to reduce the resulting integrals to massless $\theta \delta \delta$-integrals.
To compute $A_{2}$, we need to drop
the $\theta(\alpha_1 - \beta_1)$ constraint, as it is only relevant for $\alpha_1 \sim \beta_1 \sim 1$, and then simplify
a particular integral under the assumption that $\alpha_1 \sim m^2 \gg \alpha_2,\alpha_3,\beta_1,\beta_2,\beta_3$.
We note that in this case all the relevant integrals can be straightforwardly
computed in a closed form in terms of hypergeometric
functions.

Finally, we  require integrals with two $\theta$-functions and one $\delta$-function\footnote{As we explained
  earlier, there are no master integrals with three theta-functions in case of  the $nnn$ configuration.} 
\be
I_{\theta \theta \delta} = \int
\frac{\DIFFL     \Phi^{nnn}_{\theta \theta  \delta }}{
(k_{123}^2 + m^2)^i \dots},
\ee
 The asymptotic $m \to \infty$ mass-dependence
of such  integrals reads 
\be
\lim_{m \to \infty} I_{\theta \theta \delta}(m) \sim m^{-2 i_2 -4\ep} A_{3} +  m^{-2 i_1-2\ep} A_2 + m^{-2i} A_{1}.
\label{eqbc.4}
\ee
In Eq.~(\ref{eqbc.4})  $A_{3}$ receives contributions from the integration region with
$\alpha_1 \sim \alpha_2 \sim m^2$,
$A_{2}$ -- from regions where either $\alpha_1 \sim m^2,\; \alpha_2 \sim 1$
or   $\alpha_2 \sim m^2, \;\alpha_1 \sim 1$  and $A_{1}$ --  from the region where
$\alpha_1 \sim \alpha_2 \sim 1$. 
We obtain $A_1$ upon Taylor expansion of an integrand in $k_{123}^2/m^2$  and $A_2$ upon Taylor
expanding in ``small'' $\alpha$- and $\beta$-parameters and
neglecting the corresponding $\theta$-function constraint. 

To illustrate how contributions of different regions can be computed,  we  consider one of the  integrals
with $1/(k_{123}^2+m^2)$ propagator 
\be
B_{1,\delta \theta \theta  }  =  \int
\frac{\DIFFL     \Phi^{nnn}_{\delta  \theta  \theta }}{(k_{123}^2 + m^2) (k_1 k_3) (k_{12} n) (k_3 \bar n) }.
\label{eqbc.5a}
\ee
We would like to compute   the  leading contribution to 
${\cal O}(m^{-2\ep})$ branch of this integrals in the limit $m \to \infty$. 
We consider two distinct contributions  $\alpha_2 \sim m^2 \gg 1$ and $\alpha_3 \sim m^2 \gg 1$.  In the first case,
$[{\rm d} k_2] \sim {\rm d} \alpha_2 \alpha_2^{-\ep}$, $k_{123}^2 + m^2 \sim \alpha_2(\beta_1 + \beta_3) \cdots + m^2$,
$k_1 k_3 \sim 1, k_{12} n \sim \beta_{12} \sim 1, k_3 \bar n \sim \alpha_3 \sim 1$. 
Hence,  we conclude that 
\be
\lim_{m \to \infty}
B_{1,\delta \theta \theta  }\Bigg |_{\alpha_2 \sim m^2}  \sim \int \frac{{\rm d} \alpha_2 \;
  \alpha_2^{-\ep} }{\alpha_2( \beta_1 + \beta_3) + m^2} \sim m^{-2\ep}.
\ee
However, in the case $\alpha_3 \sim m^2$, there are additional dependences of the integrand on $k_3$ including
$k_3 k_1 \sim \alpha_3 \beta_1 \sim m^2  $ and $k_3 \bar n \sim \alpha_3 \sim m^2$. This implies that 
\be
\lim_{m \to \infty} B_{1,\delta \theta \theta }\Bigg |_{\alpha_3 \sim m^2}  \sim \int \frac{{\rm d} \alpha_3 \; \alpha_3^{-\ep} }{
  ( \alpha_3( \beta_1 + \beta_2) + m^2) \alpha_3 \alpha_3 } \sim m^{-2\ep - 4}.
\ee
We conclude that the leading $m \to \infty$ contribution to the ${\cal O}(m^{-2\ep})$ branch
of the integral $B_{1,\delta \theta \theta}$ in  Eq.~(\ref{eqbc.5a}) arises  from the region where $\alpha_2 \sim m^2$ and all
other variables are of order one.

To compute this  contribution, we notice that, upon taking the limit $\alpha_2 \sim m^2 \to \infty$ in $k_{123}^2 + m^2$,
the only dependence of the integrand on  relative azimuthal angles  of   massless partons resides in
the simple scalar products, e.g.  $k_1 k_3$ in $B_1$.   This integration can be easily performed following the
discussion in the  previous sections. We find 
\be
\begin{split} 
B_{1,\delta \theta \theta  }\Bigg |_{\alpha_2 \sim m^2}
= & \; 2   \;
\int \frac{{\rm d} \beta_1 {\rm d} \beta_2 {\rm d} \beta_3 \; \beta_1^{-2\ep} \beta_2^{-\ep} \beta_3^{-\ep}
  \delta(1-\beta_{123}) \; {\rm d} \alpha_3 \alpha_3^{-\ep}
  \theta(\alpha_3 - \beta_3) }{(\beta_1 + \beta_2) \beta_1 \alpha_3^2  }
\\
 & \times \HYPGF{1}{1+\ep}{1-\ep}{\frac{\beta_3}{\alpha_3}}
\int \limits_{0}^{\infty} 
\frac{ {\rm d} \alpha_2 \; \alpha_2^{-\ep} }{\alpha_2 \beta_{13} + m^2}.
\end{split} 
\ee
We  integrate over  $\alpha_2$ and obtain 
\be
\int \limits_{0}^{\infty} 
\frac{ {\rm d} \alpha_2 \; \alpha_2^{-\ep} }{\alpha_2 \beta_{13} + m^2}
= m^{-2\ep} ( \beta_{13} )^{\ep-1} \Gamma(\ep) \Gamma(1-\ep)\;. 
\ee
We then change variables $\alpha_3 \to \xi$,  $\alpha_3 = \beta_3/\xi$, observe that integration over $\xi$ factorizes and
find
\be
\begin{split} 
B_{1,\delta \theta \theta} \Bigg |_{\alpha_2 \sim m^2}
 = &     \; 2  \;
m^{-2\ep}  
\; \frac{\Gamma(1-\ep) \Gamma^2(1+\ep)}{\ep \Gamma(2+\ep)} \;
\GENHYPGF{3}{2}{1,1+\ep,1+\ep}{1-\ep,2+\ep}{1}
\\
& \times \int  {\rm d} \beta_1 {\rm d} \beta_2 {\rm d} \beta_3 \; \delta(1-\beta_{123})
  \; \beta_1^{-2\ep-1} \beta_2^{-\ep} \beta_3^{-\ep} \beta_{12}^{-1} \beta_{13}^{\ep-1} \beta_3^{-\ep-1}.
\end{split} 
\ee
To compute the remaining integral, we remove  the $\delta$-function
by integrating over $\beta_2$ and change variables as follows 
$\beta_1 = xy$ and $\beta_3 = x(1-y)$. The integral over $x$ and $y$ is then of a hypergeometric type and
we obtain the result for the required branch
\be
\begin{split} 
& B_{1,\delta \theta \theta }\Bigg |_{\alpha_2 \sim m^2}
 =   \; 2  \;
m^{-2\ep}  
\; \frac{
  \Gamma^2(1-\ep) \Gamma^2(1+\ep) \Gamma(-1-3\ep) \Gamma^2(-2\ep) }{ \ep \Gamma(2+\ep) \Gamma^2(-4\ep) }  
\\
& \;\;\;\; \times \GENHYPGF{3}{2}{1,1+\ep,1+\ep}{1-\ep,2+\ep}{1}
\; \GENHYPGF{3}{2}{1,-1-3\ep,-2\ep}{-4\ep,-4\ep}{1}.
\end{split}
\ee
The computation of the asymptotic behavior of the ${\cal O}(m^{-2\ep})$ branch that we just
described is typical. In fact, all  such
boundary conditions can be calculated in terms of hypergeometric functions in a straightforward manner.

\vspace*{0.3cm}
It turns out that the   most difficult boundary contributions to compute are  the 
${\cal O}(m^{-4\ep})$ branches of  $I_{\theta \theta \delta}$ integrals.  Indeed, for 
such branches, we  need to consider
asymptotic limits $\alpha_1 \sim \alpha_2 \sim m^2 \gg \alpha_3, \beta_{1},  \beta_{2}, \beta_{3}$.
  Inspecting $k_{123}^2$ in this limit, we find that we cannot simplify the scalar product
  $k_1 k_2$ so that $k_{123}^2+m^2$ still depends  on  the relative orientation of $k_1$ and $k_2$
  in the transverse plane, even if $m$ is taken to infinity. 

  To understand how such integrals can be calculated, we note that for the computation
  of the ${\cal O}(m^{-4\ep})$  branch, we need to 
  neglect all $\theta$-functions in the definition of  $\theta \theta \delta$ integrals
  since they are only relevant for $\alpha_{1,2} \sim \beta_{1,2}
  \sim 1$ which violates the assumption $\alpha_{1,2} \sim m^2$. 
  Hence, a  ${\cal O}(m^{-4\ep})$ branch of any
  $\theta \theta \delta$ integral  can be computed by writing
  \begin{align}
  \begin{split}
  m^{-2i_1 - 4\ep} A_{3} \sim & \int {\rm d}^d q  \; [{\rm d} k_3]
  \frac{ \delta (1 - qn -k_3 n) \delta( k_3 \bar n - k_3 n ) }{( (k_3 + q)^2 + m^2)^i}  \\
  & ~ \times  \int [{\rm d} k_1] [{\rm d} k_2 ] \delta ( q - k_1 - k_2) \dots,
  \end{split}
  \end{align}
  where ellipses stand for various scalar products of momenta that appear in a particular integral.
  All these scalar
  products have to be simplified under the  assumption
  that $k_{1} \bar n \sim k_2 \bar n \sim m^2 \gg k_{1} n , k_2 n,  k_{3} n, k_3 \bar n$.
  We can further write
  \be
k_{123}^2 = (q+k_3)^2 = q^2 + 2 k_3 q  \to q^2 + \beta_3 (q \bar n) = q^2 + (1-qn) (q \bar n).
  \ee
  Therefore, to determine the ${\cal O}(m^{-4\ep})$ branch of any $I_{\theta \theta \delta}$ integral, we need to compute 
  \be
  \begin{split} 
  &  m^{-2i_1 - 4\ep} A_3 \sim \int \frac{ {\rm d}^d q }{( q^2 + (1-qn)(q \bar n) +m^2)^i } \; F(q^2,qn, q \bar n),
  \\
& F(q^2,qn, q\bar n) = \int [{\rm d} k_3] \delta (1 - qn -k_3 n) \delta( k_3 \bar n - k_3 n ) 
  \int [{\rm d} k_1] [{\rm d} k_2 ] \delta ( q - k_1 - k_2) \dots.
\end{split} 
  \ee

  At this point,   we can again use the integration-by-parts method
  and express  all relevant integrals $F(q^2,qn, q\bar n)$ as linear combinations of five
  master integrals. They read 
  \be
\begin{split} 
& \{{\cal B}_1,{\cal B}_2,{\cal B}_3,{\cal B}_4,{\cal B}_5\} \\
& = \int [{\rm d} k_3] \delta (1 - qn -k_3 n) \delta( k_3 \bar n - k_3  n ) \int [{\rm d} k_1] [{\rm d} k_2 ] \delta ( q - k_1 - k_2)  \\
& ~~~~ \times  \left \{1, \frac{1}{k_2 \bar n}, \frac{1}{(k_2n)(k_2 \bar n)}, \frac{1}{(k_2 n) (k_1 \bar n) }, \frac{1}{1-k_1 n}    \right \}.
\end{split}
\label{eqbc.7}
  \ee
  Note that these integrals do not contain scalar products of the gluon four-momenta. This happens because
  such  scalar products  are either simple as e.g. in case of $k_1 k_2 = q^2/2$
  or can be simplified for large $\alpha_{1,2}$ as  e.g. in  $k_3 k_1 \to (k_3 n) (k_1 \bar n)$,
  $k_3 k_2 \to (k_3 n) (k_2 \bar n)$.
  
  It is convenient to compute the  integrals ${\cal B}_{1,..,5}$ in the rest frame of $q$ where $k_{1,2}$ are back-to-back. Then,
  using the following result for the angular integral
  \be
  \int \frac{{\rm d} \Omega_k^{(d-1)}}{(1 - \vec n_k \vec n) (1 - \vec n_k \vec {\underline n} ) }
  = 2^{-2\ep} \Omega^{(d-2)} \frac{ \Gamma(1-\ep) \Gamma(-\ep)}{\Gamma(1-2\ep)} \HYPGF{1}{1}{1-\ep}{\frac{1 + \vec n \vec {\underline n}}{2}},
  \ee
  we easily  find 
  \be
\begin{split} 
  & {\cal B}_1 = \tilde N_\ep (q^2)^{-\ep}(1-qn)^{-2\ep},
  \\
  & {\cal B}_2 = \tilde N_\ep \frac{(1-2\ep)}{(-\ep)} \frac{ (q^2)^{-\ep}(1-qn)^{-2\ep}}{q \bar n},
  \\
    & {\cal B}_3 = \tilde N_\ep  \frac{(1-2\ep)}{(-\ep)} \frac{ 2 (q^2)^{-\ep}(1-qn)^{-2\ep}}{(qn) (q \bar n)^2}
  \HYPGF{1}{1}{1-\ep}{1-\frac{q^2}{(qn)(q\bar n)}},\\
   & {\cal B}_4 = \tilde N_\ep  \frac{(1-2\ep)}{(-\ep)} \frac{ 2 (q^2)^{-\ep}(1-qn)^{-2\ep}}{(qn) (q \bar n)^2}
  \HYPGF{1}{1}{1-\ep}{\frac{q^2}{(qn)(q\bar n)}}, \\
     & {\cal B}_5 = \tilde N_\ep  q^{-2\ep}(1-qn)^{-2\ep}  \HYPGF{1}{1-\ep}{2-2\ep}{qn},
  \end{split} 
  \ee
  where
  \be
\tilde N_\ep = \frac{1}{(2\pi)^{d-1}}  \left [
     \frac{\Omega^{(d-2)}}{4 (2\pi)^{d-1}}
     \right ]^2\; \frac{\Gamma^2(1-\ep)}{\Gamma(2-2\ep)} .
  \ee

  A  typical function $F(q^2,qn,  q\bar n)$ is given by  a linear combination of the
  integrals ${\cal B}_{1,..,5}$. For  example, for one of the boundary integrals that we will  refer to
  as $B_2$    this function reads
  \be
\begin{split} 
  F_{2}(q^2,qn, q \bar n)  & =  \left [  \frac{ 4 (1 - 2 \ep)^2 (2 - qn) (1 - qn)}{(1 + \ep) m^6 (q n)^2} \;{\cal B}_1
  + \frac{ 4 \ep^2}{(1 + \ep) m^4}{\cal B}_2
  \right. 
\\
& \left. 
 - \frac{ 2 (1 + 2 \ep) (1 - qn) q \bar n }{m^2 q^2} \;{\cal B}_4
\right ].
\end{split} 
  \ee
Hence, to determine the $m^{-4\ep}$ branch  of the corresponding integral, we need to compute 
  \be
B_{2} =   \int \frac{ {\rm d}^d q \; \theta(1-qn)}{q^2 + q \bar n (1 - qn) + m^2} F_{2}(q^2,qn,q \bar n).
  \label{eqbc.13}
  \ee

  To calculate  the integral in Eq.~(\ref{eqbc.13}),
  we need to choose a convenient parameterization to integrate over $q$. We do  this in the following
  way. We  introduce the Sudakov decomposition for the vector $q$ and write
  $q = \frac{1}{2} \alpha_q n + \frac{1}{2} \beta_q \bar n + q_\perp$. It follows that 
  \be
     {\rm d}^d q \; \theta(1-qn) = \frac{1}{4} {\rm d} \alpha_q
     {\rm d} \beta_q {\rm d} q_\perp^2 (q_\perp^2)^{-\ep} {\rm d} \Omega^{(d-2)} \;
     \theta(1-\beta_q).
  \ee
In addition, the four-vector $q$ needs to be time-like, $q^2 > 0$. This implies 
  \be
q^2 = \alpha_q \beta_q - q_\perp^2 > 0.
  \ee
  We would like to simplify the mass-dependent denominator in Eq.~(\ref{eqbc.13}). To do that, we first re-write it using  Sudakov
  variables
  \be
q^2 + q \bar n (1 - qn) + m^2 = \alpha_q - q_\perp^2 + m^2 = t + m^2,
  \ee
  where we introduce a new variable $t = \alpha_q - q_\perp^2$. Positivity of $q^2$ requires
  \be
\alpha_q > \frac{q_\perp^2}{\beta_q}. 
  \ee
  This implies
  \be
 \begin{split} 
  & t = \alpha_q - q_\perp^2 > \frac{q_\perp^2 (1-\beta_q) }{\beta_q} > 0,\;\;\; \\
  & q^2 > 0 \Rightarrow t - \alpha_q(1-\beta_a) > 0.
  \end{split} 
  \ee
  Hence, we can write
  \be
 {\rm d}^d q \theta(1-qn) = \frac{1}{4} {\rm d} \alpha_q {\rm d} \beta_q {\rm d} t  (\alpha_q - t)^{-\ep} {\rm d} \Omega^{(d-2)}
     \theta(1-\beta_q), 
  \ee
  where the integration boundaries are $ 0 <  \beta_q < 1$, $0 < \alpha_q < \infty$ and $ \alpha_q(1-\beta_q) < t < \alpha_q$.
  Using the fact that
  the integrand in Eq.~(\ref{eqbc.13}) does not depend on the azimuthal angle, we
  integrate over $ {\rm d} \Omega^{(d-2)}$ and write
  \be
  \begin{split}
B_{2} =   & \frac{\Omega^{(d-2)}}{4} \int \frac{ {\rm d} \alpha_q {\rm d} \beta_q {\rm d} t  (\alpha_q - t)^{-\ep} }{t + m^2} 
 \theta(t -  \alpha_q(1-\beta_q)) \times \\
&  \left [  \frac{ 4 (1 - 2 \ep)^2 (2 - \beta_q) (1 - \beta_q)}{(1 + \ep) m^6 \beta_q^2} \;{\cal B}_1  + \frac{ 4 \ep^2}{(1 + \ep) m^4}{\cal B}_2 -
\frac{ 2 (1 + 2 \ep) (1 - \beta_q) \alpha_q }{m^2 (t -\alpha_q(1-\beta_q)) } \;{\cal B}_4
\right ].
  \end{split}
  \label{eqbc.20}
  \ee
  To integrate further, we change variables $\alpha_q \to \xi$ with  $\alpha_q = t/\xi$ and $0 < \xi < 1$.  Since 
  \be
q^2 = t - \alpha_q(1-\beta_q) = \frac{t}{\xi}(\xi - (1-\beta_q) ) >0, 
  \ee
  the integration boundary for $\beta_q$ becomes $ 1-\xi < \beta_q < 1$.  To accommodate these
  boundaries in a natural way,  we change variables $\beta \to r$, with 
  $\beta = 1-\xi r$,  $0 < r < 1$. 

Upon changing variables and using explicit expressions for integrals ${\cal B}_{1,2,4}$, we note  that integration
over $t$ factorizes and can be performed easily. We obtain
\be
\begin{split} 
  & B_{2} = -\frac{m^{-4-4\ep} \Gamma^2(1 - \ep) \Gamma(1 + 2 \ep)}{\ep^2}
  \int \limits_{0}^{1} {\rm d} \xi \;  {\rm d} r \;
W_{2}(\xi,r),
\\
& W_{2}(\xi,r) = 
  \Bigg \{   
 2 \ep  \frac{
  \left( r(1+r \xi) +
  \ep \left ( 1 -2  r(1+\xi) -2 r^2 \xi(1-\xi/2) \right ) \right)}{(1+ \ep
  ) (1-r)^{\ep} r^{2 \ep} (1-\xi)^{\ep} \xi^{\ep}  (1- r \xi)^2}
\\
& 
+ 2 (2 \ep+1) \frac{  r^{1-2 \ep}
   \xi^{1-\ep} (1-r \xi)^{\ep}}{(1-r)^{1+\ep}  (1-\xi)^{1+2 \ep}}
  \, \HYPGF{-\ep}{-\ep}{1-\ep}{\frac{(1-r) \xi}{1-r \xi}}
  \Bigg \}.
  \label{eqbc.22}
\end{split}
\ee
Integrating over $\xi$ and $r$, we find
\be
\begin{split} 
B_{2} =& 
m^{-4 \ep-4} \Bigg (
-\frac{3}{2 \ep^4}  -\frac{6}{\ep^3}-\frac{12}{\ep^2}
+\frac{15 \zeta_3-36}{\ep}+60 \zeta_3+\frac{\pi ^4}{4}-84
+\ep \left(120 \zeta_3
\right. 
\\
& \left. +81 \zeta_5+\pi ^4-204\right)
+ \ep^2 \left(-75 \zeta_3^2+360 \zeta_3+324 \zeta_5+\frac{11 \pi ^6}{63}+2 \pi ^4-468
\right)
\Bigg ).
\end{split} 
\ee

Calculation of   ${\cal O}( m^{-4\ep} ) $ branches for other integrals that are needed to determine  boundary
conditions proceeds along similar lines. Integration over  $t$ can always be performed exactly and the subsequent
integration over $\xi$ and $r$ is then completed  by constructing subtraction terms of end-point singularities to facilitate
expansion of integrands in $\ep$.

\subsubsection{Numerical solution of the differential equations}
\label{ssnum}

Having discussed the computation of  the boundary conditions, we need to explicitly write down
and solve the differential equations. We note that we need to consider about two hundred
mass-dependent master integrals in total to close the system of differential equations. Since it is  impossible
to discuss such a huge system of differential equations in any  detail, we decided to choose a small eleven-by-eleven
sub-system and discuss  it in a comprehensive way.

We consider eleven integrals 
\begin{align}
    {\cal J}_{1}
    & =  \int \DIFFL \Phi^{nnn}_{\delta \delta\delta  },\;\;\;\;\;\;\;\;\;\;\;\;\;\;\;\;\;\;\;\;\;\;\;\;\;\;
    {\cal J}_{2}
     =
    \int \frac{ \DIFFL \Phi^{nnn}_{\delta \delta\delta  }}{k_{1 2 3}^{2} + m^2},
\nonumber     \\
    {\cal J}_{3}
    & =
    \int \frac{ \DIFFL \Phi^{nnn}_{\delta \delta\delta  }}{(k_{1 2 3}^{2} + m^2) k_{1} \cdot n},\;\;\;\;\;\;\;\;
    {\cal J}_{4} =   \int \frac{ \DIFFL \Phi^{nnn}_{\delta \delta \theta  } }{k_{1 2 3}^{2} + m^2},
 \nonumber   \\
    {\cal J}_{5}
    & = \int \frac{ \DIFFL \Phi^{nnn}_{\delta \delta \theta  } }{(k_{1 2 3}^{2} + m)^{2}},\;\;\;\;\;\;\;\;\;\;\;\;\;\;\;\;
       {\cal J}_{6}
     = \int \frac{ \DIFFL \Phi^{nnn}_{\delta \delta' \theta  } }{k_{1 2 3}^{2} + m^2},
\label{eq5.85} \\
    {\cal J}_{7}
    & =
    \int \frac{ \DIFFL \Phi^{nnn}_{\delta \delta \theta  } }{(k_{1 2 3}^{2} + m^2) ( k_{1} n )},  \;\;\;\;\;\;\;\;\;
       {\cal J}_{8}
   =
    \int \frac{ \DIFFL \Phi^{nnn}_{\delta \delta \theta  } }{ (k_{1 2 3}^{2} + m^2) (k_{3} \bar n )},
   \nonumber  \\
    {\cal J}_{9}
    & = \int 
    \frac{ \DIFFL \Phi^{nnn}_{\delta \theta \theta  } }{k_{1 2 3}^{2} + m^2},\;\;\;\;\;\;\;\;\;\;\;\;\;\;\;\;\;\;\;
    {\cal J}_{10}
     =
    \int \frac{ \DIFFL \Phi^{nnn}_{\delta \theta \theta  } }{(k_{1 2 3}^{2} + m^2) (k_{2} \bar n)},
\nonumber \\
    {\cal J}_{11}
    & =
    \int \frac{ \DIFFL {\tilde \Phi}^{nnn}_{\delta \theta \theta  } }{(k_{1 2 3}^{2} + m^2) (k_2 \bar n)}.
    \nonumber 
    \end{align}
We note that somewhat different notations for the phase-space measures in ${\cal J}_6$ and ${\cal J}_{11}$,
compared to
other integrals,  imply  
that derivatives of $\delta$-functions, namely  ${\rm d}\delta ( k_2 \bar n - k_2  n)/{\rm d} (k_2  n)$  and 
${\rm d}\delta (1 - k_{123} n) /{\rm d} k_{123}n $ appear in ${\cal J}_6$ and ${\cal J}_{11}$ respectively.
It is straightforward
to compute these derivatives if the corresponding $\delta$-functions are represented using Eq.~(\ref{eq2.9new})
in the spirit of reverse  unitarity. 

To derive differential equations, we consider a vector of eleven integrals introduced in Eq.~(\ref{eq5.85}), differentiate
it with respect to $m^2$, use integration-by-parts identities 
to perform a reduction of the resulting integrals back to the integrals
${\cal J}_{1,...,11}$  and obtain 
\begin{equation}
    \frac{\partial}{\partial m^2}
    \boldsymbol{{\cal J}}
    =
    \left[
        \frac{\boldsymbol{M}_{1}}{m^2 + 1}
        +
        \frac{\boldsymbol{M}_{2}}{m^2 + \frac{1}{4}}
        +
        \frac{\boldsymbol{M}_{3}}{m^2}
        +
        \boldsymbol{M}_{4}
        +
        m^2
        \boldsymbol{M}_{5}
    \right]
    \boldsymbol{{\cal J}}.
    \label{eq5.83}
\end{equation}
The matrices $\boldsymbol{M}_{1,..,5}$ are independent of $m^2$ but depend on $\ep$. They can be found in Appendix \ref{app:deq}.

We have already explained how to compute the boundary conditions in the previous Section; explicit results for integrals ${\cal J}_{1,..,11}$ that appear
in Eq.~(\ref{eq5.85}) can be found in the ancillary file. With the boundary
conditions at hand, we solve the differential equation in the following way.   
At $m^2 \to i \infty$,
all master integrals are  written as power-logarithmic series  in $y = 1 / m^2$
\be
\boldsymbol{{\cal J}} = \sum \boldsymbol{C_\infty}(k,n) y^k \ln^n y.
\label{eq5.84}
\ee
Coefficients of these series  solutions are  fixed with the help of  boundary conditions and the differential equations.
We can use these  series solutions to evaluate integrals  in the upper complex
half-plane within its radius of convergence. The radius of convergence follows
from Eq.~(\ref{eq5.83}) where we observe singularities at $m^2 = 0,-1/4,-1$. 

Suppose we take the point $m^2 = m_0^2=1/y_0$ where the solution Eq.~(\ref{eq5.84}) is valid. 
We construct another series solution at this point
\be
\boldsymbol{{\cal J}} = \sum \boldsymbol{C_{y_0}}(k) (y-y_0)^k,
\label{eq5.85a}
\ee
and find coefficients $\boldsymbol{C_{y_0}}(k)$ by matching the above equation to Eq.~(\ref{eq5.84}) at around $y=y_0$.
The solution in Eq.~(\ref{eq5.85a}) has its own radius of convergence and allows us to move past the radius of convergence
of the original solution Eq.~(\ref{eq5.84}). 
We then repeat this procedure  at another point $m^2 = m^2_{1}$ closer to the physical point at $m^2 = 0$ and iterate. 

After sufficient number of steps,
we  arrive at a point $m^2 = m^2_{f}$ within the radius of convergence of the formal solution at $m^2 = 0$.
As follows from the differential equation Eq.~(\ref{eq5.83})  $m^2 = 0$ is a regular
singular point of the differential equation; to determine coefficients of the solution constructed as an expansion around
$m^2=0$ we proceed as follows. 

The formal solution in the neighborhood of the point $m^2 = 0$ can be written as
\begin{equation}
    \boldsymbol{{\cal J}}(m^2)
    =
    \boldsymbol{P}(\varepsilon, m^2)
    (m^2)^{\boldsymbol{M}(\varepsilon)}
    \boldsymbol{{\cal J}}_{0}(\varepsilon)
    \text{,}
\end{equation}
where $\boldsymbol{P}$ is a matrix that can be computed as a series
in $m^2$ and the matrix $\boldsymbol{M}$ is related to $\boldsymbol{M}_3$  defined in Eq.~(\ref{eq5.83}).

The matrix  $\boldsymbol{P}$ can be found by constructing a series solution around  $m^2 = 0$. 
The vector $\boldsymbol{{\cal J}}_{0}$ corresponds to the boundary conditions at $m = 0$; we can determine 
it  by matching the formal solution against the evaluation at a finite-$m$ point $m^2_{f}$. We write 
\begin{equation}
    \boldsymbol{{\cal J}}_{0}(\varepsilon)
    =
    \left[
        \boldsymbol{P}(\varepsilon, m^2_{f})
        ( m_{f}^2) ^{\boldsymbol{M}(\varepsilon)}
    \right]^{-1}
    \boldsymbol{{\cal J}}(m^2_{f})
    \text{.}
\end{equation}

The above procedure allows us to obtain the complete solution $\boldsymbol{{\cal J}}$ in the neighborhood of the point $m^2 = 0$, 
and the only   thing left to do is to select integrals we are interested in.

By construction, these integrals correspond to a situation when the limit $m^2 \to 0$ is taken while keeping
$\ep$ fixed; this means that all contributions to
integrals ${\cal \boldsymbol{J} }$ that scale as $ (m^2 )^{i \ep + i_1}  \ln^k m^2$, where $i,i_1$ and $k$
are non-vanishing rational numbers, should be  set to zero. 

To isolate those terms, we  determine 
eigenvectors  of the matrix $\boldsymbol{M}(\ep)$ with zero eigenvalues
\be
\boldsymbol{M}(\ep) \boldsymbol{\xi}_a = 0,\;\;\;\;a =1,..,N_0,
\ee
and write
\be
   {\cal \boldsymbol{ J} }_{m^2 = 0}  =  \sum \limits_{a=1}^{N} ( \widetilde{\boldsymbol{\xi}}^{\dagger}_a \cdot \boldsymbol{{\cal J}}_{0} )
   \; \boldsymbol{P}(\varepsilon, 0)
           \; \boldsymbol{\xi}_a,
\ee
where dual normalized eigenvectors $\{ \widetilde{\boldsymbol{\xi}}^{\dagger}_a \}$ are orthogonal
to $\{ \boldsymbol{\xi}_a \}$,
i.e $\widetilde{\boldsymbol{\xi}}^{\dagger}_b \cdot \boldsymbol{\xi}_a = \delta_{ab}$. 
The above equation provides the desired results for integrals ${\cal J}_{1,..,11}$ at $m^2 = 0$.

We emphasize that the algorithm described above
can be used to obtain numerical solutions of the differential equations   with \emph{arbitrary} precision.
To illustrate this point, we present the results for  two integrals that appear
in Eq.~(\ref{eq5.85}) at the physical point $m^2 = 0$ computed up to at least  $15$ significant digits
through weight six. They read 
\begin{align}
{\cal J}_{9}|_{m^2=0}
    & = \int 
    \frac{ \DIFFL \Phi^{nnn}_{\delta \theta \theta  } }{k_{1 2 3}^{2}}  =
    -
    \frac{0.5}{\ep}
    -
    5
    -
    28.17026373260709 \;\ep
    -
    119.43143332972728 \;\ep^{2}
    \nonumber
    \\
    & \phantom{= {}}
    -
    430.4404286909044 \;\ep^{3}
    -
    1410.1679482808422 \;\ep^{4}
    \nonumber
    \\
    & \phantom{= {}}
    -
    4372.111524529197 \;\ep^{5}
    -
    13148.701437210732 \;\ep^{6}
    +
    \mathcal{O}(\;\ep^{7}),
    \\
  {\cal J}_{10}|_{m^2=0}     & =
    \int \frac{ \DIFFL \Phi^{nnn}_{\delta \theta \theta  } }{(k_{1 2 3}^{2}) (k_{2} \bar n)}
     =
    \frac{0.5}{\varepsilon^{2}}
    +
    \frac{5.5}{\varepsilon}
    +
    34.34926305180546
        +
    170.0583525098628 \;\ep
    \nonumber
    \\
    & \phantom{= {}}
    +
    758.7443815516605 \;\ep^{2}
    +
    3238.222100561864 \;\ep^{3}
    \nonumber
    \\
    & \phantom{= {}}
    +
    13535.346184323936 \;\ep^{4}
    +
    \mathcal{O}(\;\ep^{5}).
\end{align}
Numerical  results for remaining nine integrals shown in Eq.(\ref{eq5.85}) can be found in the ancillary file.

%% file: sections/06_conclusion.tex
\section{Conclusion}
\label{sec:conclusion}
In this paper we discussed computation of real-emission integrals for observables that
contain Heaviside functions. This is an interesting problem  because reverse unitarity \cite{Anastasiou:2002yz}
cannot be immediately applied  to map such integrals onto   multi-loop integrals, preventing
straightforward  use of integration-by-parts identities in such cases.

We discussed a way to re-introduce  integration-by-parts technology into the 
computation of  such integrals and showed that
the resulting IBP relations
have a clear hierarchical structure since, in addition to original integrals, 
there   appear integrals with 
Heaviside functions replaced with $\delta$-functions. 
Integrals with $\delta$-functions
are, however, much simpler since they can be dealt with using reverse unitarity and, thus, IBP equations 
for them are self-contained.  In addition, IBP relations provide
a foundation for deriving differential equations for real-emission  master integrals with Heaviside functions 
that can be solved numerically even if analytic integration becomes too difficult. 

We have shown the efficacy  of this approach by  computing the real-emission contribution to
the zero-jettiness soft function at NNLO. We have also discussed several non-trivial contributions
to the zero-jettiness soft function at N3LO and used them to illustrate all the different aspects of the proposed
computational techniques.  We believe that theoretical methods discussed in this paper will be useful  for computing other
phase-space integrals that involve Heaviside functions. In particular, we hope that their application
will allow us to complete the computation of all  real-emission contributions to zero-jettiness soft function at N3LO.

\section*{Acknowledgments}
We would like to thank Arnd Behring for fruitful discussions and help related to
multivariate partial fractioning.
We are grateful to Dimitri Colferai for clarifying correspondence regarding the soft eikonal function
for triple-gluon emission and for providing us with a digital version of the results of 
Ref.~\cite{Catani:2019nqv}.
We are indebted to  Fabian Lange for his help with  \texttt{Kira}.

This research is partially supported by the Deutsche Forschungsgemeinschaft
(DFG, German Research Foundation) under grant 396021762 - TRR 257 and by Karlsruhe School of
Particle  and Astroparticle physics (KSETA). MD is  supported by the Excellence Cluster \textsc{ORIGINS} funded by the
Deutsche Forschungsgemeinschaft (DFG, German Research Foundation) under Germany's Excellence Strategy - EXC-2094 - 390783311 and
by the ERC Starting Grant 949279 \textsc{HighPHun}.

%% file: sections/07_appendix.tex
\section{Master integrals for $nnn$ contribution to the soft function}
\label{app:master_integrals}

The master integrals that arise in the computation of the $nnn$ contribution to the soft function
are defined as follows:
\begin{itemize}
\item integrals for $\int \DIFFL \Phi^{nnn}_{\theta\theta\theta } \;  \omega_{n  \bar n}^{(3),a}\;$:
\begin{footnotesize}
\begin{equation}
\begin{alignedat}{3}
I_1 ={} & \int  \DIFFL \Phi^{nnn}_{\delta\delta\delta}  \,,
& ~~ I_2 ={} &  \int   \frac{ \DIFFL\Phi^{nnn}_{\delta\delta\delta}}{(k_{12}n)(k_{13}n)}  \,, \\
I_3 ={} & \int  \frac{\DIFFL   \Phi^{nnn}_{\delta\delta\theta}}{(k_{13}\bar{n})} \,, 
&~~ I_4 ={} & \int \frac{\DIFFL   \Phi^{nnn}_{\delta\delta\theta} }{(k_{123}\bar{n})} \,, \\
I_5 ={} & \int  \frac{\DIFFL   \Phi^{nnn}_{\delta\delta\theta}}{(k_{13}n)(k_{123}\bar{n})} \,,  
&~~ I_6 ={} & \int  \frac{\DIFFL \Phi^{nnn}_{\delta\theta\theta}}{(k_{123}\bar{n})} \,.
\end{alignedat}
\label{eqn:mi_3g_nnn_a}
\end{equation}
\end{footnotesize}%

\item additional integrals for $\int \DIFFL \Phi^{nnn}_{\theta\theta\theta } \;  \omega_{n  \bar n}^{(3),b}\; $:
\begin{footnotesize}
\begin{equation}
\begin{alignedat}{4}
& I_{ 7}  =  \int  \frac{\DIFFL     \Phi^{nnn}_{\delta\delta\theta}}{(k_1k_3)(k_3\bar{n})}  \,,
& & ~~ I_{ 8}  =  \int  \frac{\DIFFL     \Phi^{nnn}_{\delta\delta\theta}}{(k_1k_3)(k_3n)(k_{13}\bar{n})}  \,, \\
& I_{ 9}  =  \int \frac{ \DIFFL     \Phi^{nnn}_{\delta\delta\theta}}{(k_1k_3)(k_3n)(k_{23}\bar{n})} \,, 
& &~~ I_{10}  =  \int  \frac{\DIFFL     \Phi^{nnn}_{\delta\delta\theta}}{(k_1k_3)(k_{12}n)(k_{13}\bar{n})} \,, \\
& I_{11}  = \int  \frac{\DIFFL     \Phi^{nnn}_{\delta\delta\theta}}{(k_1k_3)(k_3n)(k_{123}\bar{n})} \,,  
& &~~ I_{12}   = \int  \frac{\DIFFL    \Phi^{nnn}_{\delta\theta\delta}}{(k_1k_2)(k_{12}n)(k_2\bar{n})(k_{13}n)} \,, \\
& I_{13}   = \int  \frac{\DIFFL    \Phi^{nnn}_{\delta\theta\delta}}{(k_1k_2)(k_{12}n)(k_2\bar{n})(k_{123}\bar{n})}  \,,
&&~~ I_{14}  =  \int  \frac{\DIFFL     \Phi^{nnn}_{\theta\delta\delta}}{(k_1k_3)(k_{12}n)(k_{13}\bar{n})}  \,, \\
& I_{15}  = \int  \frac{\DIFFL     \Phi^{nnn}_{\theta\delta\delta}}{(k_1k_2)(k_{12}n)(k_{13}\bar{n})}  \,,
&&~~ I_{16}   = \int \frac{\DIFFL    \Phi^{nnn}_{\delta\theta\theta} }{(k_1k_2)(k_{23}\bar{n})}  \,, \\
& I_{17}   = \int \frac{\DIFFL    \Phi^{nnn}_{\delta\theta\theta} }{(k_1k_2)(k_2\bar{n})(k_{13}\bar{n})}  \,,
&&~~ I_{18}   = \int \frac{\DIFFL    \Phi^{nnn}_{\delta\theta\theta} }{(k_1k_2)(k_2\bar{n})(k_{123}\bar{n})}  \,, \\
& I_{19}   = \int \frac{\DIFFL    \Phi^{nnn}_{\delta\theta\theta} }{(k_1k_2)(k_2n)(k_{123}\bar{n})}  \,,
&&~~ I_{20}  =  \int \frac{ \DIFFL     \Phi^{nnn}_{\delta\theta\theta}}{(k_1k_3)(k_{12}n)(k_{123}\bar{n})}  \,, \\
& I_{21}  = \int \frac{ \DIFFL     \Phi^{nnn}_{\delta\theta\theta}}{(k_1k_2)(k_{23}n)(k_{123}\bar{n})}  \,,
&&~~ I_{22}  = \int \frac{ \DIFFL     \Phi^{nnn}_{\theta\delta\theta}}{(k_1k_3)(k_{12}n)(k_{123}\bar{n})}  \,.
\end{alignedat}
\label{eqn:mi_3g_nnn_b}
\end{equation} 
\end{footnotesize}%

\item additional integrals for $\int \DIFFL \Phi^{nnn}_{\theta\theta\theta } \;  \omega_{n  \bar n}^{(3),c}\;$:
\begin{footnotesize}
\begin{align}
& I_{23}  =  \int\! \frac{\DIFFL     \Phi^{nnn}_{\theta\delta\delta}}{(k_1k_2)(k_1k_3)}\,,
&&~~ I_{ 24}  =  \int\!  \frac{\DIFFL     \Phi^{nnn}_{\theta\delta\delta}}{(k_1k_2)(k_1 k_3 )(k_{12}\bar{n})}\,,\notag \\
& I_{25}  =  \int\!  \frac{\DIFFL     \Phi^{nnn}_{\theta\delta\delta}}{(k_1k_2)(k_1k_3)(k_{123} \bar n)}\,,
&&~~  I_{ 26}  =  \int\!  \frac{\DIFFL     \Phi^{nnn}_{\theta\delta\delta}}{(k_1k_2)(k_1 k_3 )(k_{12} n) (k_{13}\bar{n})}\,,\notag \\
& I_{27}  =  \int\!  \frac{\DIFFL     \Phi^{nnn}_{\theta\delta\delta}}{(k_1k_2)(k_1k_3)(k_{12} n) (k_{123} \bar n)}\,,
&&~~  I_{ 28}  =  \int\!  \frac{\DIFFL     \Phi^{nnn}_{\theta\delta\theta }}{(k_1k_2)(k_1 k_3 ) (k_3 \bar{n})}\,,\notag \\
& I_{29}  =  \int\!  \frac{\DIFFL     \Phi^{nnn}_{\theta\delta\theta}}{(k_1k_2)(k_1k_3)(k_{12} \bar n) (k_{3} \bar n)}\,,
&&~~  I_{ 30}  =  \int\!  \frac{\DIFFL     \Phi^{nnn}_{\theta\delta\theta }}{(k_1k_2)(k_1 k_3 ) (k_{12} n) (k_{13} \bar{n})}\,, \notag \\
& I_{31}  =  \int \! \frac{\DIFFL     \Phi^{nnn}_{\theta\delta\theta}}{(k_1k_2)(k_1k_3)(k_{123} \bar n) (k_{3} \bar n)}\,,
&&~~  I_{32}  =  \int \! \frac{\DIFFL     \Phi^{nnn}_{\theta\delta\theta }}{(k_1k_2)(k_1 k_3 ) (k_{3} n) (k_{123} \bar{n})}\,, \label{eqn:mi_3g_nnn_c} \\
& I_{33}  =  \int \! \frac{\DIFFL     \Phi^{nnn}_{\theta\delta\theta}}{(k_1k_2)(k_1k_3)(k_{12} n) (k_{123} \bar n)} \,,
&&~~ I_{34}  =  \int \! \frac{\DIFFL     \Phi^{nnn}_{\theta\delta\theta }}{(k_1k_2)(k_1 k_3 ) (k_{3} n) (k_{123} \bar{n})^2}\,, \notag \\
& I_{35}  =  \int \! \frac{\DIFFL     \Phi^{nnn}_{\delta\theta\theta}}{(k_1k_2)(k_1k_3)(k_2 n) (k_{123} \bar n) (k_3 \bar n) }\,,
&&~~ I_{36}  =  \int \! \frac{\DIFFL     \Phi^{nnn}_{\delta \theta \theta }}{(k_1k_2)(k_1 k_3 ) (k_{12} n) (k_2 \bar n) (k_{13} \bar{n})}\,, \notag \\
& I_{37}  =  \int \! \frac{\DIFFL     \Phi^{nnn}_{\delta \theta\theta }}{(k_1k_2)(k_1k_3)(k_{12} n) (k_{123} \bar n) (k_2 \bar n) }\,,
&&~~   I_{38}  =  \int \! \frac{\DIFFL     \Phi^{nnn}_{\theta \theta \delta  }}{(k_1k_2)(k_1 k_3 ) (k_{12} n) (k_2 \bar n) (k_{13} \bar{n})}\,, \notag \\
& I_{39}  =  \int \! \frac{\DIFFL     \Phi^{nnn}_{\theta\theta \delta }}{(k_1k_2)(k_1k_3)(k_{12} n)(k_{123} \bar n) (k_2 \bar n) } \notag \,.
\end{align}
\end{footnotesize}%

\item additional integral for $\int \DIFFL \Phi^{nnn}_{\theta\theta\theta } \;  \omega_{n  \bar n}^{(3),d}$ without $1/k_{123}^2$ propagator:
\begin{footnotesize}
\begin{equation}
I_{40}  = \!  \int \! \frac{\DIFFL     \Phi^{nnn}_{\theta\delta\theta}}{(k_1k_3)(k_1 n)(k_{123} \bar n) (k_{3} \bar n)}\,,
\label{eqn:mi_3g_nnn_d}
\end{equation}
\end{footnotesize}%

\item additional integrals for  $\int \DIFFL \Phi^{nnn}_{\theta\theta\theta } \;  \omega_{n  \bar n}^{(3),d}$ with $1/k_{123}^2$ propagator:
\begingroup
\allowdisplaybreaks
\begin{footnotesize}
\begin{align}
&I_{41} = \!\int\!\frac{\DIFFL\Phi^{nnn}_{\delta\delta\delta}}{k_{123}^2 (k_{12} n)} \,, 
&&~~I_{42} = \!\int\!\frac{\DIFFL\Phi^{nnn}_{\delta\delta\delta}}{k_{123}^2 (k_1 n)(k_2 n)} \,, \notag \\
&I_{43} = \!\int\!\frac{\DIFFL\Phi^{nnn}_{\delta\delta\theta}}{k_{123}^2 } \,, 
&&~~I_{44} = \!\int\!\frac{\DIFFL\Phi^{nnn}_{\delta\delta\theta}}{k_{123}^2 (k_3 \bar{n})} \,, \notag \\
&I_{45} = \!\int\!\frac{\DIFFL\Phi^{nnn}_{\delta\delta\theta}}{k_{123}^2 (k_{13} \bar{n})} \,, 
&&~~I_{46} = \!\int\!\frac{\DIFFL\Phi^{nnn}_{\delta\delta\theta}}{k_{123}^2 (k_{23} n)} \,, \notag \\
&I_{47} = \!\int\!\frac{\DIFFL\Phi^{nnn}_{\delta\delta\theta}}{k_{123}^2 (k_{13} \bar{n})^2} \,, 
&&~~I_{48} = \!\int\!\frac{\DIFFL\Phi^{nnn}_{\delta\delta\theta}}{k_{123}^2 (k_1 n)(k_3 \bar{n})} \,, \notag \\
&I_{49} = \!\int\!\frac{\DIFFL\Phi^{nnn}_{\delta\delta\theta}}{k_{123}^2 (k_1 k_3)(k_3 \bar{n})} \,, 
&&~~I_{50} = \!\int\!\frac{\DIFFL\Phi^{nnn}_{\delta\delta\theta}}{k_{123}^2 (k_1 k_3)(k_3 n)} \,, \notag \\
&I_{51} = \!\int\!\frac{\DIFFL\Phi^{nnn}_{\delta\delta\theta}}{k_{123}^2 (k_{13} \bar{n})(k_2 n)} \,, 
&&~~I_{52} = \!\int\!\frac{\DIFFL\Phi^{nnn}_{\delta\delta\theta}}{k_{123}^2 (k_{13} \bar{n})(k_1 n)} \,, \notag \\
&I_{53} = \!\int\!\frac{\DIFFL\Phi^{nnn}_{\delta\delta\theta}}{k_{123}^2 (k_1 k_3)(k_{23} \bar{n})} \,, 
&&~~I_{54} = \!\int\!\frac{\DIFFL\Phi^{nnn}_{\delta\delta\theta}}{k_{123}^2 (k_{12} n)(k_1 k_3)} \,, \notag \\
&I_{55} = \!\int\!\frac{\DIFFL\Phi^{nnn}_{\delta\delta\theta}}{k_{123}^2 (k_1 k_3)(k_{23} n)} \,, 
&&~~I_{56} = \!\int\!\frac{\DIFFL\Phi^{nnn}_{\delta\delta\theta}}{k_{123}^2 (k_{123} \bar{n})(k_1 n)} \,, \notag \\
&I_{57} = \!\int\!\frac{\DIFFL\Phi^{nnn}_{\delta\delta\theta}}{k_{123}^2 (k_{123} \bar{n})(k_3 n)} \,, 
&&~~I_{58} = \!\int\!\frac{\DIFFL\Phi^{nnn}_{\delta\delta\theta}}{k_{123}^2 (k_{123} \bar{n})(k_{13} n)} \,, \notag \\
&I_{59} = \!\int\!\frac{\DIFFL\Phi^{nnn}_{\delta\delta\theta}}{k_{123}^2 (k_1 k_3)(k_{23} \bar{n})^2} \,, 
&&~~I_{60} = \!\int\!\frac{\DIFFL\Phi^{nnn}_{\delta\delta\theta}}{k_{123}^2 (k_{123} \bar{n})^2(k_{13} n)} \,, \notag \\
&I_{61} = \!\int\!\frac{\DIFFL\Phi^{nnn}_{\delta\delta\theta}}{k_{123}^2 (k_{12} n)(k_1 k_3)(k_3 \bar{n})} \,, 
&&~~I_{62} = \!\int\!\frac{\DIFFL\Phi^{nnn}_{\delta\delta\theta}}{k_{123}^2 (k_{123} \bar{n})(k_1 n)(k_3 n)} \,, \notag \\
&I_{63} = \!\int\!\frac{\DIFFL\Phi^{nnn}_{\delta\delta\theta}}{k_{123}^2 (k_{123} \bar{n})(k_1 k_3)(k_3 n)} \,, 
&&~~I_{64} = \!\int\!\frac{\DIFFL\Phi^{nnn}_{\delta\delta\theta}}{k_{123}^2 (k_{123} \bar{n})(k_1 k_3)(k_3 \bar{n})} \,, \\
&I_{65} = \!\int\!\frac{\DIFFL\Phi^{nnn}_{\delta\delta\theta}}{k_{123}^2 (k_{123} \bar{n})(k_{13} n)(k_1 n)} \,, 
&&~~I_{66} = \!\int\!\frac{\DIFFL\Phi^{nnn}_{\delta\delta\theta}}{k_{123}^2 (k_{123} \bar{n})(k_1 k_3)(k_{23} n)} \,, \notag \\
&I_{67} = \!\int\!\frac{\DIFFL\Phi^{nnn}_{\delta\theta\theta}}{k_{123}^2 (k_2 \bar{n})} \,, 
&&~~I_{68} = \!\int\!\frac{\DIFFL\Phi^{nnn}_{\delta\theta\theta}}{k_{123}^2 (k_{13} \bar{n})} \,, \notag \\
&I_{69} = \!\int\!\frac{\DIFFL\Phi^{nnn}_{\delta\theta\theta}}{k_{123}^2 (k_{13} \bar{n})(k_1 k_2)} \,, 
&&~~I_{70} = \!\int\!\frac{\DIFFL\Phi^{nnn}_{\delta\theta\theta}}{k_{123}^2 (k_{12} n)(k_3 \bar{n})} \,, \notag \\
&I_{71} = \!\int\!\frac{\DIFFL\Phi^{nnn}_{\delta\theta\theta}}{k_{123}^2 (k_{123} \bar{n})(k_2 n)} \,, 
&&~~I_{72} = \!\int\!\frac{\DIFFL\Phi^{nnn}_{\delta\theta\theta}}{k_{123}^2 (k_{123} \bar{n})(k_{12} n)} \,, \notag \\
&I_{73} = \!\int\!\frac{\DIFFL\Phi^{nnn}_{\delta\theta\theta}}{k_{123}^2 (k_1 k_2)(k_2 \bar{n})(k_3 \bar{n})} \,, 
&&~~I_{74} = \!\int\!\frac{\DIFFL\Phi^{nnn}_{\delta\theta\theta}}{k_{123}^2 (k_1 k_2)(k_2 n)(k_3 \bar{n})} \,, \notag \\
&I_{75} = \!\int\!\frac{\DIFFL\Phi^{nnn}_{\delta\theta\theta}}{k_{123}^2 (k_{13} \bar{n})(k_2 n)(k_2 \bar{n})} \,, 
&&~~I_{76} = \!\int\!\frac{\DIFFL\Phi^{nnn}_{\delta\theta\theta}}{k_{123}^2 (k_{13} \bar{n})(k_1 k_2)(k_2 n)} \,, \notag \\
&I_{77} = \!\int\!\frac{\DIFFL\Phi^{nnn}_{\delta\theta\theta}}{k_{123}^2 (k_1 k_2)(k_{23} \bar{n})(k_2 \bar{n})} \,, 
&&~~I_{78} = \!\int\!\frac{\DIFFL\Phi^{nnn}_{\delta\theta\theta}}{k_{123}^2 (k_{123} \bar{n})(k_2 n)(k_2 \bar{n})} \,, \notag  \\
&I_{79} = \!\int\!\frac{\DIFFL\Phi^{nnn}_{\delta\theta\theta}}{k_{123}^2 (k_{123} \bar{n})(k_{12} n)(k_1 k_3)} \,, 
&&~~I_{80} = \!\int\!\frac{\DIFFL\Phi^{nnn}_{\theta\delta\theta}}{k_{123}^2 (k_{12} n)(k_1 k_3)(k_3 \bar{n})} \,, \notag \\
&I_{81} = \!\int\!\frac{\DIFFL\Phi^{nnn}_{\theta\delta\theta}}{k_{123}^2 (k_1 k_3)(k_1 n)(k_{23} \bar{n})} \,, 
&&~~I_{82} = \!\int\!\frac{\DIFFL\Phi^{nnn}_{\theta\delta\theta}}{k_{123}^2 (k_1 k_3)(k_1 n)(k_{23} \bar{n})^2} \,, \notag \\
&I_{83} = \!\int\!\frac{\DIFFL\Phi^{nnn}_{\delta\theta\theta}}{k_{123}^2 (k_{123} \bar{n})(k_1 k_3)(k_2 n)(k_3 \bar{n})} \,, 
&&~~I_{84} = \!\int\!\frac{\DIFFL\Phi^{nnn}_{\delta\theta\theta}}{k_{123}^2 (k_{123} \bar{n})(k_{13} n)(k_1 k_2)(k_2 \bar{n})} \,, \notag \\
&I_{85} = \!\int\!\frac{\DIFFL\Phi^{nnn}_{\delta\theta\theta}}{k_{123}^2 (k_{123} \bar{n})(k_1 k_3)(k_{23} n)(k_2 n)} \,, 
&&~~I_{86} = \!\int\!\frac{\DIFFL\Phi^{nnn}_{\theta\delta\theta}}{k_{123}^2 (k_{123} \bar{n})(k_1 k_3)(k_1 \bar{n})(k_{23} n)} \notag \,.
\label{eqn:mi_3g_nnn_d_k123}
\end{align}
\end{footnotesize}%
\endgroup  

\item additional integral for $\int \DIFFL \Phi^{nnn}_{\theta\theta\theta } \;  \omega_{n  \bar n}^{(3),d}$ with $1/k_{123}^2$ propagator and $1/\nu$ behaviour:
\begin{footnotesize}
\begin{equation}
J_{\nu}^{(d)}  = \!  \int \! \frac{\DIFFL     \Phi^{nnn}_{\theta\delta\theta} \, (k_1 n)^\nu (k_2 n)^\nu (k_3 n)^\nu }{k_{123}^2 (k_1k_3)(k_1 n)(k_{12} \bar n) (k_{3} \bar n)}\,.
\label{eqn:mi_3g_nnn_d_nu}
\end{equation}
\end{footnotesize}%

\end{itemize}

\begin{landscape}
\section{Matrices for the differential equations in Section~\ref{ssnum}}
\label{app:deq}

In Section~\ref{ssnum}, we discussed an example of a differential equation w.r.t. the auxiliary mass parameter.
The differential equation for eleven selected integrals is given in Eq.~(\ref{eq5.83});
it involves five eleven-by-eleven matrices whose explicit form is shown below. 

\vspace*{0.5cm}
  \begin{footnotesize}
\begin{align}
\boldsymbol{M}_{1} & =
\begin{pmatrix}
0 & 0 & 0 & 0 & 0 & 0 & 0 & 0 & 0 & 0 & 0 \\
1-3 \ep & -\ep & 0 & 0 & 0 & 0 & 0 & 0 & 0 & 0 & 0 \\
-\frac{(3 \ep-1) (6 \ep-1)}{2 \ep} & \frac{1}{2} (1-6 \ep) & 0 & 0 & 0 & 0 & 0 & 0 & 0 & 0 & 0 \\
0 & 0 & 0 & 0 & 0 & 0 & 0 & 0 & 0 & 0 & 0   \\
        0 & 2 \ep-1 & 0 & -\frac{1}{2} (5 \ep-1) (6 \ep-1) & 1-6 \ep & -3 \ep & 0 & 0 & 0 & 0 & 0
        \\
        0 & 1-2 \ep & 0 & \frac{1}{2} (5 \ep-1) (6 \ep-1) & 6 \ep-1 & 3 \ep & 0 & 0 & 0 & 0 & 0
        \\
        0 & 0 & 0 & 0 & 0 & 0 & 0 & 0 & 0 & 0 & 0
        \\
        \frac{1-3 \ep}{\ep} & \frac{1-3 \ep}{\ep} & 0 & \frac{(5 \ep-1) (6 \ep-1)}{2 \ep} & \frac{6 \ep-1}{\ep} & 3 & 0 & 0 & 0 & 0 & 0
        \\
        0 & 0 & 0 & 0 & 0 & 0 & 0 & 0 & 0 & 0 & 0
        \\
        0 & 0 & 0 & 0 & 0 & 0 & 0 & 0 & 0 & 0 & 0
        \\
        -\frac{2 (3 \ep-1)}{\ep} & -\frac{2 (3 \ep-1)}{\ep} & 0 & \frac{30 \ep^2-11 \ep+1}{\ep} & \frac{2 (6 \ep-1)}{\ep} & 6 & 0 & 0 & 0 & 0 & 0
    \end{pmatrix}
\end{align}

\begin{align}
    \boldsymbol{M}_{2}
    & =
    \begin{pmatrix}
        0 & 0 & 0 & 0 & 0 & 0 & 0 & 0 & 0 & 0 & 0
        \\
        0 & 0 & 0 & 0 & 0 & 0 & 0 & 0 & 0 & 0 & 0
        \\
        0 & 0 & 0 & 0 & 0 & 0 & 0 & 0 & 0 & 0 & 0
        \\
        0 & 0 & 0 & 0 & 0 & 0 & 0 & 0 & 0 & 0 & 0
        \\
        0 & 0 & 0 & 0 & 0 & 0 & 0 & 0 & 0 & 0 & 0
        \\
        0 & 0 & 0 & 0 & 0 & 0 & 0 & 0 & 0 & 0 & 0
        \\
        0 & 0 & 0 & 0 & 0 & 0 & 0 & 0 & 0 & 0 & 0
        \\
        \frac{3 \ep-1}{\ep} & \frac{3 \ep-1}{\ep} & \frac{1}{4} & -\frac{(5 \ep-1) (6 \ep-1)}{2 \ep} & -\frac{3 (6 \ep-1)}{8 \ep} & -3 & 0 & -3 \ep & 0 & 0 & 0
        \\
        0 & 0 & 0 & 0 & 0 & 0 & 0 & 0 & 0 & 0 & 0
        \\
        0 & 0 & 0 & 0 & 0 & 0 & 0 & 0 & 0 & 0 & 0
        \\
        \frac{2 (3 \ep-1) (7 \ep-2)}{\ep (5 \ep-2)} & \frac{11 \ep-4}{2 \ep} & 0 & \frac{-238 \ep^3+157 \ep^2-32 \ep+2}{\ep (5 \ep-2)} & \frac{-25 \ep^2+11 \ep-1}{\ep (5 \ep-2)} & \frac{34-97 \ep}{2 (5 \ep-2)} & \frac{7 \ep}{2} & \frac{1}{2} (2 \ep-1) & 4 (2 \ep-1) (3 \ep-1) & -\frac{1}{2} (2 \ep-1) (6 \ep-1) & \frac{1}{2} (1-8 \ep)
    \end{pmatrix}
    \\
    \boldsymbol{M}_{3}
    & =
    \begin{pmatrix}
        0 & 0 & 0 & 0 & 0 & 0 & 0 & 0 & 0 & 0 & 0
        \\
        3 \ep-1 & -2 \ep & 0 & 0 & 0 & 0 & 0 & 0 & 0 & 0 & 0
        \\
        \frac{(3 \ep-1) (6 \ep-1)}{2 \ep} & \frac{1}{2} (6 \ep-1) & \frac{1}{2} (-6 \ep-1) & 0 & 0 & 0 & 0 & 0 & 0 & 0 & 0
        \\
        0 & 0 & 0 & 0 & 0 & 0 & 0 & 0 & 0 & 0 & 0
        \\
        0 & 1-2 \ep & 0 & \frac{1}{2} (5 \ep-1) (6 \ep-1) & \frac{1}{2} (-4 \ep-1) & 3 \ep & 0 & 0 & 0 & 0 & 0
        \\
        0 & \frac{1}{2} (2 \ep-1) & 0 & -\frac{1}{2} (5 \ep-1) (6 \ep-1) & 0 & -4 \ep & 0 & 0 & 0 & 0 & 0
        \\
        0 & \frac{2 \ep-1}{2 \ep} & 0 & 0 & 0 & -2 & -3 \ep & 0 & 0 & 0 & 0
        \\
        0 & 0 & 0 & 0 & 0 & 0 & 0 & 0 & 0 & 0 & 0
        \\
        \frac{1}{2-5 \ep} & 0 & 0 & \frac{(4 \ep-1)^2}{(3 \ep-1) (5 \ep-2)} & 0 & 0 & 0 & 0 & 1-2 \ep & 0 & 0
        \\
        0 & 0 & 0 & 0 & 0 & 0 & 0 & 0 & 0 & \frac{1}{2} (1-6 \ep) & \frac{1}{2}
        \\
        -\frac{4 (3 \ep-1)}{5 \ep-2} & 0 & 0 & \frac{4 (4 \ep-1)^2}{5 \ep-2} & 0 & 0 & 0 & 0 & -4 (2 \ep-1) (3 \ep-1) & \frac{1}{2} (2 \ep-1) (6 \ep-1) & \frac{1}{2} (1-2 \ep)
    \end{pmatrix}
    \end{align}

\begin{align}
    \boldsymbol{M}_{4}
    & =
    \begin{pmatrix}
        0 & 0 & 0 & 0 & 0 & 0 & 0 & 0 & 0 & 0 & 0
        \\
        0 & 0 & 0 & 0 & 0 & 0 & 0 & 0 & 0 & 0 & 0
        \\
        0 & 0 & 0 & 0 & 0 & 0 & 0 & 0 & 0 & 0 & 0
        \\
        0 & 0 & 0 & 0 & -1 & 0 & 0 & 0 & 0 & 0 & 0
        \\
        0 & 0 & 0 & 0 & 0 & 0 & 0 & 0 & 0 & 0 & 0
        \\
        0 & 0 & 0 & 0 & 0 & 0 & 0 & 0 & 0 & 0 & 0
        \\
        0 & 0 & 0 & 0 & 0 & 0 & 0 & 0 & 0 & 0 & 0
        \\
        0 & 0 & 0 & 0 & 0 & 0 & 0 & 0 & 0 & 0 & 0
        \\
        0 & 0 & 0 & \frac{(4 \ep-1) (6 \ep-1)}{(2 \ep-1) (5 \ep-2)} & -\frac{2 (4 \ep-1)}{(3 \ep-1) (5 \ep-2)} & \frac{2 \ep (4 \ep-1)}{(2 \ep-1) (3 \ep-1) (5 \ep-2)} & -\frac{2 \ep^2}{(2 \ep-1) (3 \ep-1)} & 0 & 0 & 0 & 0
        \\
        0 & 0 & 0 & 0 & 0 & 0 & 0 & 0 & 0 & 0 & 0
        \\
        0 & 0 & 0 & 0 & \frac{2 (4 \ep-1)}{5 \ep-2} & 0 & 0 & 0 & 0 & 0 & 0
    \end{pmatrix}
    \end{align}

\begin{align}
    \boldsymbol{M}_{5}
    & =
    \begin{pmatrix}
        0 & 0 & 0 & 0 & 0 & 0 & 0 & 0 & 0 & 0 & 0
        \\
        0 & 0 & 0 & 0 & 0 & 0 & 0 & 0 & 0 & 0 & 0
        \\
        0 & 0 & 0 & 0 & 0 & 0 & 0 & 0 & 0 & 0 & 0
        \\
        0 & 0 & 0 & 0 & 0 & 0 & 0 & 0 & 0 & 0 & 0
        \\
        0 & 0 & 0 & 0 & 0 & 0 & 0 & 0 & 0 & 0 & 0
        \\
        0 & 0 & 0 & 0 & 0 & 0 & 0 & 0 & 0 & 0 & 0
        \\
        0 & 0 & 0 & 0 & 0 & 0 & 0 & 0 & 0 & 0 & 0
        \\
        0 & 0 & 0 & 0 & 0 & 0 & 0 & 0 & 0 & 0 & 0
        \\
        0 & 0 & 0 & 0 & -\frac{2 (4 \ep-1)}{(2 \ep-1) (5 \ep-2)} & 0 & 0 & 0 & 0 & 0 & 0
        \\
        0 & 0 & 0 & 0 & 0 & 0 & 0 & 0 & 0 & 0 & 0
        \\
        0 & 0 & 0 & 0 & 0 & 0 & 0 & 0 & 0 & 0 & 0
    \end{pmatrix}
\end{align}
\end{footnotesize}%

  \end{landscape}